\documentclass[12pt]{article}
\usepackage{graphicx}
\usepackage{multicol} 
\usepackage{color}
\definecolor{rosso}{cmyk}{0,1,1,0.4}
\definecolor{rossos}{cmyk}{0,1,1,0.55}
\definecolor{rossoc}{cmyk}{0,0.5,1,0.2}
\definecolor{blu}{cmyk}{1,1,0,0.3}
\definecolor{blus}{cmyk}{1,1,0,0.6}
\definecolor{blucc}{cmyk}{1,0.4,0.2,0}
\definecolor{viola}{cmyk}{0,1,0,0.6}
\definecolor{viola2}{cmyk}{0,1,0.2,0.6}
\definecolor{verde}{cmyk}{0.92,0,0.59,0.25}
\definecolor{verdec}{cmyk}{0.92,0,0.59,0.15}
\definecolor{verdes}{cmyk}{0.92,0,0.59,0.4}
\font\tenrsfs=rsfs10 at 12pt
\font\sevenrsfs=rsfs7
\font\fiversfs=rsfs5
\newfam\rsfsfam
\textfont\rsfsfam=\tenrsfs
\scriptfont\rsfsfam=\sevenrsfs
\scriptscriptfont\rsfsfam=\fiversfs
\def\mathscr#1{{\fam\rsfsfam\relax#1}}

\def\Ham{\mathscr{H}}

\newcommand{\riga}[1]{\noalign{\hbox{\parbox{\textwidth}{#1}}}\nonumber}

\newcommand{\fig}[1]{~\ref{fig:#1}}
\newcommand{\eq}[1]{~{\rm (\ref{eq:#1})}}
\newcommand{\sys}[1]{~{\rm (\ref{sys:#1})}}

\newcommand{\TeV}{\,{\rm TeV}}
\def\circa#1{\,\raise.3ex\hbox{$#1$\kern-.75em\lower1ex\hbox{$\sim$}}\,}

\newcommand{\NP}{Nucl. Phys.}
\newcommand{\PRL}{Phys. Rev. Lett.}
\newcommand{\PL}{Phys. Lett.}
\newcommand{\PR}{Phys. Rev.}
\newcommand{\beq}{\begin{equation}}
\newcommand{\eeq}{\end{equation}}
\newcommand{\diag}{\hbox{diag}\,}

\def\circa#1{\,\raise.3ex\hbox{$#1$\kern-.75em\lower1ex\hbox{$\sim$}}\,}
\makeatletter

\def\art{\@ifnextchar[{\eart}{\oart}}
\def\eart[#1]#2#3#4#5#6{{\rm #2}, {\em #3 \rm #4} {\rm (#6) #5} ({\em #1})}
\def\hepart[#1]#2{{\rm #2, \em#1}}
\newcommand{\oart}[5]{{\rm #1}, {\em #2 \rm #3} {\rm (#5) #4}}

\newcounter{alphaequation}[equation]
\def\thealphaequation{\theequation\hbox to
0.6em{\hfil\alph{alphaequation}\hfil}}
\def\eqnsystem#1{
\def\@eqnnum{{\rm (\thealphaequation)}}
\def\@@eqncr{\let\@tempa\relax \ifcase\@eqcnt \def\@tempa{& & &} \or
  \def\@tempa{& &}\or \def\@tempa{&}\fi\@tempa
  \if@eqnsw\@eqnnum\refstepcounter{alphaequation}\fi
\global\@eqnswtrue\global\@eqcnt=0\cr}
\refstepcounter{equation} \let\@currentlabel\theequation \def\@tempb{#1}
\ifx\@tempb\empty\else\label{#1}\fi
\refstepcounter{alphaequation}
\let\@currentlabel\thealphaequation
\global\@eqnswtrue\global\@eqcnt=0 \tabskip\@centering\let\\=\@eqncr
$$\halign to \displaywidth\bgroup \@eqnsel\hskip\@centering
$\displaystyle\tabskip\z@{##}$&\global\@eqcnt\@ne
\hskip2\arraycolsep\hfil${##}$\hfil& \global\@eqcnt\tw@\hskip2\arraycolsep
$\displaystyle\tabskip\z@{##}$\hfil
\tabskip\@centering&\llap{##}\tabskip\z@\cr}
\def\endeqnsystem{\@@eqncr\egroup$$\global\@ignoretrue} \makeatother

\newcommand{\MeV}{\,\hbox{\rm MeV}}
\newcommand{\eV}{\,\hbox{\rm eV}}

\newcommand{\gcm}{\,\mathrm{g}\,\mathrm{cm}^{-3}}
\newcommand{\erg}{\,\mathrm{erg}}
\renewcommand{\sec}{\,\mathrm{sec}}

\newcommand{\nue}{\nu_e}
\newcommand{\nueb}{\bar\nu_e}
\newcommand{\num}{\nu_\mu}
\newcommand{\numb}{\bar\nu_\mu}
\newcommand{\nut}{\nu_\tau}
\newcommand{\nutb}{\bar\nu_\tau}

\newcommand{\nusb}{\bar\nu_{\rm s}}
\newcommand{\nB}{n_{B}}
\newcommand{\GF}{G_{\rm F}}
\newcommand{\km}{\,\mathrm{km}}
\newcommand{\Feo}{\Phi^0_{\nueb}}
\newcommand{\Fmuo}{\Phi^0_{\numb}}
\newcommand{\Ftauo}{\Phi^0_{\nutb}}
\newcommand{\Fso}{\Phi^0_{\nusb}}
\newcommand{\Fe}{\Phi_{\nueb}}
\newcommand{\Fmu}{\Phi_{\numb}}
\newcommand{\Ftau}{\Phi_{\nutb}}
\newcommand{\Fs}{\Phi_{\nusb}}

\begin{document}
\thispagestyle{empty}
\setcounter{page}{0}

{hep-ph/0403158\hfill IFUP--TH/2004--2}
\vspace{1cm}

\begin{center}
{\LARGE \bf \color{rossos}
Probing oscillations into sterile neutrinos\\[3mm]
with cosmology, astrophysics and experiments
}\\[1cm]

{
{\large\bf M. Cirelli}$^1$,
{\large\bf G. Marandella}$^2$,
 {\large\bf A. Strumia}$^{3}$,  {\large\bf F. Vissani}$^{4}$
}  
\\[7mm]
{\it $^1$ Physics Department, Yale University, New Haven, USA} \\[3mm]
{\it $^2$ Scuola Normale Superiore and INFN, Pisa,
Italia } \\[3mm]
{\it $^3$ Dipartimento di Fisica dell'Universit\`a di Pisa,
Italia}\\[3mm]
{\it $^4$ INFN, Laboratori Nazionali del Gran Sasso, Italia}\\[1cm]
\vspace{1cm}
{\large\bf\color{blus} Abstract}
\end{center}
\begin{quote}
{\large\noindent\color{blus}
We perform a thorough analysis of oscillation signals generated by one extra sterile neutrino,
extending previous analyses done in simple limiting cases
and including the effects of established oscillations among active neutrinos. 
We consider the following probes:
solar, atmospheric, reactor and beam neutrinos,
Big-Bang Nucleosynthesis (helium-4, deuterium), Cosmic Microwave Background, 
Large Scale Structure, supernov\ae,
neutrinos from other astrophysical sources.
We find no evidence for a sterile neutrino in present data,
identify the still allowed regions, and study which future experiments can best probe them:
sub-MeV solar experiments,
more precise studies of CMB or BBN,
future supernova explosions, etc.
We discuss how the LSND hint is strongly disfavoured by the constraints of (standard) cosmology.}
\end{quote}

\vfill

\hrule

\smallskip

\noindent
$^1$ e-mail: Marco.Cirelli@yale.edu\\
$^2$ e-mail: Guido.Marandella@sns.it\\
$^3$ e-mail: Alessandro.Strumia@df.unipi.it\\
$^4$ e-mail: Francesco.Vissani@lngs.infn.it

\newpage

\tableofcontents

\setcounter{page}{1}

\section{Introduction}

High-energy colliders are the tool to discover new heavy particles with sizable couplings.
New  light particles with small couplings can be searched for in many different ways.
Neutral fermions with eV-scale mass, called `sterile neutrinos' in the jargon,
are typically stable enough to give effects in cosmology,
and can affect neutrino oscillation experiments.\footnote{Small fermions masses
are stable under quantum corrections.
Cosmology, astrophysics and neutrino experiments are also sensitive to
light scalars or vectors, which give
different kinds of signals.}
Both fields recently discovered new physics 
(in both cases around meV energies),
but so far no sterile neutrinos.

Today the most powerful cosmological probe of sterile effects
is standard Big-Bang Nucleosynthesis (BBN)~\cite{BBN}, which constrains
the number of thermalized neutrinos present at $T\sim 0.1\MeV$
to be $N_\nu = 2.5\pm 0.7$.
Since the uncertainty is controversial it is not clear
how much $N_\nu=4$ is disfavoured.

The established solar and atmospheric neutrino anomalies
seem produced by oscillations among the three SM neutrinos,
with at most minor contributions from possible extra sterile neutrinos.
The present 99\% C.L.
bounds on sterile mixing,
computed {\em assuming} that the initial active neutrino $|\nu_{\rm a}\rangle$ 
 oscillates with a large $\Delta m^2$ into
an energy-independent mixed neutrino $\cos\theta_{\rm s}|\nu'_{\rm a}\rangle + \sin\theta_{\rm s}|\nu_{\rm s}\rangle$,
are $\sin^2\theta_{\rm s} <0.25$ in solar oscillations (where $\nu_{\rm a}=\nu_e$)~\cite{etas}
and $\sin^2\theta_{\rm s} <0.21$ in atmospheric oscillations (where $\nu_{\rm a}=\nu_\mu$)~\cite{etaa}.

\bigskip

We relax these simplifying assumptions and study
the more general $4$-neutrino context.
This demands a remarkable effort, but it is an
important task for the present, for two reasons.
First, because recent discoveries
in cosmology and neutrino physics
are stimulating  new experiments that  will
study the new phenomena with the redundancy necessary to 
test the minimal models suggested by present data.
Second, because oscillations in extra light particles
are the natural extension of the emergent
massive neutrinos scenario.

Although we list hints that might be interpreted as  sterile neutrino effects (LSND~\cite{LSND}, certain
pieces of solar or atmospheric or BBN data, ...), 
we do not focus on any of them in particular.
Rather, we compute experimental capabilities and constraints on sterile oscillations
and compare them with capabilities  and constraints 
from various present and future cosmological and astrophysical probes.
Some important constraints are based on untested assumptions
and plagued by systematic uncertainties.
In such cases, rather than performing global fits, we identify and compute
the key observables,
trying to explain the basic physics in simple terms emphasizing the
controversial issues, so that the reader can judge.

\medskip

The paper is organized as follows.
In section~\ref{osc} we briefly review theoretical motivations:
as in the case of the known light particles, some fundamental reason
would presumably be behind the lightness of an extra sterile neutrino.
We also describe the non-standard  parametrization
of active/sterile mixing that we choose (because
more convenient and intuitive than standard parametrizations)
and describe the qualitatively different kinds
of spectra on which we will focus.

In section~\ref{cosmo} we study sterile effects in cosmology,
comparing the relative sensitivities of
two BBN probes (the helium-4 and deuterium abundances),
of Cosmic Microwave Background (CMB)
and of Large Scale Structures (LSS).
In section~\ref{solar} we study sterile oscillations in solar (and KamLAND) neutrinos.
In section~\ref{SN} we study sterile oscillations in SN1987A and future supernov\ae.
We also briefly discuss other less promising probes
(relic SN background, high-energy cosmic neutrinos,\ldots).
In section~\ref{atm} we study sterile oscillations in atmospheric and reactor neutrinos,
and in short and long-baseline neutrino beams (including LSND).

In order to avoid a unreadably long paper
each section is written in a concise way and 
contains a `Results' subsection, which can be read
skipping the other more technical parts.
From a computational point of view,
exploring 
4$\nu$ oscillations is $3\div 4$ orders of magnitude more
demanding than usual 2 or 3$\nu$ fits
and therefore requires significant improvements of usual techniques.
Subsections entitled `technical details' describe how this was achieved.
In particular we describe how we employ the density matrix formalism
in neutrino oscillation computations, and explain why even in simple
$3\nu$ situations it gives significant advantages with respect to 
more direct equivalent approaches.

In section~\ref{all} we conclude by summarizing and comparing
the different probes of sterile neutrinos.

\section{Active/sterile neutrino mixing}\label{osc}
We first introduce a non-standard useful parameterization of
the most generic $4\nu$ spectrum, and later present a brief review 
of theoretical models.

\subsection{Parameterization}
A generic $4\times 4$ Majorana neutrino mass matrix is described by 4 masses, 6 mixing angles and 6 CP-violating phases;
3 of them affect oscillations. 

In absence of sterile neutrinos, we denote by
$U$  the usual $3\times 3$ mixing matrix that relates
neutrino flavour eigenstates $\nu_{e,\mu,\tau}$ to active
neutrino mass eigenstates $\nu^{\rm a}_{1,2,3}$
as $\nu_\ell = U_{\ell i} \nu_i^{\rm a}$
($i=\{1,2,3\}$, $\ell = \{e,\mu,\tau\}$).
The extra sterile neutrino can mix with one arbitrary combination
of active neutrinos,  
$$\vec{n}\cdot \vec{\nu} = n_e \nu_e + n_\mu \nu_\mu + n_\tau \nu_\tau = 
n_1\nu_1^{\rm a} + n_2 \nu_2^{\rm a} + n_3 \nu_3^{\rm a}\qquad (n_i = U_{\ell i}n_\ell).$$
Mixing of the sterile neutrino can be therefore fully described
by a complex unit 3-versor $\vec{n}$
(containing two CP-violating phases)
and by one mixing angle $\theta_{\rm s}$.
With this parameterization the 4 neutrino mass eigenstates are
\begin{eqnsystem}{sys:param}
\label{eq:nu1234}
\left\{ \begin{array}{l}
\nu_4 =  \nu_{\rm s} ~\cos\theta_{\rm s} + n_\ell \nu_\ell ~\sin\theta_{\rm s} \cr
\nu_i = U_{\ell i}^* [\delta_{\ell\ell'}- n_{\ell}^* n_{\ell'}(1-\cos\theta_{\rm s}) ] \nu_{\ell'} - \sin\theta_{\rm s} 
n_\ell^* U_{\ell i}^*\nu_{\rm s}
\end{array}\right.\\
\riga{i.e.\ the $4\times 4$ neutrino mixing matrix $V$ as
that relates flavour to mass eigenstates as $\nu_{e,\mu,\tau,s} = V\cdot \nu_{1,2,3,4}$ is}\\
 \label{eq:U}
V = \pmatrix{1-(1-\cos\theta_{\rm s}) \vec{n}^* \otimes \vec{n} &
\sin\theta_{\rm s} \vec{n}^* \cr
-\sin\theta_{\rm s} \vec{n}   & \cos\theta_{\rm s}} \times \pmatrix{U&0\cr 0 & 1} \\
\riga{or, more explicitly,}\\
 V = \bordermatrix{& \nu_{i} & \nu_4 \cr
\nu_{\ell} & U_{\ell i} - n_i n_\ell^*(1-\cos\theta_{\rm s})
& n_\ell^* \sin\theta_{\rm s} \cr 
\nu_{\rm s} & - n_i \sin\theta_{\rm s} & \cos\theta_{\rm s}}.
\end{eqnsystem}
In summary, $\vec{n}$ identifies which combination of active neutrinos mixes
with $\nu_{\rm s}$ with mixing angle $\theta_{\rm s}$.

\begin{figure}
$$\includegraphics[width=7cm]{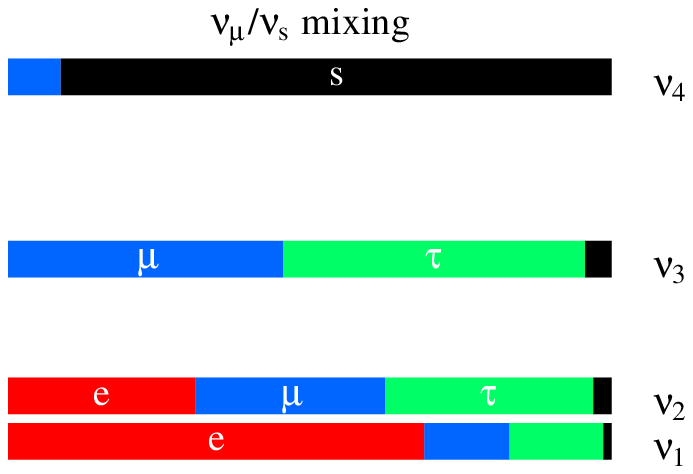}\qquad
\includegraphics[width=7cm]{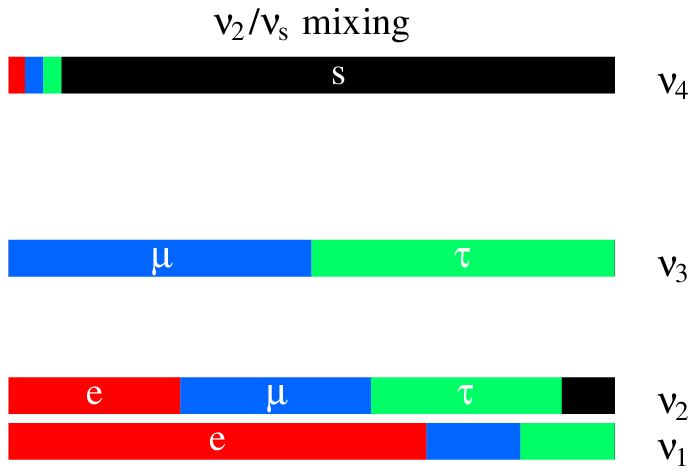}$$
\caption{\label{fig:spettrias}\em {\bf 
Basic kinds of four neutrino mass spectra}. 
Left: sterile mixing with a flavour eigenstate ($\nu_\mu$ in the picture).
Right: sterile mixing with a mass eigenstate ($\nu_2$ in the picture).
}
\end{figure}

In order to understand how neutrinos oscillate in the generic case,
it is convenient to focus on two different kinds of limiting cases,
pictorially exemplified in fig.\fig{spettrias}:
\begin{itemize}
\item  {\bf Mixing with a flavour eigenstate}  (fig.\fig{spettrias}a):
$\vec{n}\cdot\vec{\nu}=\nu_\ell$ ($\ell =e$ or $\mu$ or $\tau$).
The sterile neutrino oscillates into a well defined flavour
at 3 different $\Delta m^2$
(which cannot all be smaller than the observed splittings $\Delta m^2_{\rm sun,atm}$).

\item {\bf Mixing with a mass eigenstate}  (fig.\fig{spettrias}b):
$\vec{n}\cdot\vec{\nu}=\nu_i$ ($i=1$ or 2 or 3).\footnote{From here on, 
we omit the superscript `a' (that stands for 
active): it should be clear that whenever we speak of $\nu_2/\nu_{\rm s}$ 
mixing, this is just a short-hand for $\nu_2^{\rm a}/\nu_{\rm s}$.}
The sterile neutrino oscillates into a neutrino of mixed flavour
at a single $\Delta m^2$, which can be arbitrarily small.
\end{itemize}
We think that our parametrization of sterile mixing, in eq.~(\ref{sys:param}),
makes physics more transparent than
other frequently used choices\footnote{When studying sterile mixing with a flavour eigenstate
our expression is directly related to the `standard' parameterization
$$ V= R_{34}R_{24}R_{14}\cdot U_{23}U_{13} U_{12}$$
where $R_{ij}$ represents a rotation in the $ij$ plane by angle $\theta_{ij}$
and $U_{ij}$ a complex rotation in the $ij$ plane.
 $\theta_{14}$ or $U_{e4}$ gives rise to $\nu_e/\nu_{\rm s}$ mixing,
 $\theta_{24}$ or $U_{\mu 4}$ to $\nu_\mu/\nu_{\rm s}$ mixing,
and  $\theta_{34}$ or $U_{\tau 4}$ to $\nu_\tau/\nu_{\rm s}$ mixing.

The above  `standard' parameterization becomes inconvenient when
studying mixing with a mass eigenstate.
In such a case our parameterization is directly related
to the alternative `standard' parameterization appropriate for this case,
$$ V = U_{23}U_{13}U_{12}\cdot  R_{34}R_{24}R_{14}$$
Now $\theta_{i4}$ gives rise to $\nu_i/\nu_{\rm s}$ mixing.
Our  parameterization instead is convenient because it remains simple in both cases.}.

The  oscillation probabilities
among active neutrinos
in the limit where the active/sterile mass splitting dominates,
and active/active mass splittings can be neglected, are
\beq\label{eq:Pij} P(\nu_\ell \to \nu_{\ell'})=P(\bar\nu_\ell \to\bar\nu_{\ell'})
=\left\{\begin{array}{ll}
1-4|V_{\ell 4}^2|(1-|V_{\ell 4}^2|)  \sin^2 ({\Delta m^2_{14} L}/{4E_\nu})    &  \hbox{for $\ell=\ell'$}  \\
4 |V_{\ell 4}^2||V_{\ell' 4}^2| \sin^2 ({\Delta m^2_{14} L}/{4E_\nu} )     &    \hbox{for $\ell\neq\ell'$}
\end{array}\right.\eeq
and in our parametrization $V_{\ell 4} = n_\ell^* \sin\theta_{\rm s}$.

\medskip

Older papers studied 
active/sterile mixing in 2 neutrino approximation.
In such a case $\theta_{\rm s} = \pi/2$ gives no oscillation effect.
On the contrary, in the full $4$ neutrino case
 $\theta_{\rm s} =\pi/2$ swaps the sterile neutrino
with one active neutrino.
(e.g.\   $\nu_\mu$ in  fig.\fig{spettrias}a or
$\nu_2$ in fig.\fig{spettrias}b,
if $\theta_{\rm s}$ were there increased up to $\pi/2$)
affecting solar and atmospheric oscillations in an obvious way.
Therefore large active/sterile mixing is excluded by 
experiments for all values of $\Delta m^2_{i4} \equiv m^2_4-m^2_i$ 
(with one exception:
the sterile neutrino mixes with a mass eigenstate $\nu_i$
 and the two states form a quasi-degenerate pair.
This structure arises naturally in certain models~\cite{SterileB}).

In order to explore a more  interesting slice of parameter space
when considering sterile mixing with a mass eigenstate $\nu_i$,
for $\theta_{\rm s}>\pi/4$ we modify the spectrum of neutrino masses
and replace $(m_i^2,m_4^2)$ with $(2m_i^2- m_4^2, m_i^2)$.
In such a way, the mostly active state always keeps the same squared mass
(that we fix to its experimental value), so that
in the limit $\theta_{\rm s}= \pi/2$
the sterile neutrino gives no effect rather than giving an already excluded effect.
Physically, in our $\nu_{\rm s}/\nu_i$ plots the 
mostly sterile neutrino is heavier (lighter)
than the mass eigenstate $\nu_i$ to which it mixes when  
$\theta_{\rm s}<\pi/4$ ($\theta_{\rm s}>\pi/4$).
When studying mixing with a flavour eigenstate we do not
modify the spectra at $\theta_{\rm s}>\pi/4$ in order to obtain some other
experimentally allowed configuration.
For this reason, we restrict such plots to $\theta_{\rm s}<\pi/4$.

\medskip

We do not consider `$2+2$' neutrino mixing,
namely two neutrino couples separated by a mass splitting much larger than
$\Delta m^2_{\rm sun,atm}$.
In fact this spectrum does not reduce to
active-only oscillations in any limiting case so that
 sterile effects are always sizable,
and present experiments already exclude this possibility~\cite{2+2}.
When the separation among the two couples is comparable to
$\Delta m^2_{\rm sun,atm}$, `$2+2$' is no longer
a special case qualitatively different from `$3+1$'. 

We assume that active neutrinos have normal hierarchy, $\Delta m^2_{23}>0$.
Finally, we assume $\theta_{13}=0$.
We verified that using $\theta_{13}\sim 0.2$,
the maximal value allowed by present experiments,
leads to minor (in some cases) or no (in other cases) modifications,
that we do not discuss.
Measuring $\theta_{13}$ and 
discovering sterile effects will likely be two independent issues
(however it is curious to note that both could first manifest as disappearance  of reactor $\bar\nu_e$).

These assumptions are made because we consider three active neutrinos
with normal hierarchy as the most plausible spectrum, and view
inverted hierarchy, large $\theta_{13}$ and sterile neutrinos as possible surprises:
we here study the latter one.

\subsection{Theory}
The relevant terms in the SU(2)$_L$-invariant  effective Lagrangian that describes active
neutrinos $\nu$ together with extra light singlet fermions $\nu_R$ are
\beq\label{eq:Lsterile} \frac{m_{LL}}{2v^2} (LH)^2+ \frac{m_{RR}}{2}\nu_R^2 + \frac{m_{LR}}{v} \nu_R LH + \hbox{h.c.}\eeq
H is the higgs doublet with vacuum expectation value $(0,v)$.
The first dimension-5 operator gives Majorana $\nu$ masses $m_{LL}$ and is naturally small
if lepton number is broken at a high-energy scale.
The second term gives Majorana $\nu_R$ masses $m_{RR}$,
and the third term Dirac $\nu_L\nu_R$ masses $m_{LR}$:
one needs to understand why $m_{LR}$ and $m_{RR}$ are small.

A few theoretically favoured patterns emerge from rather 
general naturalness  considerations.
We consider the most generic mass matrix with $LL$, $RR$ and $LR$ mass terms.
If $m_{LL}$ dominates one obtains light sterile neutrinos with mass $m_{\rm s}\ll m_{\rm a}$ and
active/sterile mixings
$\theta_{\rm s}^2 \sim m_{\rm s}/m_{\rm a}$.
If $m_{RR}$ dominates  sterile neutrinos are heavy with $\theta_{\rm s}^2 \sim m_{\rm a}/m_{\rm s}$.
If $m_{LR}$ dominates one obtains quasi-Dirac neutrinos that split into couples.
These `more likely' regions can be represented as lines in the
logarithmic $(\tan^2\theta_{\rm s}, \Delta m_{i4}^2)$ plane,
that we will use to present our results.

\medskip

A new light particle would probably be a discovery of fundamental importance,
because it lightness is likely related to some fundamental principle,
as it is the case for the known light particles,  the photon, the neutrinos and the graviton.
Attempts of guessing physics beyond the SM
from first principles motivate a number of fermions which might
have $\TeV^2/M_{\rm Pl}$ masses and
behave as sterile neutrinos.
A few candidates are, in alphabetic order,
axino, branino, dilatino, familino, Goldstino, Majorino, modulino,
radino.
These ambitious approaches so far do not give
useful predictions on the flavour parameters in the effective Lagrangian of eq.\eq{Lsterile}.
Therefore one needs to consider more specific ad-hoc models~\cite{Sterile}.

Unification of matter fermions into SO(10) 16 or $E_6$ 27 representations 
predicts extra singlets, which however
generically receive GUT-scale masses.
It is easy to invent ad-hoc discrete or continuous symmetries that keep a fermion light~\cite{Sterile}.
One might prefer to use only ingredients already present in the SM.
For example, the extra fermions can be forced to be light assuming that they are chiral
under some extra gauge symmetry (that could possibly become non perturbative at some
QCD-like scale, and give composite sterile neutrinos)~\cite{Sterile}.
Alternatively, the extra fermions may be light for
the same reason why neutrinos are light in the SM~\cite{Sterile}.
Following this point of view up to its extreme, one can add to the SM a set of `mirror particles',
obtaining 3 sterile neutrinos~\cite{Sterile}.
In presence of Planck-suppressed corrections that mix the two sectors,
mirror neutrinos with the same mass as SM neutrinos
give rise to quasi-maximal mixing angles between 
SM neutrino mass eigenstates and sterile neutrinos, splitted by
$\Delta m^2 \sim m_i v^2/M_{\rm Pl}\sim 10^{-8}\eV^2$
(this kind of $\nu_2/\nu_{\rm s}$ effects are disfavoured by solar data).
Mirror neutrinos with different masses from SM neutrinos
can give detectable  $\nu_1/\nu_{\rm s}$ oscillation effects.
The two sectors might instead communicate because 
coupled to the same heavy see-saw neutrinos:
this gives massless states which can be mostly sterile.

\section{Sterile effects in cosmology}\label{cosmo}
Present cosmological data seem compatible with 
the following minimal assumption (see e.g.~\cite{WMAP}): 
primordial perturbations are generated by
minimal inflation (flat space, Gaussian perturbations with
flat spectral index $n_s=1$)
and evolve as dictated by general relativity (with a small cosmological constant)
and by the Standard Model (adding some unknown Cold Dark Matter).
Global analyses performed under this {\em assumption} give
precise determinations of the cosmological parameters
and bounds on non standard properties of neutrinos.
In particular, the sensitivity to neutrino masses and to oscillations into extra sterile neutrinos
is competitive with direct experimental bounds.


Present bounds are unsafe  because
based on assumptions which have been only partially tested.
At the level of precision needed to derive bounds on sterile neutrinos stronger than experimental bounds these assumptions stand practically untested.
Since these bounds come from a few key measurements
(rather than from a redundant set of different observables)
compensations among different kinds of new physics are not unnatural.
For instance, the dominant BBN probe~\cite{BBN}, the helium-4 abundancy~\cite{He4},
 is plagued by controversial  systematic uncertainties, and
presently it is not clearly incompatible with a fourth thermalized sterile neutrino~\cite{NnuBBN}

Therefore we do not try to attribute a precise probabilistic meaning to cosmological bounds
by performing global fits.
We prefer to identify and compute the key observables.
This allows to present the underlying physics in a simple and critical way
(while global fits would indirectly involve other cosmological parameters and data).
The implications of present or future measurements can be read from our plots.
Cosmology is interesting not only because gives indications today,
but also because it will allow powerful future searches.

\medskip

BBN probes the total energy density at $T \sim (0.1 \div 1)\MeV$ 
(dominantly stored in electrons, photons and neutrinos according to the SM)
and is also directly sensitive to reactions involving neutrinos (e.g.\ $\bar\nu_e p \leftrightarrow \bar{e} n$). Given a few input parameters 
(the effective number $N_\nu$ of thermalized relativistic species,
the baryon asymmetry $n_B/n_\gamma=\eta$, 
and possibly the $\nu_\ell /\bar\nu_\ell$ lepton asymmetries)
BBN successfully predicts the abundances of several light nuclei~\cite{BBN}. 
Its non trivial success strongly indicates that primordial BBN really happened.
Today $\eta$ is best determined within minimal cosmology by CMB data to be $\eta = (6.15 \pm 0.25) 10^{-10}$~\cite{WMAP}. 
Thus, neglecting the lepton asymmetries (which is an excellent approximation
unless they are much larger than the baryon asymmetry)
one can use the observations of primordial abundances to test if $N_\nu=3$
as predicted by the SM.

Today the $^{4}$He abundancy~\cite{He4} is the most sensitive probe of $N_\nu$.
We study also the deuterium abundancy, that might have brighter prospects of future improvements~\cite{DeutQuas}.
For arbitrary values of $N_\nu$ around the SM value of 3, BBN predicts~\cite{BBN}
\begin{eqnsystem}{sys:HeD}
Y_p &\simeq& 0.248+0.0096\ln\frac{\eta}{6.15~10^{-10}} + 0.013 (N_\nu^{^4{\rm He}}-3),\\
 \frac{Y_{\rm D}}{Y_{\rm H}} &\simeq &(2.75\pm 0.13)~10^{-5}~\frac{1+0.11~(N_\nu^{\rm D}-3)}{(\eta/6.15~10^{-10})^{1.6}}
, \end{eqnsystem}
where $Y_p  \equiv n_{^4{\rm He}}/n_B$.
Both $Y_p$ and $Y_{\rm D}$  are plagued
 by controversial systematic uncertainties.
 We do not enter into these issues and refer the reader to~\cite{Dolgov,review,NnuBBN,He4,DeutQuas}.
Adopting conservative estimates we get
\beq\begin{array}{lcl}\label{eq:pD}
 Y_p = 0.24\pm 0.01     &\Rightarrow& N_\nu^{^4{\rm He}} \simeq  2.4\pm0.7,    \\
 \displaystyle
 \frac{Y_{\rm D}}{Y_{\rm H}} =  (2.8 \pm 0.5)\,10^{-5}     &\Rightarrow&   N_\nu^{\rm D}\simeq 3\pm 2.
\end{array}\eeq
In order to safely test and possibly rule out $N_\nu=4$ it is necessary to either
$i)$ somewhat improve the determination of $Y_p$, or to
$ii)$ improve on $Y_{\rm D}$, somewhat  improve on
$\eta$ and on the theoretical uncertainty on $Y_{\rm D}$.

In our computation, we assume standard cosmology plus a sterile neutrino,
and conservatively assume it has zero initial abundancy  at $T\gg \MeV$.
Oscillations produce sterile neutrinos~\cite{bbnosc}.
This is a rather robust phenomenon:
it is difficult to modify cosmology in order to avoid 
production of sterile neutrinos while keeping the success of BBN.
In fact sterile neutrinos are dominantly produced at $T\sim \MeV$,
simultaneously or after neutrino decoupling.

However, a neutrino asymmetry $\eta_\nu$ 8 orders of magnitude larger than $\eta$ in baryons
is a significant extra parameter: it affects the $\nu_e$ abundancy and consequently 
the $n/p$ ratio at freeze-out,
and finally the primordial abundances.
Also,  a relatively large lepton asymmetry, around $10^{-5}$, 
gives extra MSW effects which can 
suppress active/sterile oscillations removing cosmological signals~\cite{etarimuove}.
More generically, allowing a non vanishing $\eta_\nu$
any $N_\nu$ is compatible with the measurement of 
the helium-4 abundancy, 
just because a single measurement  cannot fix two parameters ($N_\nu$ and $\eta_\nu$).
It is important to study how the  situation improves when also $Y_{\rm D}$
will be precisely measured.
For example increasing $N_\nu $ from 3 to 4 increases 
$Y_{\rm D}$ by $11\%$ and
$Y_p$ by $5\%$.
If the latter effect were compensated by a large neutrino asymmetry,
$Y_{\rm D}$ still remains about $8\%$ higher.
Numbers can be somewhat different, depending on how $N_\nu$ gets dynamically
increased by sterile oscillations, which are also directly affected by a
large neutrino asymmetry in an important way~\cite{etarimuove,asymm}.
However the general point remains.
Of course one can make any $N_\nu$ allowed by just introducing
more than one free parameter;
but measuring two or more observables would make hard to believe
that new physics effects cancel among each other in all cases.

\medskip

In the following we stick to standard cosmology.

For each choice of oscillation parameters, 
we follow the evolution with temperature of neutrino abundances
and compute how  $Y_p$ and $Y_{\rm D}$ are modified. 
For easy of presentation, we convert their values into effective numbers of neutrinos, 
 $N_\nu^{^4{\rm He}}$ and $N_\nu^{\rm D}$,
univocally defined by the inversion of eq.\sys{HeD}.
These parameters do not necessarily lie between $3$ and $4$
and are employed just as a useful way of presenting our final results.

\medskip

 We also compute another effective number of neutrinos,
 that parameterizes the total energy density in relativistic species at photon decoupling as\footnote{Including small effects of spectral distortions
the precise SM prediction for the effective number of neutrinos is $N_\nu^{\rm CMB} = 3.04$, 
but we can ignore this subtlety.}
 \beq\label{eq:CMB}
 \rho_{\rm relativistic} = \rho_\gamma\bigg[1 + \frac78\bigg(\frac{4}{11}\bigg)^{4/3} N_\nu^{\rm CMB}\bigg].\eeq
 This quantity (together with other cosmological parameters) determines
 the pattern of fluctuations of the CMB measured by WMAP (and other experiments).
 Neutrinos affect CMB in various ways~\cite{NuCDM};
 in the future studying how 
 neutrino free-streaming shifts the acoustic peaks
 should offer a clean way of directly counting neutrinos.
Global fits at the moment imply~\cite{CMBfits}
\beq N_\nu^{\rm CMB} \approx 3\pm2\eeq
somewhat depending on which priors and on which data are included in the fit.
Future data might start discriminating $3$ from $4$ neutrinos.

\medskip

Finally, neutrinos can be studied looking at distribution of galaxies because
massive neutrinos move without interacting,
making galaxies less clustered~\cite{boundMnu}.
The effect depends on two parameters, that in absence of sterile neutrinos
are both determined by neutrino masses:
$1)$ the temperature at which neutrinos become non relativistic, 
$T_\nu\sim m_\nu/3 $ (so that active neutrinos operated when
the horizon of the universe had the size that clusters of galaxies have now);
$2)$ the energy density in neutrinos, $\Omega_\nu h^2$,
that determines how large is the effect of neutrinos
(neutrinos give at most a minor correction).
As usual, the parameter $h$ is $H_{\rm today}/(100 \hbox{km/s\,Mpc})$.

Global analyses of cosmological data are usually reported as
a bound on $\Omega_\nu$, assuming the standard correlation with $T_\nu$~\cite{WMAP,boundMnu}.
However, in presence of a non-thermal population of sterile neutrinos these two parameters
are no longer universally related.
E.g.\ there could be a little number of sterile neutrinos
with few eV mass (heavy neutrinos affect also CMB,
 behaving as cold dark matter).
 This scenario has not yet been compared with present data (see~\cite{HR} for closely related work).
 We assume that the bound can be approximated
 with the standard one on the following quantity:\footnote{We are neglecting CP-violation, 
 and assuming that neutrinos
 and anti-neutrinos have equal density matrices.}
  \beq\label{eq:LSS}\Omega_\nu h^2=\frac{\hbox{Tr} [m\cdot \rho]}{93.5\eV}\eeq
where $m$ is the $4\times4$ neutrino mass matrix and $\rho$ is the $4\times4$ 
neutrino density matrix, as discussed below.
We approximate the present bound with $\Omega_\nu h^2< 0.01$
(e.g.\ the WMAP global fit gives $\Omega_\nu h^2<0.76 ~10^{-2}$ at $95\%$ C.L.\
in the standard case~\cite{WMAP}. 
The bound becomes slightly weaker if $N_\nu^{\rm CMB}=4$,
or if more conservative priors or estimates of systematic uncertainties are adopted~\cite{boundMnu}).
The atmospheric mass splitting guarantees $\Omega_\nu\circa{>} 0.5~10^{-3}$ so that
future attempts to reach this level of sensitivity are guaranteed.
We assume that a $0.001$ sensitivity in $\Omega_\nu h^2$ will be reached.

\subsection{Technical details}
Before discussing results, we present the main technical details.
In order to compute the observables discussed above one needs to
set up the network of relevant Boltzmann equations
and study neutrino oscillations in the early universe~\cite{OscUniverse,DolgovReview}.
We use our BBN code that includes all main effects, 
relying on more accurate public codes only to precisely fix the central SM values.
This is a complicated computation because many processes
proceed at $T\sim \MeV$: active (and maybe sterile) 
neutrino oscillations, neutrino decoupling,
neutron decay, electron decoupling and finally nucleosynthesis.
The various processes have been accurately studied in the past.
The extension from 2 neutrinos oscillations
to 4 neutrinos oscillations
does not involve new ingredients,
but rewriting and implementing old ones
in an appropriate way is not completely trivial
due to the presence e.g.\ of oscillations at different frequencies.

Brute force alone would not allow
to explore the key observables in a vast parameter space:
it is necessary to employ accurate approximations
that one can invent understanding the physics of BBN.
Neglecting spectral distortions
possibly induced by active/sterile oscillations,
the Boltzmann equation for the $4\times 4$ 
neutrino (and anti-neutrino) density matrix is~\cite{OscUniverse,DolgovReview}
\beq \label{eq:dotrho}\dot\rho =  zZH \frac{d\rho}{dz} = i [\Ham,\ \rho] - \{\Gamma,(\rho-\rho^{\rm eq})\}.\eeq
where  $H$ is the Hubble constant at temperature $T$, $z = m_e/T$.
The factor
 $Z = -3\; \textrm{d} \ln z / \textrm{d} \ln s$ (where $s$ is the entropy density)
differs from 1 when the temperature of the universe does not decrease
as the inverse of its comoving radius,
namely during electron decoupling ($z\sim 1$) and when  sterile neutrinos thermalize.
In the flavour basis~\cite{OscUniverse}
\beq \Ham =\frac{m m^\dagger}{2E_\nu }- \frac{7\pi^3\,T_{\nu }\,{{\alpha }_2}}
  {15\,M_W^4} \bigg[T_\nu^4\cos^2\theta_{\rm W} \; \diag(\rho_{ee},\rho_{\mu \mu},\rho_{\tau \tau},0) + 2 \; T^4 \; \diag(1,0,0,0)\bigg]\eeq
We can neglect contributions from the off-diagonal elements of $\rho$~\cite{OscUniverse}.
 The usual MSW effect~\cite{MSW} gives an additional subdominant term
 with different sign for neutrinos and antineutrinos.
 The average over the neutrino energy spectra is performed
 using a Fermi-Dirac distributions, and therefore 
 neglecting spectral distortions possibly caused by oscillations.

In eq.\eq{dotrho} we use the standard `anticommutator' approximation for the
 collision terms $\Gamma$, that describe weak $\nu e$ and $\nu\nu$ interactions
 that tend to thermalize neutrinos, driving 
 their matrix density to its thermal equilibrium value, 
$\rho^{\rm eq} = {\rm diag}(1,1,1,0)$.
A detailed comparison with the full equations~\cite{DolgovReview}
reveals that they are accurately mimicked by inserting the following
values of the damping coefficients.
In the equations for the off-diagonal components of $\rho$, we insert the
 total scattering rate~\cite{DolgovReview}
$$
  \Gamma_{\rm tot}  \approx  3.6 \; G_{\rm F}^2 \; T^5 \quad \textrm{for $\nu_e\qquad$ and}\qquad
  \Gamma_{\rm tot}  \approx  2.5 \; G_{\rm F}^2 \; T^5 \quad\textrm{for $\nu_{\mu,\tau}$}
$$
because all scatterings damp the coherent interference
between different flavours.
 In the equations for the diagonal components of  $\rho_{ii}$,
we insert the annihilation rate~\cite{DolgovReview}
$$
  \Gamma_{\rm ann} \approx 0.5 \; G_{\rm F}^2 \; T^5 \quad \textrm{for $\nu_e\qquad$ and}\qquad
  \Gamma_{\rm ann}\approx  0.3 \; G_{\rm F}^2 \; T^5 \quad\textrm{for $\nu_{\mu,\tau}$}
$$
since annihilations are needed to change the number of neutrinos.
However this procedure~\cite{DolgovReview} introduces an artificial choice of basis,
giving equations which are no longer invariant under
rotations in the $(\mu,\tau)$ sector.
In order to correctly maintain important coherencies among
$\nu_\mu$ and $\nu_\tau$, we introduce the distinction
between $\Gamma_{\rm tot}$ and $\Gamma_{\rm ann}$
in the $\nu_\mu \pm \nu_\tau$ basis
(assuming maximal atmospheric mixing; otherwise the generalization is immediate).

\medskip

After determining neutrino evolution we can study the relative $n/p$ abundancy,
that evolves according to~\cite{BBN,bbnosc}
$$
\dot{r} = zHZ\frac{dr}{dz} = \Gamma_{p\to n} (1-r)- r \Gamma_{n\to p}\qquad
r = \frac{n_n}{n_n+n_p} $$
where $\Gamma_{p\to n}$ is the total
$pe\bar\nu_e\to n$, $pe\to n\nu_e$ and $p\bar\nu_e\to n\bar{e}$
reaction rate
(in thermal equilibrium the inverse process would satisfy
$\Gamma_{n\to p} = \Gamma_{p\to n} e^{(m_n-m_p-m_e)/T}$). 
The production of sterile neutrinos affects $n/p$ by~\cite{bbnosc} 
$1)$ increasing the Hubble parameter $H$;
$2)$ modifying the $\Gamma_{p\to n},\Gamma_{n\to p}$ rates, 
if the $\nu_e$ population is depleted by oscillations. 

\medskip

Finally a network of Boltzmann equations describes how
electroweak, strong and electromagnetic processes
control the evolution of the various nuclei:
$p$, $n$, D, T, $^3$He, $^4$He,\ldots
Rather than recalling here the main features of these equations,
we just state (without explanation) the approximation we use
(see also~\cite{BBNapprox}).
At a sufficiently low temperature $T^*\sim 0.08\MeV$ almost all neutrons wind up in $^4{\rm He}$, 
so that its mass abundancy is given by
$ Y_p\simeq 2r(T^*)$ with $T^*$ obtained solving
$$180 H = \Gamma_{{\rm DD}\to p{\rm T}} (\Gamma_{pn\to {\rm D}\gamma}/\Gamma_{{\rm D}\gamma\to pn})^2.$$
The precise numerical value is fixed in such a way that in the SM case our simplified code
precisely agrees with state of the art codes 
(that include thermal, radiative and other corrections corrections,
smaller than the present experimental uncertainty).\footnote{A recent paper~\cite{DV}
studied, in a four-neutrino context, how active/sterile oscillations affect the 
$^4$He abundancy.
The authors of~\cite{DV} take into account spectral distortions of $\nu_{\rm s}$ (that we neglect),
and neglects other minor corrections (that we include), such as those related to the electron mass.
They compute an `effective BBN neutrino number' $N_\nu^{\rm BBN}$,
which, unlike our $N_\nu^{^4{\rm He}}$ and $N_\nu^{{\rm D}}$,
is not directly related to the observable helium-4 and deuterium abundances.
However, in the parameter range
covered by their plots, their $N_\nu^{\rm BBN}$ should be
an approximation to $N_\nu^{^4{\rm He}}$.
The plots that can be compared show a reasonable level of agreement.

At $\Delta m^2\sim 10^{-8}\eV^2$ spectral distortions of electron neutrinos
are not negligible: in the two-neutrino limit they make the helium-4
abundancy more sensitive to sterile oscillations
(see the papers by Kirilova et al.\ in~\cite{asymm}).}
The deuterium abundancy is obtained with a similar technique.

\begin{figure}[t]
$$\includegraphics[width=17cm]{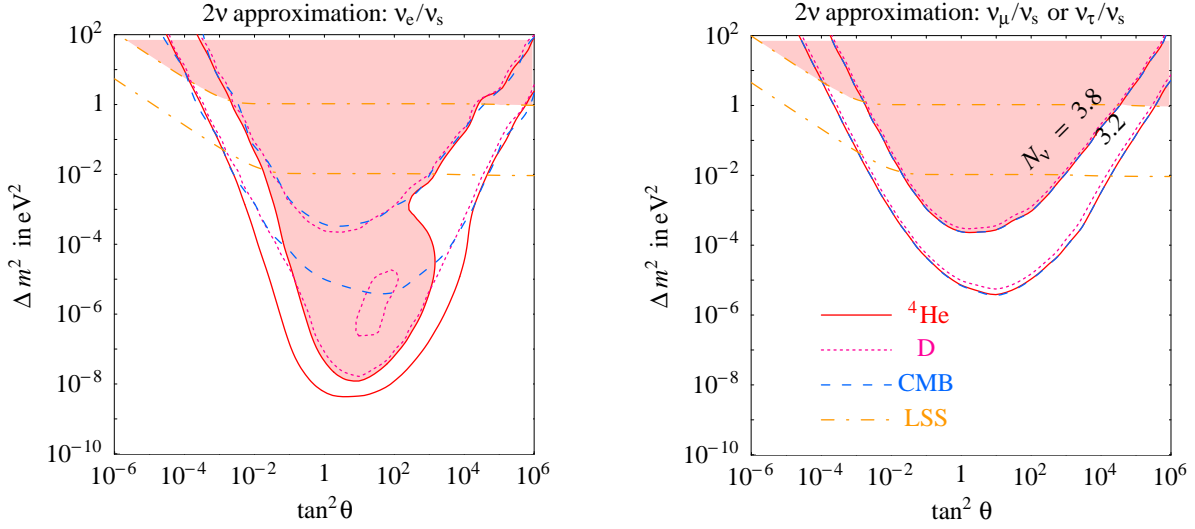}$$
\caption{\label{fig:2nu}\em Isocurves to the effective
number of neutrinos produced by 2 neutrino oscillations in the cases
$\nu_e/\nu_{\rm s}$ (left plot) and $\nu_{\mu,\tau}/\nu_{\rm s}$.
Solar and atmospheric oscillations are included
in the 3 neutrino plots of fig.\fig{BBN}, where the meaning of the various isolines
is precisely explained.}
\end{figure}

\begin{figure}[p]
$$\hspace{-8mm}\includegraphics[width=18cm]{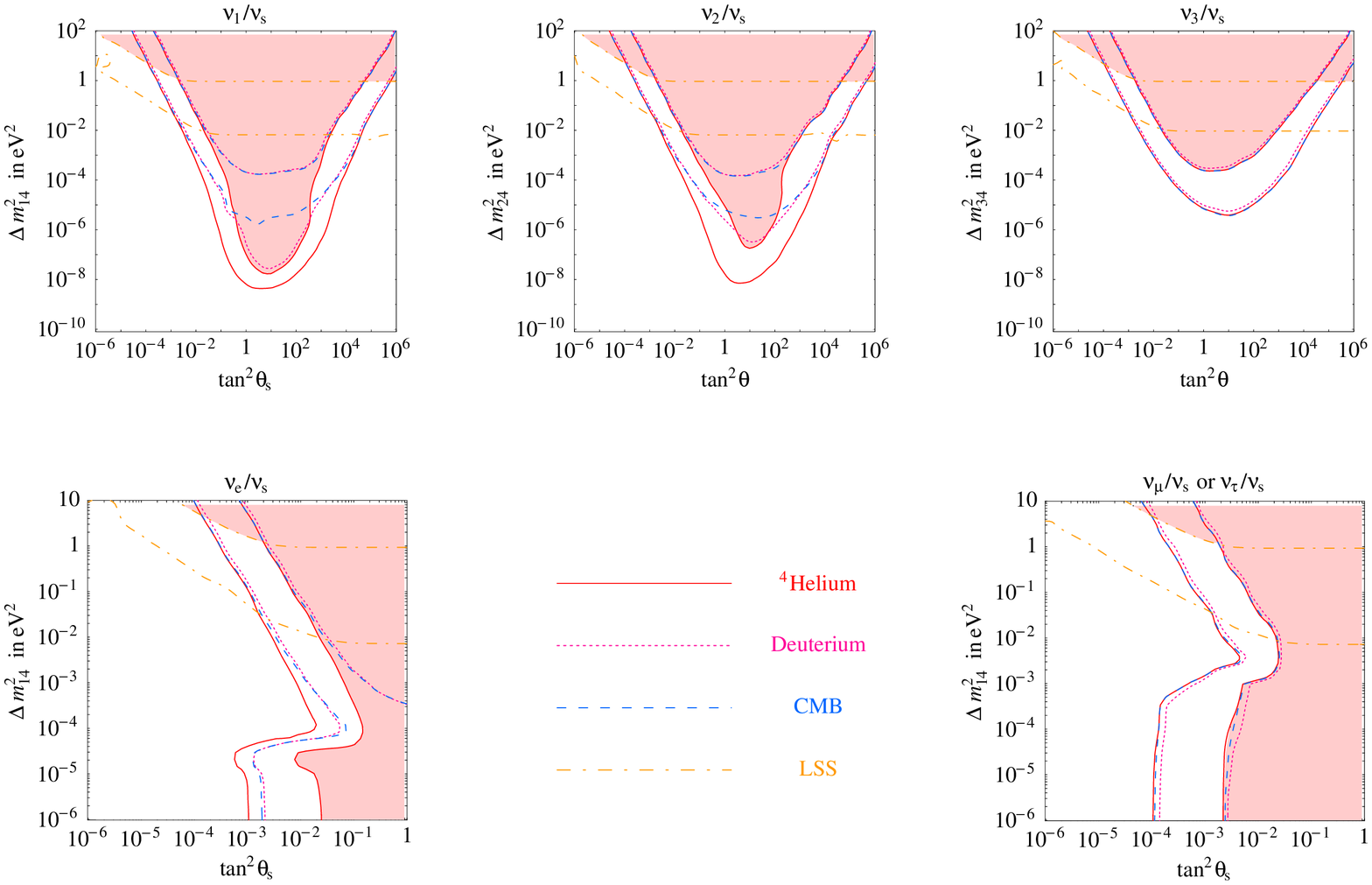}$$
\caption{\label{fig:BBN}\em {\bf Cosmological effects of sterile neutrino oscillations}.
We compare four different signals.
{\color{rosso} The continuous red line refers to the $^4\hbox{\rm He}$ abundancy}
(we shaded as `disfavoured' regions where its value corresponds to $N_\nu > 3.8$),
{\color{viola} the violet dotted line to the deuterium abundancy}, and
{\color{blu} the dashed blue line to the effective number of neutrinos at recombination}.
We plotted isolines of these three signals corresponding to an effective number of
neutrinos $N_\nu=3.2$ and $3.8$.
The precise meaning of the parameter $N_\nu$ in the three cases is  explained in the text.
{\color{rossoc}
The upper (lower) dot-dashed orange lines corresponds to $\Omega_\nu h^2= 10^{-2}$ $(10^{-3})$},
where $\Omega_\nu$ is the present energy density  in neutrinos.}
\end{figure}

\subsection{Results}
We plot the effective numbers $N_\nu$ of neutrinos defined in terms of
the physical observables (the $^4$He and D abundances and the energy density
at recombination)
from eq.s~(\ref{sys:HeD}) and \eq{CMB}.
We also plot the value of the present energy density in neutrinos $\Omega_\nu $,
probed by observations of Large Scale Structure
together with CMB constraints.

The plots have the following meaning:
shaded regions have
$N^{^4{\rm He}}_\nu> 3.8$
or $\Omega_\nu h^2>10^{-2}$ and are therefore `disfavoured' or `excluded'
(depending on how conservatively one estimates systematic uncertainties)
within minimal cosmology.
The other lines indicate the sensitivity that future experiments might reach.
More precisely we plot contour-lines corresponding to
$N_\nu = 3.2$ and $3.8$
and to $\Omega_\nu h^2 = 10^{-2}$ and $10^{-3}$.

\medskip

It is useful to start discussing the unrealistic but simple cases considered in old papers~\cite{bbnosc}.
In fig.\fig{2nu} we show the effects produced by
2 neutrino mixing: $\nu_{\rm s}/\nu_e$ in fig.\fig{2nu}a and
$\nu_{\rm s}/\nu_{\mu}$ or $\nu_{\rm s}/\nu_\tau$ mixing in fig.\fig{2nu}b.\footnote{Previous papers
studied the $^4$He abundancy and we agree with their results.
We however use as a variable $\tan^2\theta$ rather than $\sin^22\theta$,
so that we unify in a unique plot the non-resonant ($0<\theta < \pi/4$)
and the resonant ($ \pi/4<\theta<\pi/2$) case.}
The red dashed line shows the total number of neutrinos, $N_\nu^{\rm CMB}$:
it essentially does not depend on which flavour ($\nu_e$, $\nu_\mu$ or $\nu_\tau$)
mixes with $\nu_{\rm s}$ and is not affected by oscillations with $\Delta m^2 \circa{<} 10^{-5}\eV^2$
that are too slow and start only after neutrino decoupling,
when the total number of neutrinos is frozen.\footnote{To be
precise we should say `the total entropy in neutrinos per comoving volume
remains constant'. 
For simplicity we will adopt such loose abbreviations.}
At this stage neutrinos can still change flavour.
The difference between fig.\fig{2nu}a and\fig{2nu}b
 is due to the fact that only
electron neutrinos are involved in the reactions that control the $n/p$ ratio.
Therefore $\nu_{\rm s}/\nu_e$ oscillations that occur after neutrino freeze-out and that do not
affect the total number of neutrinos ($\nu_{\rm s}$ are created by depleting $\nu_e$)
affect $n/p$ and consequently the  $^4$He abundancy\footnote{These region
are strongly disfavoured because have a helium-4 abundancy
corresponding to $N_\nu^{^4{\rm He}} >4$, up to about 5.} (continuous line)~\cite{bbnosc}
and, to a lesser extent, the D abundancy.
This happens down to $\Delta m^2 \sim 10^{-8}\eV^2$:
oscillations with $\Delta m^2 \circa{<} 10^{-8}\eV^2$ occur after decoupling of electroweak scatterings,
when the relative $n/p$ abundancy is only affected by neutron decay.

Effects are larger at $\theta > \pi/4$ (i.e.\ $\tan\theta>1$) because this corresponds
to having a mostly sterile state lighter than the mostly active state,
giving rise to MSW resonances in neutrinos and anti-neutrinos
(like in cosmology, also
supernova $\bar\nu_e$ feel a MSW resonance for $\theta_{\rm s}\circa{>}\pi/4$.
On the contrary solar neutrinos feel a resonance for $\theta_{\rm s}\circa{<}\pi/4$).
In the past years it has been debated about
if a neutrino asymmetry and/or large inhomogeneity
develop as a consequence of non-linear effects,
and this issue has not yet been fully clarified.
Our Boltzmann equations assume that both these effects can be neglected.
At $\tan\theta>1$ the bound from $\Omega_\nu$ holds
even for very small mixing, $\theta\simeq\pi/2$ just because these region correspond
to heavy active neutrinos.

\bigskip

We now discuss how the above picture changes taking into account oscillations among active
neutrinos.
Our results are shown in fig.\fig{BBN}:
the upper row refers to sterile mixing with mass eigenstates $\nu_{1,2,3}$ and
the lower row  to mixing with flavour eigenstates $\nu_e$ and $\nu_{\mu,\tau}$.

An inspection of the upper row shows that their main features can
be understood in terms of the (unrealistic) results in the case of $2$ neutrino mixing,
fig.\fig{2nu}.
Having assumed  $\theta_{13}=0$, $4\nu$ sterile mixing with $\nu_3$
gives no new effects with respect to $2\nu$ sterile mixing with $\nu_{\mu,\tau}$.
Due to solar and atmospheric oscillations, $\nu_e$ depletion due to
oscillations into sterile neutrinos now happens in all other cases
and becomes milder, because
no longer confined to $\nu_e$ but shared among 
all active neutrinos.
Fig.\fig{BBN}b shows the effects of
$\nu_{\rm s}/\nu_2$ mixing (this kind of neutrino spectrum is plotted in fig.\fig{spettrias}b):
since $\nu_2$ contains some $\nu_e$ component, 
electron neutrinos are in part directly affected.
Fig.\fig{BBN}a shows the effects of
$\nu_{\rm s}/\nu_1$ mixing: $\nu_e$ depletion effects are largest in this last case
because $\nu_1$ is the neutrino eigenstate with the largest
$\nu_e$ component.
In summary: depletion gets transferred to all $\nu$ flavours and diluted.

Mixing with flavour eigenstates is qualitatively different,
for the general reasons explained in section~\ref{osc}.
We can see the effects of the solar (atmospheric)
mass splitting as bumps in fig.\fig{BBN}d (\ref{fig:BBN}e)
where cosmological effects of $\nu_{\rm s}/\nu_e$ ($\nu_{\rm s}/\nu_{\mu,\tau}$) mixing are computed.
In the case of $\nu_{\rm s}/\nu_e$ mixing
$N_\nu^{\rm D}, N_\nu^{\rm CMB} =4$ is reached
only if the sterile neutrino has a large enough $\Delta m^2$:
the solar $\Delta m^2_{\rm sun}\approx 0.7~10^{-4}\eV^2$ alone
is not sufficient, as also indicated by the $2\nu$ limit plotted in fig.\fig{2nu}a.
On the contrary, fig.\fig{2nu}b shows that
in the case of $\nu_{\rm s}/\nu_{\mu,\tau}$ mixing a
$\Delta m^2\sim \Delta m^2_{\rm atm}\approx 2~10^{-3}\eV^2$ is large enough
to reach $N_\nu\approx 4$ for any value of the sterile mass.
We have verified that setting $\theta_{13}\sim 0.2$ 
fig.s\fig{BBN} do not vary in a significant way.

\subsection{Hints and anomalies: cosmology}
To conclude we list cosmological data that
do not fit well into the scheme  presently considered as standard,
and that can be interpreted as manifestations of sterile neutrino effects:
\begin{itemize}

\item 
Various determinations of the primordial helium-4 abundancy $Y_p$
 try to reduce uncertainties by appropriately choosing and modeling the  astrophysical systems 
 used for the observation~\cite{He4}.
 Some analyses find lower values of $Y_p$
 corresponding to less than 3 neutrinos
 (an effect which could come from active-sterile oscillations),
 even after a detailed examination of the systematic uncertainties.
 Other determinations give higher values 
of $Y_p$, more compatible with $N_\nu=3$.

\item $X$-ray cluster data seem to prefer a lower value of the  parameter $\sigma_8$
than the other sets of CMB and LSS data, considered in~\cite{WMAP,boundMnu}.
A degenerate spectrum of active neutrinos can alleviate this discrepancy~\cite{Allen}.
Otherwise, one can invoke a contribution to $\Omega_\nu$ from sterile neutrinos.

\item Decays of sterile neutrinos with mass $\sim 200\MeV$ 
provide an interpretation of the reionization
at large redshift observed by WMAP, 
alternative to the standard one
(early formation of massive stars)~\cite{reionization}.

\item Decays of sterile neutrinos with mass $\sim 10\MeV$ 
and abundancy $\Omega_\nu=10^{-5\div 9}$
can be a non standard source of recently observed galactic positrons~\cite{Picciotto}.

\end{itemize}

\section{Sterile effects in solar (and KamLAND) neutrinos}\label{solar}
We compare sterile effects with present and future solar neutrino experiments 
and with KamLAND reactor anti-neutrino data.
While computing sterile effects in reactor $\bar\nu_e$ is straightforward,
solar neutrinos are detected after a long trip from
the center of the sun during which they can 
experience sterile effects in several different ways.
This is what makes solar neutrinos a powerful probe of sterile effects.
In section~\ref{suntec} we describe how we precisely compute these effects.
Section~\ref{sunfit} describes how we fit data, dealing with the complication
that these data contain a positive evidence for active/active oscillations.
Results and a qualitative understanding are presented in section~\ref{solarres}.
In section~\ref{compare} we compare our results with  previous analyses
performed in limiting cases.

\subsection{Technical details}\label{suntec}
In order to understand what happens when a sterile neutrino is added
and to write a sufficiently fast numerical code,
we must develop an analytical approximation.
It is convenient to employ some more formalism
than in the $2\times 2$ case
and study the evolution of the $4\times4$
neutrino density matrix $\rho_m$ written in the $m$ass basis of instantaneous mass eigenstates.
A $\nu_e$ produced with energy $E_\nu$ at radius $r=r_0$ inside the sun is described by
 $\rho_m  = V_m^\dagger\cdot
\hbox{diag}\,(1,0,0,0)\cdot V_m $
where $V_m$ depends on $E_\nu$ and $r_0$.
Mixing matrices in matter ($V_m$) and vacuum ($V$)
are computed diagonalizing the Hamiltonian~\cite{MSW}
$$\Ham = \frac{mm^\dagger}{2E_\nu} + \sqrt{2}G_{\rm F}\diag(N_e-\frac{N_n}{2},-\frac{N_n}{2},-\frac{N_n}{2},0)$$
and ordering eigenstates according to their eigenvalues $H_i \equiv m_{\nu_{mi}}^2/2E_\nu$:
$\nu_{m1}$ ($\nu_{m4}$) is the lightest (heaviest) neutrino mass eigenstate in matter.
The evolution up to the detection point is described by
a $4\times4$ unitary evolution matrix $\mathscr{U}$ so that
at detection point the density matrix in the basis of flavour eigenstates is
$\rho =\langle V\cdot \mathscr{U}\cdot\rho_m(r,E_\nu)\cdot \mathscr{U}^\dagger \cdot V^\dagger\rangle$.
where $\langle\cdots\rangle$ denotes average over the production point.
The various observables involve additional averages over neutrino energy
and time (and consequently over different paths in the earth and in vacuum).
The oscillation probabilities are given by $P(\nu_e\to \nu_e)=\rho_{ee}$, 
$P(\nu_e\to \nu_{\rm s})= \rho_{ss}$, etc.

We briefly describe the main steps in the computation of $\mathscr{U}$, focussing on the subtle ones.
The evolution matrix can be decomposed as
$$
\mathscr{U} = \mathscr{U} _{\rm earth}\cdot \mathscr{U} _{\rm vacuum}\cdot \mathscr{U} _{\rm sun}.$$
Evolution in vacuum is given by
$\mathscr{U} _{\rm vacuum} = \diag\exp (-i L m_{\nu_i}^2/2E_\nu)$.
Combined with average over neutrino energy it suppresses
the off-diagonal element $\rho_m^{ij}$ when
the phase differences among eigenstates $i$ and $j$ are large.

Evolution in the earth is computed in 
mantle/core approximation, improved by
using the average density appropriate for each 
trajectory as predicted in~\cite{PREM}.
Therefore we use $ \mathscr{U} _{\rm earth}=1$ when the earth is not crossed,
$ \mathscr{U} _{\rm earth}
=P\cdot\diag \exp (-i L_{\rm mantle} m_{\nu_{mi}}^2/2E_\nu)\cdot P^\dagger$
when only the mantle is crossed (for a length $L_{\rm mantle}$),
and the obvious generalization
when both mantle and core are crossed.
$P=V^\dagger_B V_A$ is a non-adiabaticity factor
that takes into account the sharp flavour  variation of mass eigenstates
when passing from medium $A$ (vacuum) into medium $B$ (the earth mantle)
\label{Psun}.

Evolution in the sun is more complicated because there can be various
$P$ factors at non-adiabatic level crossings at radii $r_n$ ($n=1,2,\ldots, n_{\rm max}$), that give
$$ \mathscr{U}_{\rm sun} = P_{r_n}\cdots P_{r_2}\cdot 
\diag\exp(-i\int_{r_1}^{r_2} ds~\frac{ m_{\nu_{mi}}^2}{2E_\nu})\cdot  P_{r_1}\cdot
\diag\exp(-i\int_{r_0}^{r_1} ds~\frac{ m_{\nu_{mi}}^2}{2E_\nu}) $$
The number of level crossings $n_{\rm max}$ ranges between 0 and a few,
e.g.\ a $\nu_e$ produced in the side of the sun
farther from us can experience 4 crossings.
When levels $i$ and $j$ cross in an adiabatic way, $P_{r_n}=1$.
If instead level crossing is fully non adiabatic
$P_{r_n} = V_m^\dagger(r\circa{<}r_n)\cdot V_m(r\circa{>}r_n)$ 
is a rotation with angle $\alpha = 90^\circ$ in the $(ij)$ plane.
In general the rotation angle is given by $\tan^2\alpha = P_C/(1-P_C)$,
where $P_C$ is the level crossing probability.

\medskip

So far we only presented the well known formalism~\cite{MSW,Petcov,KPreview}
in a non standard way appropriate for applying it in numerical computations with
multiple and overlapping level crossings.
We now need to compute $P_C$. 
This last step turns out to be non trivial. 
By generalizing well known results valid in the simpler $2\times2$ case~\cite{Petcov,KPreview},
we find that in all the parameter range $P_C$ can be accurately
approximated analytically (i.e.\ no need of numerically solving the 
differential neutrino evolution equation $i\,d\rho/ds=[\mathscr{H},\rho]$).
Such a simple result is possible because,
in the relevant neutrino energy range, level crossings are non adiabatic only for
$\theta^2_{\rm s}\cdot \Delta m^2/10^{-8}\eV^2\gg1$
i.e.\ when either $\theta$ is small or the active/sterile
$\Delta m^2$ is much smaller than the LMA splitting.




To compute $P_C$ it is convenient to consider the 
basis of mass eigenstates in absence of active/sterile mixing
(i.e.\ $\theta_{\rm s}\to 0$ or $\theta_{\rm s}\to \pi/2$, depending
on which limit is closer to the value of $\theta_{\rm s}$ under examination.
Here we focus on $\theta_{\rm s}\to 0$.).
This limit allows to precisely define level-crossings.
When $\nu_{\rm s}$ crosses one of the active  eigenstates, $\nu_{a}^m$ ($a=1,2,3$)
(see fig.\fig{spettrias}a for one example), 
the level crossing probability $P_C$ is well approximated by
\begin{equation}\label{eq:PC}
 P_C = \frac{e^{\tilde\gamma \cos^2\theta^m_{as}}-1}{e^{\tilde\gamma} -1}\qquad
\gamma =\frac{ 4 \mathscr{H}_{as}^2}{d H_a/dr}
\equiv
\tilde\gamma \cdot \frac{\sin^2 2\theta^m_{as}}{2\pi| \cos2\theta_{as}^m|}
\quad\hbox{where}\quad \sin\theta_{as}^m = \vec{n}\cdot \vec{\nu}_a^m \sin\theta_{\rm s} .
\end{equation}
The above equation might seem a complicated way of rewriting
the well known
expression for $P_C$ valid in the simpler $2\nu$ case~\cite{Petcov,KPreview},
but there is one important difference.
In the $2\nu$ case one can write the result in an analogous way,
which contains the mixing angle in vacuum
and the off-diagonal elements of the Hamiltonian in vacuum.
On the contrary,
our $\gamma$ and $\theta^m_{as}$ must be computed
around the resonance, where $\mathscr{H}_{aa}=\mathscr{H}_{ss}$ (or around the point where adiabaticity is maximally violated, in cases where there is no resonance).
We emphasize that reducing the full $4\times 4$ Hamiltonian to the effective 
$2\times 2$ Hamiltonian of the 2 states that cross
and computing $\mathscr{H}_{as}$ is non trivial,
since sterile mixing sometimes redefines
the flavour of the active neutrino involved in the crossing.\footnote{Neglecting
this subtlety 
would give a qualitatively wrong result e.g.\ in the following situation:
$\nu_{\rm s}$ is mixed with $\nu_e$ and is quasi degenerate to $\nu_1$.}
In these situations it is useful to know the
physical meaning of $2\mathscr{H}_{as}$: it is the minimal difference between the eigenvalues
of the two states that cross.
In order to elucidate the physical meaning of
the crossing angle $\theta^m_{as}$, we emphasize that
it can also be extracted from
the scalar product between the flavour vectors of the two matter eigenstates
$i$ and $i+1$
that cross: $\sin \theta^m_{as} = \nu_{m_i}^*(r\circa{<} r_n)\cdot \nu_{m_i}(r\circa{>} r_n)$
and  $\cos \theta^m_{as} = \nu_{m_i}^*(r\circa{<} r_n)\cdot \nu_{m_{i+1}}(r\circa{>} r_n)$.

\smallskip

When the active/sterile mixing is large,
$P_C$ is sizable only if $\Delta m^2_{as}\ll \Delta m^2_{\rm LMA}$:
in this case our expression\eq{PC} reduces to the standard 2-neutrino formula.
In the narrow resonance limit (e.g.\ for $\theta_{as}^m\ll 1$),
eq.\eq{PC} reduces to a Landau-Zener form 
$P_C \simeq e^{-\pi\gamma/2}$: 
our expression for the adiabaticity factor $\gamma$ 
holds for a generic $2\times2$ Hamiltonian.
In the sun, resonances with $\theta^m_{as}$ close to $\pi/4$
happen only in the quasi-vacuum region at the border of the sun.
This will be no longer true when studying supernova neutrinos,
that will necessitate an extension of eq.\eq{PC}.


\begin{figure}
$$\includegraphics[width=8cm]{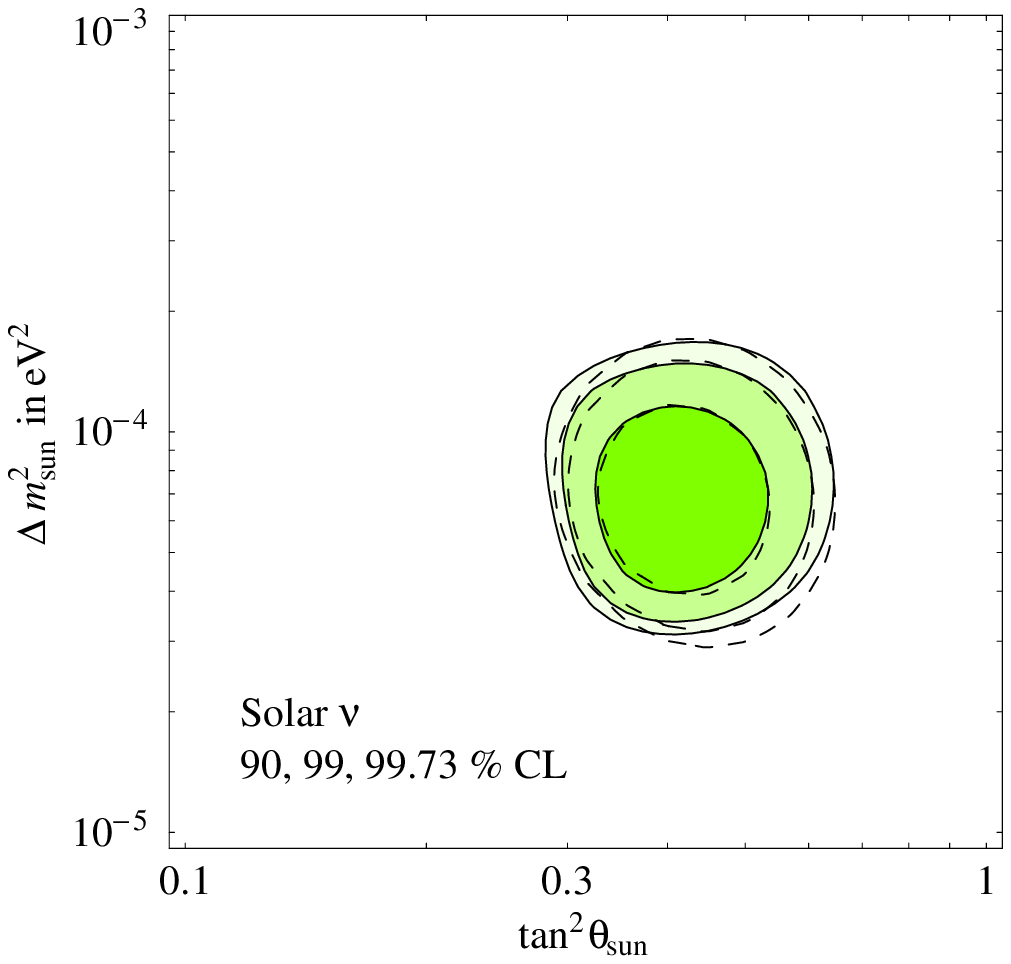}\hspace{1cm}
\includegraphics[width=8cm]{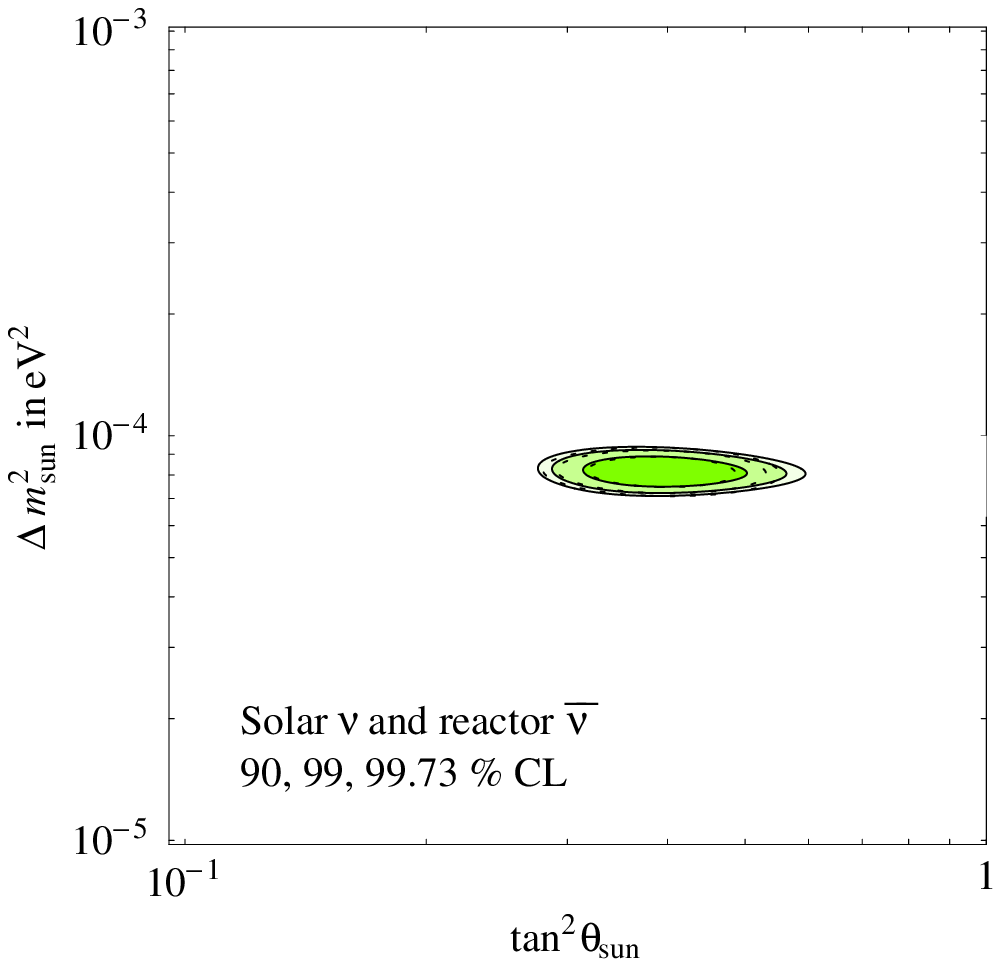}$$
\caption{\label{fig:sunfits}\em 
We compare the usual `active only'
global fit of solar (shaded regions in the left plot) and of
solar plus KamLAND data (shaded regions in the right plot),
with the Gaussian approximation employed in this paper (dashed lines)}
\end{figure}

\subsection{Experimental data and fit procedure}\label{sunfit}
We fit all available latest data:
 \begin{itemize}

\item The SNO CC+NC+ES spectra~\cite{SNOlast}, divided in 34 bins (17 energy bins times 2 day/night bins).

\item The total CC, NC and ES rates measured by SNO with enhanced NC sensitivity~\cite{SNOsalt}.

\item The Super-Kamiokande ES spectra~\cite{SKlast}, divided
in 44  zenith-angle and energy bins.

\item The Gallium rate~\cite{Galliumlast}, $R_{\rm Ga} = (68.0\pm3.8)\,{\rm SNU}$, obtained averaging
the most recent  SAGE, Gallex and GNO data.

\item The Chlorine rate~\cite{Chlorinelast},
$R_{\rm Cl} = (2.56 \pm 0.23)\,{\rm SNU}$.

\item The KamLAND reactor anti-neutrino
data, divided in 13 energy bins  with prompt energy higher than $2.6\MeV$~\cite{KamLAND}.

\end{itemize}
We revised solar model predictions and uncertainties~\cite{BP}
including the recent measurement of the
$^{14}$N$(p,\gamma)^{15}$O nuclear cross section~\cite{LUNA},
which reduces the predicted CNO fluxes by roughly $50\%$.

Data are compared with predictions forming a $\chi^2$ that takes
into account statistical, systematic and theoretical uncertainties
(on the total solar neutrino fluxes and on the $^8$B spectrum)
and their correlations~\cite{BP}.
We plot the $\chi^2$ as function of the 2 parameters that
describe sterile oscillations,
marginalizing the full $\chi^2$ with respect to
all other sources of uncertainty {\em including the LMA parameters}
$\Delta m^2_{\rm sun}$ and $\theta_{\rm sun}$.
This step is of course not performed in usual analyses.
Proceeding in a fully numerical way, it would be too demanding for present computers.
We can however approximate all observables with a first order Taylor expansion
around the best-fit LMA point,
since experiments allow only relatively minor shifts from it.
In this way marginalization over $\Delta m^2_{12}$ and $\theta_{\rm sun}$
can be performed analytically, using the same Gaussian techniques
commonly employed for other `systematic' parameters.
Fig.a\fig{sunfits} shows that performing this linearization (dashed lines) we obtain 
a satisfactory approximation to the usual active-only global fit of solar neutrinos
(continuous lines).
With more data this approximation will become more and more accurate.

Solar $\nu$ and reactor $\bar\nu$ data cannot be fit independently,
since both sets of data depend on solar oscillation parameters.
Inclusion of KamLAND data presents a slight complication:
due to poor statistics, the Poissonian distribution must be used.
Nevertheless small deviations from the LMA best fit
(due to sterile effects and to systematic uncertainties)
can be taken into account in Gaussian approximation, obtaining
$$\chi^2 = 2(t-r\ln t)\simeq 2(t_0-r\ln t_0)+2\epsilon(t_0-r) +r\epsilon^2$$
where $r$ are the observed rates and $t = t_0(1 + \epsilon)$ the predicted rates.
In our analysis $\epsilon$ is the correction due to sterile neutrinos.
Fig.\fig{sunfits}b shows that we can accurately  analytically approximate the
usual active-only global fit of solar plus KamLAND data.



\begin{figure}[t]$$
\includegraphics[width=8cm]{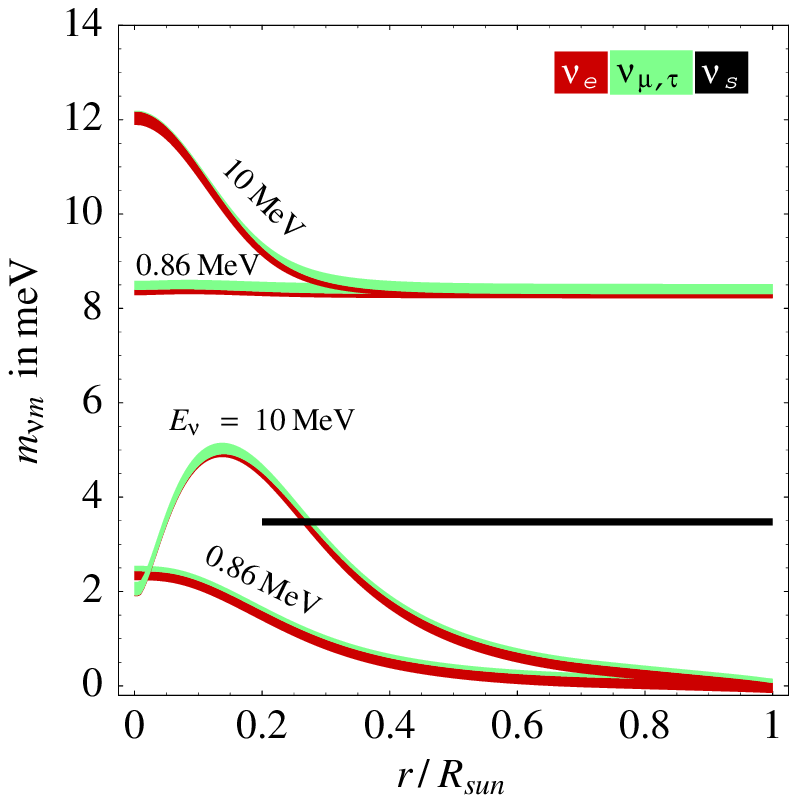}\hspace{1cm}
\raisebox{1mm}{\includegraphics[width=8cm]{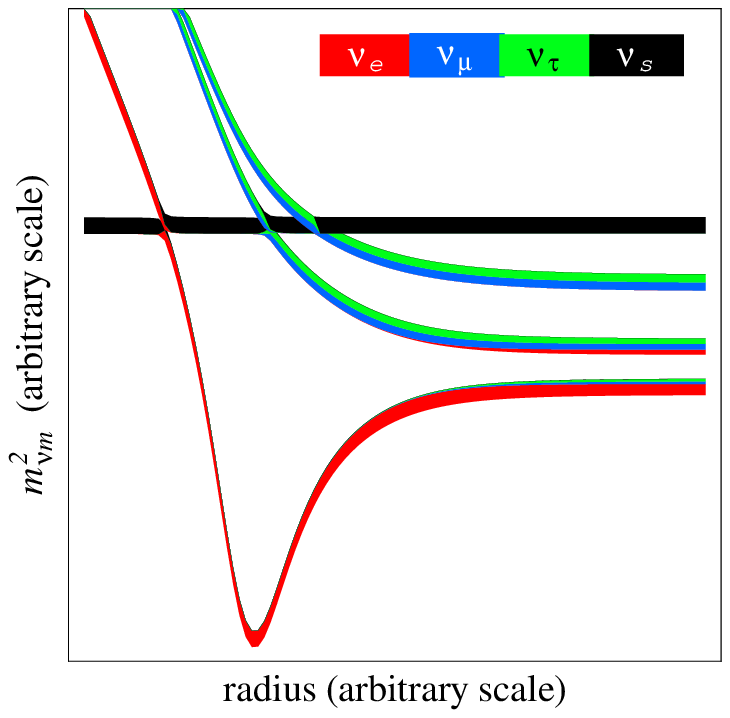}}$$
\caption{\label{fig:levels}\em {\bf Level crossing schemes}.
The right plot shows qualitatively the effective anti-neutrino masses in a supernova.
The left plot shows the effective neutrino masses in the sun.
We assumed hierarchical active neutrinos
(i.e.\ $m_1 = 0$ and $m_2 = (\Delta m^2_{\rm sun})^{1/2}$)
and plotted the two matter eigenstates that give rise to LMA oscillations
for two different values of $E_\nu$:  
$10\MeV$ and $0.86\MeV$, the energy of the main Beryllium line.
Colors indicate the flavour composition. 
An extra sterile neutrino with small mixing is represented by an  horizontal line
with height equal to its mass.
}
\end{figure}

\begin{figure}[p]
$$\hspace{-5mm}\includegraphics[width=8cm]{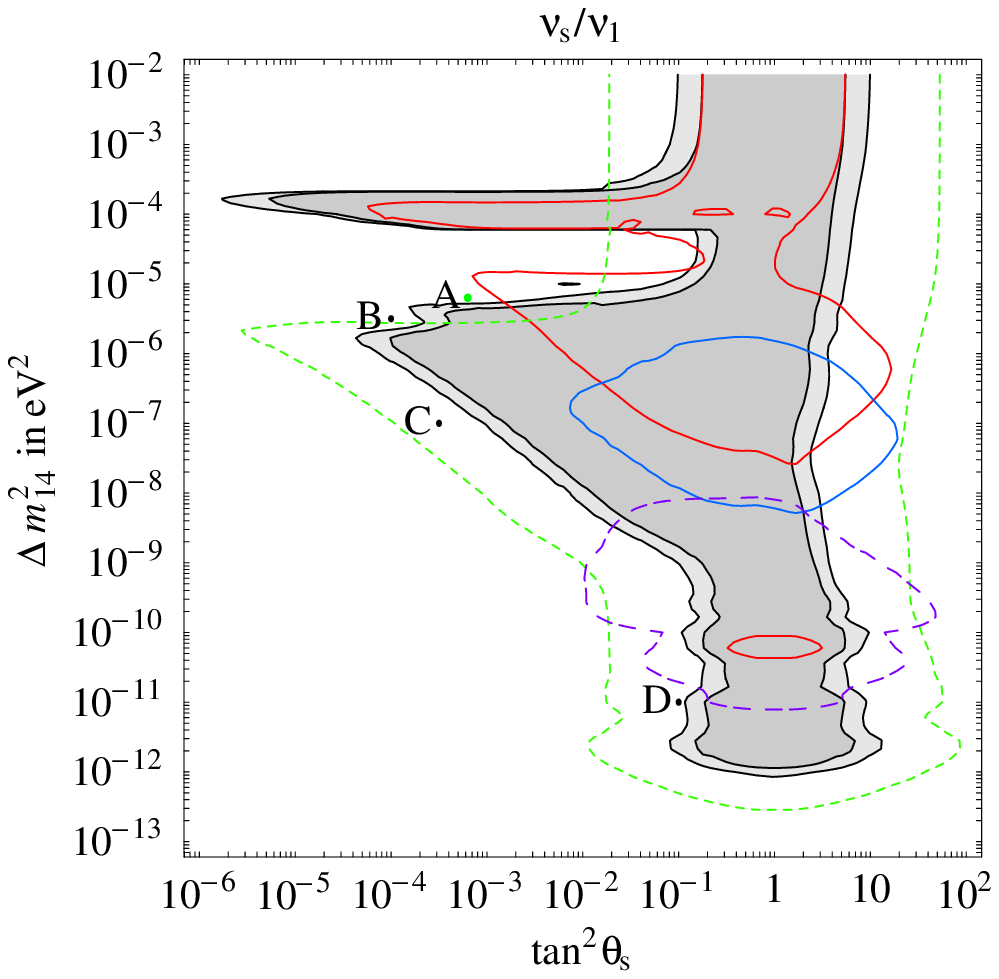}\hspace{5mm}
\includegraphics[width=8cm]{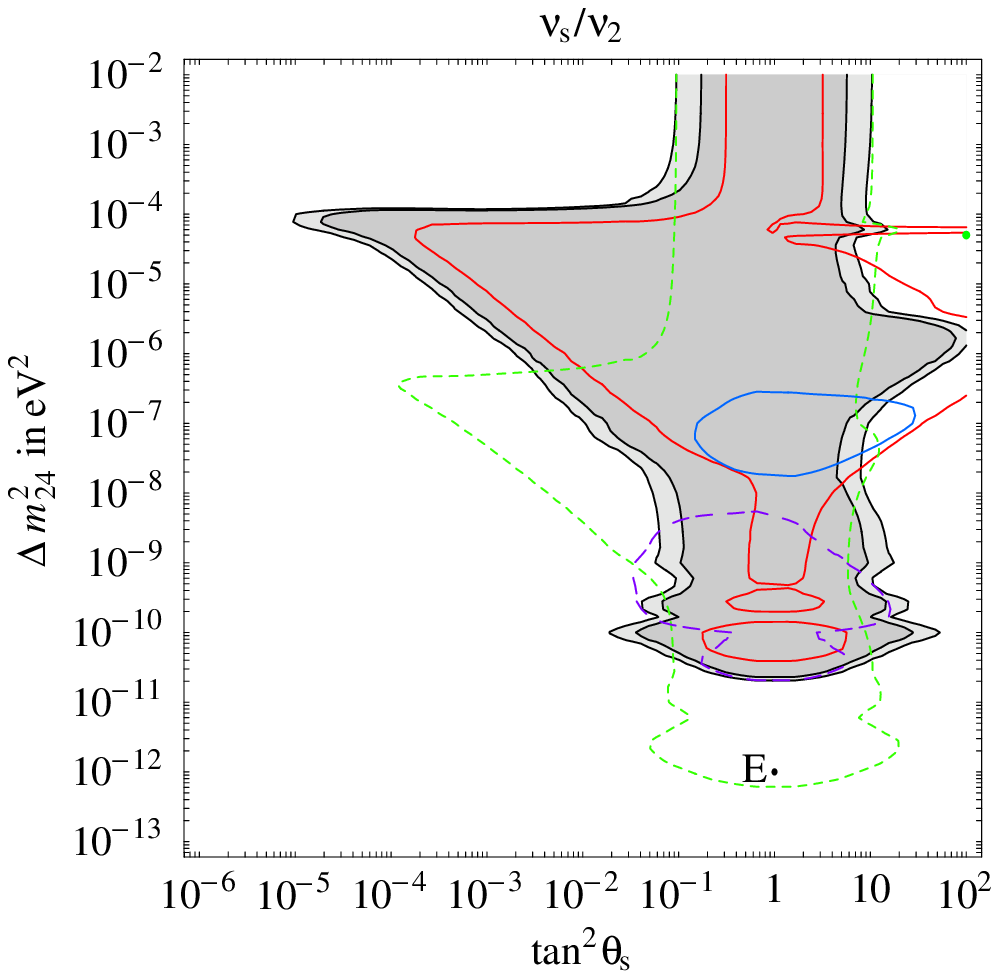}$$
$$\hspace{-5mm}\includegraphics[width=8cm]{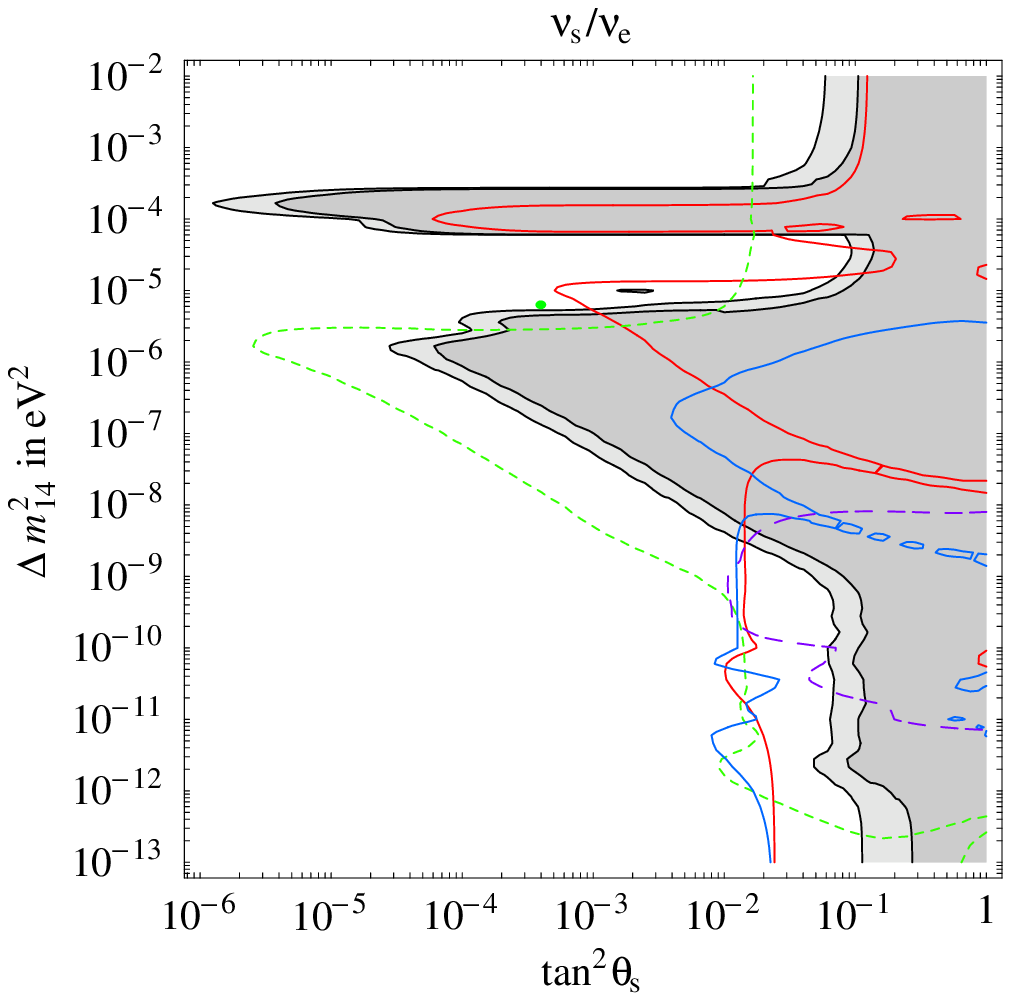}\hspace{5mm}
\includegraphics[width=8cm]{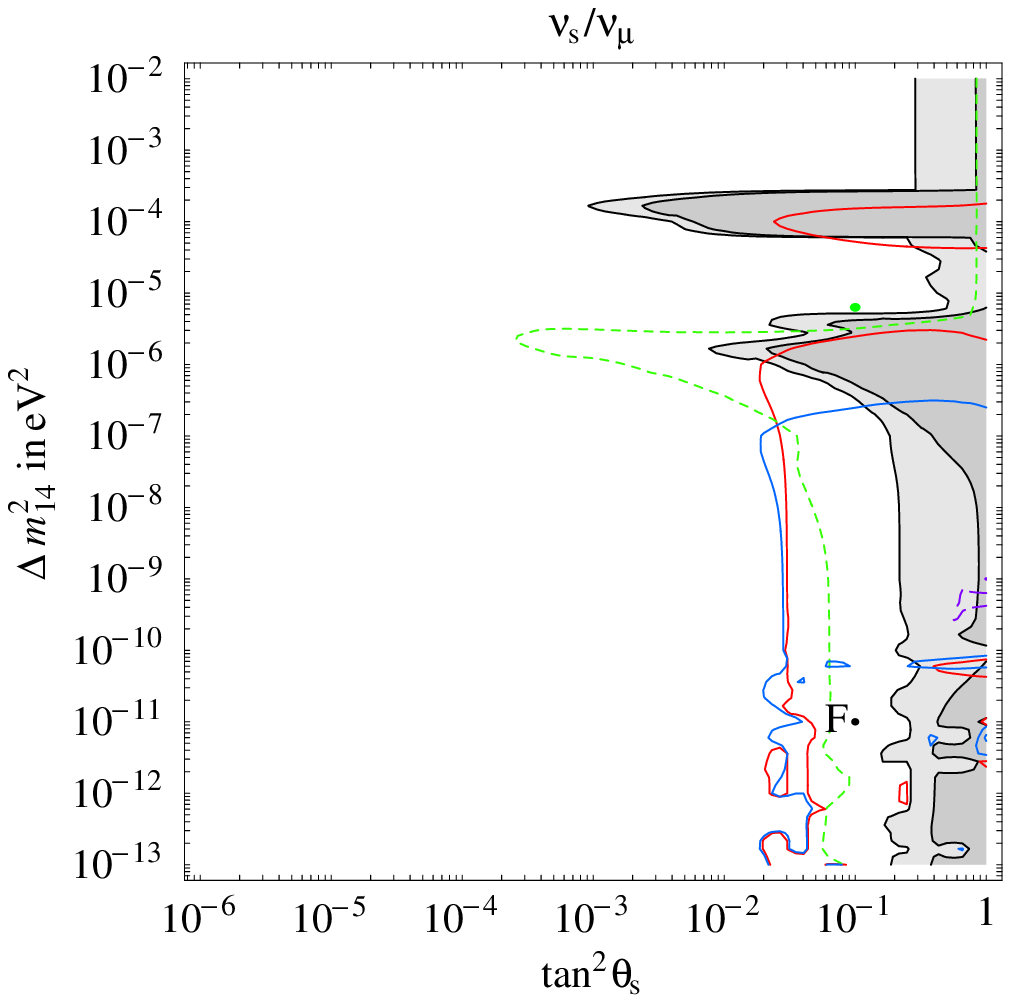}$$
\caption{\label{fig:sun}\em {\bf Sterile mixing:
effects in solar neutrinos.} No statistically significant evidence is found.
Shaded regions: excluded at $90,\,99\%$ C.L.
Coloured lines are iso-curves of a few promising signals.
{\color{rossos} Continuous red line: $A_{\rm d/n}^{\rm ES}$ differs from LMA by 0.005}.
{\color{blu} Continuous blue line: 0.02 day/night asymmetry at Borexino}.
{\color{viola} Dashed violet line: seasonal variation at Borexino with 0.02 amplitude}.
{\color{verdes} Short-dashed green line: $P_{ee}$ at sub-MeV energies differs from LMA by 0.02}.
Other signals are discussed in the text.
The letters {\rm A},\ldots,{\rm F} indicate sample points, studied in detail in fig.\fig{samples}.
}
\end{figure}

\begin{figure}
$$\hspace{-5mm}\includegraphics[width=18cm]{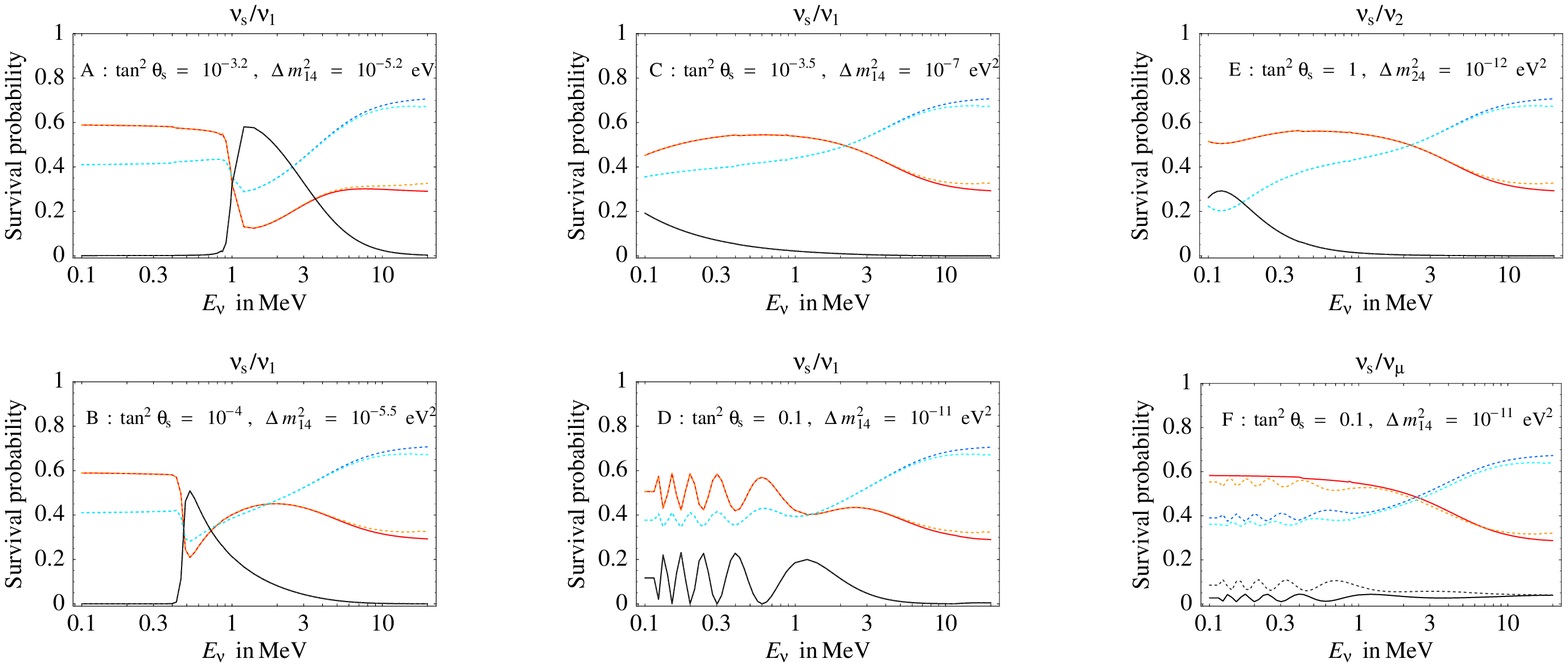}$$
\caption{\label{fig:samples}\em 
A few samples of still allowed sterile effects in solar neutrinos.
We plot, as function of the neutrino energy, 
{\color{rossos} $P(\nu_e\to\nu_e)$ (decreasing red curve)},
{\color{blus} $P(\nu_e\to\nu_{\mu,\tau}$) (increasing blue curve)} and
$P(\nu_e\to \nu_{\rm s})$ (lower black curve).
The continuous (dotted) curve are the values during day (night).
The sample points {\rm A},\ldots, {\rm F} are drawn in fig.\fig{sun} as dots.
}
\end{figure}

\subsection{Results}\label{solarres}
We start recalling how LMA oscillations behave in absence of sterile neutrinos~\cite{MSW}.
Fig.\fig{levels}a shows the composition of the two neutrino mass eigenstates.
For sake of illustration we assumed normal hierarchy, 
$m_1^2\ll m_2^2 \simeq \Delta m^2_{\rm sun}
\ll m_3^2\simeq\Delta m^2_{\rm atm}$.
At  higher neutrino energies,
$$E_\nu\gg E_* \approx 
\Delta m^2_{\rm sun}/G_{\rm F}N_e^{\rm sun}\sim\hbox{few MeV},$$ 
matter effects dominate around the center of the sun, $r\circa{<}0.2 R_{\rm sun}$ 
where neutrinos are produced as $\nu_e \approx \nu_{2m}$. 
The LMA level crossing at $r\approx 0.2 R_{\rm sun}$ is adiabatic
(in fact the solar mixing angle  $\theta_{\rm sun}$ is so large that
fig.\fig{levels} does not look like a level crossing)
so that neutrinos produced as $\nu_{2m}$ exit from the sun as
$\nu_2 = \sin\theta_{\rm sun} ~\nu_e + \cos\theta_{\rm sun}~\nu_{\mu,\tau}$
i.e. $P_{ee} = \sin^2 \theta_{\rm sun}$.
This limit roughly holds at energies probed by SNO and SK:
e.g.\ their total rate is $P_{ee} \approx 1.15 \sin^2 \theta_{\rm sun}$.
At $E_\nu\circa{<} E_*$ $P_{ee}$ increases and
$\nu_{1m}$
contains some $\nu_e$ because matter effects are no longer dominant.
At $E_\nu \ll E_*$ matter effects are negligible and
one gets averaged vacuum oscillations,
$P_{ee} = 1 - \frac12 \sin^22\theta_{\rm sun}$.
This energy range has been explored by Gallium experiments.

\medskip

We now discuss how to understand qualitatively the sterile/active  mixing effects~\cite{MSW}.
If sterile/active mixing is small, the mostly $\nu_{\rm s}$ state is represented by adding
one quasi-horizontal line to fig.\fig{levels}a.
Depending on its height (determined by the mass of the sterile neutrino) 
the mostly sterile level
crosses one or none of the two mostly active neutrinos
(for all relevant neutrino energies the sterile state does not cross both active neutrinos)
after or before the LMA resonance (or after and before).
In each case one can understand the behavior of the survival probabilities
from the level-crossing scheme:
in the example plotted in fig.\fig{levels}a
a neutrino produced at $r_0\approx 0.2$ experiences a single level 
crossing at $r_1\approx 0.3$.
This example corresponds to the case considered in~\cite{SmirnovSterile}:
a sterile neutrino weakly mixed with $\nu_1$,
with mass splitting $\Delta m^2_{14}$ somewhat smaller than $\Delta m^2_{12}$.
A solar $\nu_e$ produced at $r\sim 0.2 R_{\rm sun}$ contains a $\nu_{1m}$ component that
crosses the $\nu_{\rm s}$ state once, getting partially converted into $\nu_{\rm s}$.
This gives a dip in the survival probability at intermediate energies:
at low energies one has averaged vacuum oscillations (negligibly affected by  the small sterile mixing angle), at high energies matter effects dominate so that
$\nu_e \simeq\nu_{2m}$ that does not cross the sterile level.
Fig.\fig{samples}A,B show examples of this behavior.
Even sticking to the case of $\nu_{\rm s}/\nu_1$ mixing,
qualitatively different effects are present for other values of the oscillation parameters,
studied in fig.\fig{sun}a.
E.g.\ the example in  fig.\fig{samples}D illustrates the case discussed in~\cite{freq,Sterile}: 
large $\theta_{\rm s}$ and small $\Delta m^2_{14}\sim 10^{-12}\eV^2$.
In all these examples {\em sterile neutrinos manifest at low energy},
$E_\nu \circa{<} E_*$.  
There is a general reason for this behavior: 
in absence of sterile effects, only at such energies
LMA oscillations allow a $\nu_1$ component
in the solar neutrino flux.

\smallskip

We also consider a sterile neutrino mixed with $\nu_2$ or $\nu_3$.
$\nu_{\rm s}/\nu_1$ mixing and $\nu_{\rm s}/\nu_2$ mixing
affect solar neutrinos  in similar ways.
In fact the $\nu_1^m$ and $\nu_2^m$  neutrino eigenstates in matter
 both typically contain significant  $\nu_1$ and $\nu_2$ components.
There are some differences in ${\cal O}(1)$ factors, which produce
the difference between fig.\fig{sun}a ($\nu_{\rm s}/\nu_1$) and b
($\nu_{\rm s}/\nu_2$).
The most evident difference in the shape of the excluded region
is due to our choice of the parameterization:
when studying $\nu_{\rm s}/\nu_2$ mixing
we produce plots with $\Delta m^2_{24}$ on the vertical axis,
so that small values of $\Delta m^2_{24}$
correspond to $\nu_{\rm s}$ quasi-degenerate with $\nu_2$
(rather than with $\nu_1$).

A sterile neutrino mixed with $\nu_3$ gives much smaller effects in solar neutrinos,
so that we do not show the corresponding plot.
This happens because matter effects negligibly mix $\nu_3$ with $\nu_{1m}$ or $\nu_{2m}$,
so that MSW resonances are highly non adiabatic i.e.\ ineffective.\footnote{Narrow dips in $P_{ee}(E_\nu)$ are possible close to specific energies such that
the mostly sterile state crosses active neutrinos  
when their position-dependent eigenvalues are maximal.
A look at fig.\fig{levels}a might help to understand this issue.}

\medskip

As discussed in section~\ref{osc}, 
sterile mixing with one active mass eigenstate is a special configuration.
Therefore we also consider a sterile neutrino mixed
with a flavour eigenstate: $\nu_e$, $\nu_\mu$ or $\nu_\tau$.
In such a case there are active/sterile oscillations at multiple
$\Delta m^2$ values, which (in view of the observed mass differences
among active neutrinos) cannot be all small.
This is the main difference with respect to the previous case,
and implies that sterile oscillation effects are present even for  $\Delta m^2_{i4}=0$.
To understand better this point let us consider the case $\Delta m^2_{14}\to 0$.
In this limit neutrinos exit from the sun as an incoherent mixture of $\nu_2$
and $\nu_1$, which both contain some sterile component.
At $E_\nu\gg E_*$ LMA is fully effective and neutrinos exit as pure $\nu_2$,
so that vacuum oscillations related to the small $\Delta m^2_{14}$ have no effect.
On the contrary, at $E_\nu \circa{<} E_*$ there is a $\nu_1$ component 
which experiences vacuum oscillations
with the mostly sterile state.
These vacuum oscillations affect the $\nu_e$ flux 
(if $\nu_{\rm s}$ is mixed with $\nu_e$ as in fig.\fig{sun}c),
or the $\nu_{\mu,\tau}$ flux
(if $\nu_{\rm s}$ is mixed with $\nu_{\mu,\tau}$ as in fig.\fig{sun}d).
In general, a mostly sterile state significantly mixed with $\nu_1$
and almost degenerate to it,
$\Delta m^2_{14}\sim 10^{-10}\eV^2$,
gives significant spectral distortions, {\em but only 
below the energy threshold of SK and SNO}.
This is illustrated in fig.\fig{samples} by the sample points D, E and F.

Fig.\fig{sun}c studies $\nu_{\rm s}/\nu_e$ oscillations for generic values of $\Delta m^2_{14}$,
and is similar to fig.\fig{sun}a,b apart from the difference discussed above.
$\nu_{\rm s}/\nu_{\mu}$ mixing
affects solar neutrinos in the same way
as $\nu_{\rm s}/\nu_\tau$ mixing, and both cases
give relatively mild effects, as shown in fig.\fig{sun}d.

It is interesting to notice that there are sizable 
and specific earth matter effects
if the sterile neutrino has a large mixing  e.g.\ with $\nu_e$,
and is quasi-degenerate e.g.\ to $\nu_1$.
In fact, the two quasi-degenerate neutrinos have different amounts of
active and sterile components;
due to matter effects, in the earth the two states are no longer quasi-degenerate,
giving oscillations with wave-length
$\sqrt{2}\pi/G_{\rm F} N_e^\oplus\sim 6000\,\hbox{km}$,
comparable to the size of the earth

\bigskip

In all cases fig.s\fig{sun} show the regions already excluded
at 90 and $99\%$ C.L. (2 dof) by present data
(these regions are precisely defined by $\chi^2>\chi^2_{\rm min}+4.6$ and $9.2$).\footnote{Using
instead $\chi^2>\chi^2_{\rm LMA}+4.6$ and $9.2$
would give slightly weaker constraints.
One might think that this latter procedure should be preferred
because gives more `robust' bounds.
It gives instead under-constraining bounds.
More robust bounds should be obtained 
demanding a higher C.L.\ to the correct statistical test,
rather than inventing `more robust' tests.
A correct $90\%$ C.L.\ bound does not hold with more than $90\%$ probability.}
We emphasize that most of the excluded regions are
disfavoured only by combining  high-energy (SNO, SK) with low-energy (Gallium) solar data:
each kind of data-sets alone tests only a minor region.
This means that present data are already able of fixing oscillation parameters
with some redundancy, partially testing the LMA hypothesis.
Only little regions are excluded if we fit only high-energy data, or only low-energy data.
In fact, sterile oscillations that crucially involve $\nu_1$
affect solar neutrinos only at $E_\nu \circa{<}E_*$ because only in this energy range
 the sun emits $\nu_1$.

\medskip

The other contours in fig.s\fig{sun} show which regions can be explored by
some future experiments.\begin{itemize}
\item In view of the previous comment it should be not surprising
that the relatively more powerful future measurement (dotted green line) is an
improved measurement of $P_{ee}$ at sub-MeV energies.
Taking into account solar model uncertainties and
plans of future experiments,
we assumed ---  maybe optimistically ---
that it will be possible to see 0.02 shifts from the LMA value of $P_{ee}$.\footnote{As 
in~\cite{subMeV} we averaged $P_{ee}$ assuming that $pp$ neutrinos 
will be detected by elastic scattering on electrons; 
similar results hold for other possible techniques.
Assuming that LMA is the end of the story,
ref.~\cite{subMeV} found that
feasible sub-MeV measurements would not have a significant impact
on the determination of the oscillation parameters.
We here find that sub-MeV measurements are instead crucially important
for probing sterile oscillations.}

\item
The red continuous line shows what can be achieved by measuring the
day/night asymmetry 
(normalized as $2(R_{\rm day}-R_{\rm night})/(R_{\rm day}+R_{\rm night})$)
at Boron energies with $0.01$ precision
in a future Mton water \v{C}erenkov experiment.
We have considered this test because the needed experiment
seems highly motivated by various other considerations.

\item
Coming to near-future experiments,
the other lines show what Borexino (and/or possibly KamLAND) can do
by studying (mainly) Beryllium neutrinos.
We do not show the impact of a measurement of the total
rate, which has a relatively large theoretical error~\cite{BP}\footnote{For values of the oscillation parameters 
around the `vertex' of the MSW `triangle' (at $\Delta m^2_{14}\approx 10^{-5.5}\eV^2$
and $\theta_{\rm s}\approx 10^{-2}$)
sterile neutrinos can give a deep dip in $P_{ee}$
at $E_{\rm Be}=0.863\MeV$.
Such effects can significantly reduce the total
rate at Borexino compatibly with present data and without giving
different kind of signals, such as 
day/night or seasonal variations.}
and instead focus on signals that LMA predicts
to be unobservable.
The region inside the continuous blue line has a
day/night asymmetry in the Beryllium rate larger than $0.02$.
In the region inside the dot-dashed blue line the Beryllium rate
shows anomalous seasonal variations with amplitude larger than $0.02$.
The physics behind these effects is similar
to the one well known from discussions of 
LOW and (Q)VO $\nu_e\to \nu_{\mu,\tau}$ oscillations (now excluded).

\item
We do not show results for a few other signals, that seem
less promising than the ones discussed above.
One can measure better the energy spectrum of Boron neutrinos,
measure Beryllium neutrinos both in NC and CC scatterings,
and look for day/night or seasonal variations in $pp$ neutrino rates.
Active/sterile oscillations with
 $\Delta m^2\sim 10^{-5}$
and $\tan^2\theta_{\rm s}\sim 0.1$ can distort the
$\bar\nu_e$ energy spectrum
in KamLAND or in future reactor experiments.

\end{itemize}
In table~\ref{tab:sintesi} we qualitatively classify
how MSW resonances with a sterile neutrino affect solar neutrinos.
The main variables are: which state is crossed, how large is the mixing.
  Fig.\fig{samples} shows  a few examples of specific oscillation patterns
 compatible with present data.

\begin{table}
  \centering 
  \begin{tabular}{c|cc}
& $\theta_{\rm ad}\ll \theta_{\rm s}\ll 1$     &  $\theta_{\rm s}\sim 1$  \\  \hline
$m_{\rm s}\gg m_2$     &no effect & $P_{es}\sim 1$\\
$m_{\rm s}\circa{>} m_2$  & 
$P_{es}=1$ at large $E_\nu$, 0 below& 
$P_{es}\sim 1$ increases at $E_\nu>E_*$\\
$m_{\rm s}\circa{<} m_2$ &no effect&
$P_{es}\sim 1$ decreases at $E_\nu >E_*$\\
 $m_{\rm s}\gg m_1$ &peak $P_{es}\sim 1$ at $E\sim E_*$
  & $P_{es}\sim 1$ decreases at $E_\nu >E_*$\\
 $m_{\rm s}\circa{>} m_1$ & $P_{es}\sim 1$ decreases at $E_\nu >E_*$
  & $P_{es}\sim 1$ decreases at $E_\nu >E_*$\\
\end{tabular}
  \caption{\label{tab:sintesi}\em Rough
 classification of possible sterile MSW resonances in the sun.
 $m_{\rm s}$, $m_1$ and $m_2$ are respectively the masses
 of the mostly $\nu_{\rm s}$, $\nu_1^{\rm a}$ and $\nu_2^{\rm a}$ states;
 $E_* \sim \hbox{few MeV}$ is the LMA critical energy and 
  $\theta_{\rm ad} \sim  (\Delta m^2/10^{-9}\eV^2)^{1/2}$.
  No effect is present if $\theta_{\rm s}\ll \theta_{\rm ad}$.
  }
\end{table}

\subsection{Hints and anomalies: solar data}
Finally, we have studied if present solar data contain hints of sterile neutrino effects.
There are two  ways of  searching for new effects:
data-driven or theory-driven.

In the data-driven approach one performs a goodness-of-fit test,
that should reveal if data contain some generic indication
for new physics beyond LMA oscillations.
In practice, this means carefully looking if data contain hints of anomalous results.\footnote{In the past,
global analyses of solar data reported the result of a Person $\chi^2$ test.
Today it would tell that even the LOW  solution (which has been excluded) is perfectly acceptable.
Therefore the fact that, according to the $\chi^2$ test,  LMA gives a good fit
is not a useful information.
In fact, as discussed in~\cite{freq}, the $\chi^2$ test is only sensitive to
variations in the $\chi^2$ larger than $\sqrt{N}$, where
$N\sim 100$ is the number of data points.
In the present case $N\gg1$: this explains why the $\chi^2$ test is so inefficient
and one has to perform less standard more efficient tests.}
Present neutrino data contain the following non-statistically-significant
hints: \begin{itemize}
\item The rate measured by the Chlorine experiment is $1.4$ experimental standard
deviations lower than the best-fit value predicted by active-only oscillations.\footnote{Before
revising solar model predictions including the recent LUNA data~\cite{LUNA}, 
the Chlorine rate was lower by $1.7\sigma$ (according to our results,
in agreement with other similar analyses, see e.g.~\cite{SmirnovSterile}).
The inclusion of LUNA results negligibly affects
the global fit of solar and KamLAND data in terms of active/active oscillations.}

\item The lack of an up-turn in present $\nu_e$ energy spectra
does not give a statistically significative
hint  for new effects additional to LMA oscillations.
To get a feeling of how accurately SNO and (mainly) SK data
test the energy spectrum around $E_\nu\sim 10\MeV$ we fit these data
assuming 
$$P(\nu_e\to \nu_e)=\sin^2\theta_{\rm sun} + \alpha (\hbox{MeV}/E_\nu)^2 = 
1-P(\nu_e\to \nu_{\mu,\tau}).$$
At energies probed by SK and SNO
LMA oscillations have this form with  $\alpha \approx 2$.
Data give $\alpha = 1\pm 2.5$.

\item A $2\div 3\sigma$ indication for sterile neutrinos might appear if
analyses that suggest higher values of the
nuclear factors $S_{17}(0)$ and $S_{34}(0)$
(and therefore a  Boron flux higher than what measured by SNO)
were confirmed~\cite{S17}.

\item According to some reanalyses,
solar neutrino rates of various experiments
show some time modulation at specific frequencies.
However subsequent analyses performed by some experimental collaborations
do not corroborate these claims~\cite{Pee(t)}.

\end{itemize}
The data-driven approach cannot see new physics that manifests
giving small corrections to many observables.
These diffuse minor effects can be seen if one knows what one is looking for.
More precisely, one performs a fit assuming a theory with $n$ extra parameters
and looks if the best-fit improves by more than $n$.
Our plots in fig.s\fig{sun}
explore a few $n=2$ representative slices of the full 4-dimensional 
 parameter space of relevant active/sterile oscillation parameters:
various regions (not shown) are favoured in a non statistically significant way.
The best fit values, indicated by the green dots in fig.s\fig{sun},
have a $\chi^2$ lower than best-fit LMA oscillations at most by 3 units.
In order to study the general $4\nu$ case we also computed local minima of the $\chi^2$ using numerical minimization techniques.
We have not found statistically significant indications for an extra sterile neutrino.

\subsection{What is the bound on the sterile fraction in solar oscillations?}\label{compare}
Assuming that solar neutrinos oscillate into
$\nu_e\to \sin\alpha~\nu_{\rm s} + \cos\alpha ~\nu_{\mu,\tau}$
the question is: what is the bound on the `sterile fraction' $\eta_{\rm s} \equiv \sin^2\alpha$?
We here compare with previous results.
Oscillation effects at Boron energies can be parameterized as
\beq\label{eq:PeeEta}
\Phi_{\nu_e} = \Phi_{\rm B} P_{ee},\qquad
\Phi_{\nu_{\mu,\tau}}=\Phi_{\rm B}(1-P_{ee})(1-\eta_{\rm s}),\qquad
\Phi_{\nu_{\rm s}} = \Phi_{\rm B}(1-P_{ee})\eta_{\rm s}.\eeq
SK and SNO have measured $\Phi_{\nu_e} $ and the total
flux of active neutrinos $\Phi_{\nu_{e,\mu,\tau}}$: these two measurements alone
cannot determine the three unknown quantities
$P_{ee}$, $\eta_{\rm s}$ and $\Phi_{\rm B}$ (unoscillated total Boron flux).
Adding the solar model prediction for $\Phi_{\rm B}$ gives
\beq\label{eq:etas}
\eta_{\rm s} \approx \frac{\Phi_{\rm B} - \Phi_{\nu_{e,\mu,\tau}}}{\Phi_{\rm B} - \Phi_{\nu_e}}\approx 0\pm 0.2\eeq
where $\eta_{\rm s}$ is small because $\Phi_{\rm B}$ agrees  with the measured  total $\nu_{e,\mu,\tau}$ flux.
In line of principle this agreement might be accidental:
$\Phi_{\rm B}$ could be larger than what predicted by solar models,
and a fraction of it could oscillate into sterile neutrinos leaving a reduced $\Phi_{\nu_{e,\mu,\tau}}$.
In view of all other constraints (on spectral distortions, Gallium rates,\ldots)
it seems difficult to realize this scenario with oscillations.

Most previous analyses prefer to obtain constraints on $\eta_{\rm s}$ only from this second argument.
They add one sterile neutrino mixed only with $\nu_{\mu,\tau}$
and separated by a large mass splitting: $\Delta m^2_{14} \gg \Delta m^2_{\rm sun}$.
This special configuration gives an energy-independent $\eta_{\rm s}$
and an energy dependence of $P_{ee}$ similar to what obtained for $\eta_{\rm s} =0$.
This choice does not allow to fit low energy Gallium data together with
high energy SNO and SK data: the resulting bound on $\eta_{\rm s}$
is~\cite{etas} \beq\label{eq:etas2}\eta_{\rm s} \approx 0 \pm 0.1,\eeq
somewhat stronger than the bound in eq.\eq{etas}
obtained using only the solar model prediction for the Boron flux.
We stress that this constraint on $\eta_{\rm s}$  can be relaxed by making milder assumptions on the 
active/sterile oscillation parameters.
For example, even still maintaining $\Delta m^2 \gg \Delta m^2_{\rm sun}$
$\eta_{\rm s}$ is energy-dependent if the heavy $\nu_{\rm s}$ mixes 
with $\nu_e$, or with the mass eigenstate $\nu_{1}$
($\eta_{\rm s}$ is about 2 times larger at lower energy, see fig.\fig{samples})
or with $\nu_2$  ($\eta_{\rm s}$ is about 2 times larger at higher energy).

When we focus on the special configuration considered by previous analyses
we get $\chi^2(\eta_{\rm s} = 1/4) - \chi^2(\eta_{\rm s} =0) = 8.3$ 
(including in the fit the BP00 prediction for the total Boron flux)
and $6.9$ (not including the prediction for the Boron flux: this value 
of the $\Delta \chi^2$ agrees with other analyses~\cite{etas}).
In fact, in the special configuration
our parameter $\theta_{\rm s}$  
is related to the sterile fraction as $\sin^2\theta_{\rm s}=2\eta_{\rm s}$
(for maximal atmospheric mixing).
We produced fig.\fig{sun}
including in our data-set the solar model prediction for the total Boron flux.
Dropping it would give only minor modifications,
as a comparison of eq.\eq{etas} with eq.\eq{etas2} indicates.


\section{Sterile effects in supernov\ae{} and other astrophysical sources of neutrinos}\label{SN}
Having discussed sterile effects in the sun, it is useful to discuss the analogous effects in 
core-collapse supernov\ae{} 
(SN) focussing on the differences between the two cases and emphasizing what one loses and
what one gains studying SN neutrinos.
\begin{itemize}

\item Present detectors can only study SN neutrinos from our galaxy,
and possibly from nearby galaxies: such
	SN neutrinos have  a duty time of  ${\cal O}(10)$ seconds every ${\cal O}(10^9)$ seconds~\cite{SNgeneral,raffelt book}.
This makes backgrounds less problematic, but allowed to detect so far 
only ${\cal O}(10)$ SN1987A events~\cite{SN1987a signal}.
Running solar neutrino experiments could detect  thousands of events 
from a future SN  exploding at distance $D\sim 10$~kpc.
An even more impressive harvest of data could come from a future Mton water-\v{C}erenkov  detector or from other more SN-oriented future projects~\cite{Cei}.

\item In a SN the matter density grows
from zero up to nuclear density ($\rho \simeq 10^{14} \gcm$) in the stiff inner core:
matter effects are important for all the mass range that we consider (up to about $10^2 \eV$). 
In this mass range, and given the typical SN neutrino energy of $\sim 10 \MeV$, active/sterile MSW resonances occur outside the neutrino-sphere (roughly defined as the regions after which neutrinos freely stream, $\rho \ll 10^{12}\gcm$). In particular, conversions inside the neutrino-spheres (which could drain almost all the SN energy) do not take place.\footnote{The resonances with the sterile state would enter in the neutrino-spheres (in the inner core) for $\Delta m^2 \circa{>} 10^5 \eV^2$ ($\circa{>} 10^7 \eV^2$ respectively).}
We only receive neutrinos that come out of the neutrino-spheres on the side of the SN
closer to us.
Unlike the case of the sun, it is not necessary to average
over the production point,
because the production and the oscillation regions
are spatially separated.


\item While only $\nu_e$ are produced in the sun,
SN produce all active $\nu$ and $\bar\nu$, 
roughly in similar amounts~\cite{raffelt book}. They mix and convert among themselves,
and possibly with sterile neutrinos~\cite{SNsterile2,SNsterile1,ShiSigl,SN3+1}. 
Present experiments can accurately  study $\bar\nu_e$. In fact
in the energy range $m_e \ll E_\nu \ll m_p$ relevant for SN neutrinos,
$\bar\nu_e p\to \bar{e} n$ scatterings allows to detect $\bar\nu_e$  and to measure their energy.
Detecting other neutrinos (e.g.\ via $\nu e$ scattering, deuterium dissociation, $\nue$ absorption on carbon) is possible and future projects could give significant information.
In the following, we focus on the $\bar\nu_e$ flux.

\item  $\bar\nu_{e,\mu,\tau}$ experience no active/active matter resonance 
(unlike $\nu_{e,\mu,\tau}$;
we are assuming normal hierarchy, and equal initial fluxes of muon and tau-neutrinos):
their effective $\Delta m^2$ in matter increases in a monotonous way
when the matter density grows.
The sterile effect that can more strongly affect the $\bar\nu_e$ rate
is a low-density $\nu_{\rm s}/\nu_1$ MSW resonance, possible
when the mostly sterile neutrino is lighter than $\nu_1$
(i.e.\ $\theta_{\rm s}>\pi/4$ in our parameterization).

Furthermore,  due to the peculiar composition of the inner part of the mantle (deleptonized matter) 
 the $\nu_e$ potential changes its sign in the deep region of the mantle, and it does so in a very steep manner~\cite{raffelt book, bethe,burrowlattimer,burrows} 
 (see fig.\fig{levels}b), adding a MSW resonance also for $\theta_{\rm s}<\pi/4$.
Although details are uncertain, this is a robust prediction.
It implies that $\nu_{\rm s}$ always meets a ``sharp'' resonance with $\bar\nu_e$ in the deep region of the mantle~\cite{SNsterile1,SN3+1}.

\item SN neutrinos are emitted from galactic distances,
allowing to probe vacuum oscillations with $\Delta m^2$ as low as $10^{-18}\eV^2$ 
(the precise value depend on the SN and on its distance from the earth: different SN  probe different ranges).

SN neutrino reach energies higher than solar $\nu$, perhaps up to $100\MeV$,
and could therefore  probe earth matter effects and spectral distortions.

\item While the sun is essentially static, 
an exploding SN is a dynamical environment (neutrino light-curve evolution, passage of the shock wave...); including the time dependence in the neutrino fluxes and in the matter density profile is too much demanding and probably useless for our purposes, 
given the poor knowledge of the details. We focus on a typical SN configuration, 
which includes all the characteristic features of the SN cooling phase.

 A reliable prediction of the emitted fluxes is still lacking and the explosion mechanism is not yet under control. 
However, it is relatively easier to predict the  $\nu_{e,\mu,\tau},\bar{\nu}_{e,\mu,\tau}$ energy spectra,
which in thermal approximation only depend on their cross sections.
On the contrary their total fluxes also depend on the SN density profile;
in particular the thermal approximation does not imply equipartition among different flavours.
The underlying complexity of SN explosions makes hard to reduce 
theoretical  uncertainties, posing a threat on the usefulness of
precise SN $\nu$ experiments as tools for studying oscillations.

As a last warning, we have to recall the 
  reluctance of the neutrino data from the only SN we know 
  to fit into a simple and straightforward interpretation. 
  The main reasons are: the energy distribution of Kamiokande 
  and IMB look different; both experiments have an excess of forward 
  events; the 5 LSD events cannot be accounted for.

\end{itemize}
As for the future, there are various interesting observables, and it is
difficult to guess on which signals we should focus.
This will probably depend on  compromises among future experimental and theoretical capabilities.

Several SN-related  bounds on the active/sterile neutrino mixing have been considered in the literature~\cite{SNsterile1,SNsterile2, ShiSigl,fuller r-process}. 
Focussing on the cooling phase neutrinos, they include ({\bf i}) the detection of $\nueb$ from SN1987a in the Kamiokande and IMB experiments~\cite{SN1987a signal},
which sets a  constraint on the portion of neutrinos that oscillate into $\nu_{\rm s}$;
 ({\bf ii}) r-process nucleosynthesis: a fraction of the heavy elements in nature is supposed to be synthesized in the region surrounding the core of the exploding stars, provided that the electron fraction $Y_e < 0.5$; even small modifications of neutrino fluxes affect the process, so that the request of a successful nucleosynthesis has been used to set limits (see e.g.~\cite{fuller r-process, SNsterile1}); 
 ({\bf iii}) re-heating of the shock: in the delayed-shock/neutrino-driven picture of SN explosion, the flux of neutrinos and antineutrinos from the early stages of the accretion phase (which should carry $10\div20 ~\%$ of the total flux) are responsible for the actual explosion of the star, pushing from below the stalling shock wave; since $\nu_e$
and $\bar\nu_e$ are the most effective in this (interacting with charged and neutral
currents with baryonic matter), a cut of their flux would prevent this mechanism from working~\cite{SNsterile1,ShiSigl,sterileshock}; 
({\bf iv})...
We will only consider the bound from direct observation, which we consider
robust enough.\footnote{Indeed, for instance, alternative sites for effective nucleosynthesis have been repeatedly proposed. Moreover, the position of such a bound could depend quite heavily on the complicated details of SN dynamics and, in addition, on the interplay of sterile oscillations with it, see below. On the same footing, since the details and the nature
itself of the neutrino-driven explosion mechanism are still to be fully understood, we do not
consider it here as a robust constraint, not mentioning that it would require a demanding
understanding and simulation of the evolution of the SN mantle.}

\subsection{Technical details}

The analysis of SN neutrinos can be carried on in a way similar to that of solar neutrinos,
after adapting the computational procedure to SN  peculiarities,
which introduce some complications and allow some simplifications.
We must follow the fate of the neutrinos emitted from neutrino-spheres along their travel through the star matter, the vacuum and the earth.
The density matrix formalism, already described in the `solar' section~\ref{suntec}, 
automatically handles 
the complication that the SN initial neutrino flux has a mixed flavour composition.
In the SN case one should follow the evolution of two $4\times 4$
density matrices: $\rho(E_\nu)$ for neutrinos,
and $\bar\rho(E_\nu)$ for anti-neutrinos.
We focus on the total $\bar\nu_e$ rate, as measured
by $\bar\nu_e p$ scatterings with the cuts and efficiency of the KamiokandeII experiment.
The cross section is taken from~\cite{IBD}.

\smallskip

MSW resonances affect the neutrino density matrix, possibly introducing off-diagonal elements
which are however averaged to zero by large oscillation phases.
Therefore we can combine probabilities rather than quantum amplitudes:
the $4\times 4$ density matrix can be replaced by a $4$-vector containing its diagonal elements,
$\Phi = (\rho_{11},\rho_{22},\rho_{33},\rho_{44})$
and the evolution equation 
$\rho  = \mathscr{U}\cdot \rho^0\cdot  \mathscr{U}^\dagger$ 
($ \mathscr{U}= \mathscr{U}_n\cdots  \mathscr{U}_1$)
by~\cite{SNsterile2}
\beq\Phi = \mathscr{P}_n\cdots \mathscr{P}_1\cdot  \Phi^0
\qquad\hbox{where}\qquad (\mathscr{P}_n)_{ij} = |(\mathscr{U}_n)_{ij}|^2 \eeq
are  $4\times 4$ matrices of conversion probabilities.
At the production region, matter effects are dominant, so that
matter eigenstates coincide with flavour eigenstates
(up to a trivial permutation):
$\Phi^0 = ( \Fso,\Feo, \Ftauo,\Fmuo)$.
$\Phi$ are the fluxes of neutrino vacuum eigenstates reaching the earth surface,
related to the fluxes of flavour eigenstates by an incoherent sum 
weighted by the the neutrino mixing matrix $V$:
\begin{equation}
\left( \Fe,
 \Fmu ,
 \Ftau ,
 \Fs \right)_\alpha 
=  \sum_i |V_{\alpha i}|^2 \Phi_i  
\end{equation}
If neutrinos cross the earth, earth matter effects~\cite{SNmattereffects} may reintroduce
coherencies among the various fluxes, 
and one must return to the general quantum expression.
Similarly, if two states have $\Delta m^2\circa{<} 10^{-18}\eV^2$
vacuum oscillations do not give large phases:
evolution in the outer region of the SN and in vacuum
must be described keeping the off-diagonal components
of the matrix density.
In practice, we found convenient to evolve the density matrix
in the instantaneous mass eigenstate basis,
setting to zero its off diagonal elements when they accumulate large phases.
This procedure is described in greater detail in the `atmospheric' section~\ref{AtmDetails}.

\smallskip

The effect of active/active oscillations is well known~\cite{SNmattereffects}:
the LMA resonance is adiabatic and partially swaps active neutrinos, giving\footnote{We are 
assuming normal hierarchy and $\theta_{13}=0$.
The situation becomes  more complicated if instead $\theta_{13}\circa{>} 1^\circ$,
because the atmospheric resonance starts changing the result in eq.\eq{SN3}.
For $1^\circ\ll \theta_{13}\ll 1 $  it is adiabatic.
If $\Delta m^2_{23}>0$ (normal hierarchy) it affects only neutrinos
by  giving a full conversion, $\Phi_{\nu_e}^{\rm after} =  \Phi_{ \nu_{\mu,\tau}}^{\rm before}$.
If instead $\Delta m^2_{23}>0$ (inverted hierarchy) it affects only anti-neutrinos,
by giving  a full conversion,
 $\Phi_{\bar\nu_e}^{\rm after} =  \Phi_{ \bar\nu_{\mu,\tau}}^{\rm before}$.}
\beq\label{eq:SN3}\Phi_{\bar \nu_e}^{\rm after} = \cos^2\theta_{\rm sun} \Phi_{\bar \nu_e}^{\rm before} + \sin^2\theta_{\rm sun} 
 \Phi_{\bar \nu_{\mu,\tau}}^{\rm before}\qquad
 \Phi_{ \nu_e}^{\rm after} = \sin^2\theta_{\rm sun} \Phi_{ \nu_e}^{\rm before} + \cos^2\theta_{\rm sun} 
 \Phi_{ \nu_{\mu,\tau}}^{\rm before}\eeq
 Active/sterile MSW resonances can occur after of before the LMA resonance.

\bigskip

Concerning the initial fluxes, the accurate results of simulations are usually empirically approximated by a so called ``pinched'' Fermi-Dirac spectrum for each flavor $\alpha=\nue,\nueb, \nu_x$
($\nu_x$ collectively denotes $\nu_{\mu,\tau},\bar\nu_{\mu,\tau}$)~\cite{jankahillebrandt, raffelt book} 
\begin{equation}\label{eq:SNspettro}
\frac{d\Phi_{\nu_\alpha}}{dt\,dE_\nu}(E_\nu,t) = N(\eta_\alpha)
\frac{120}{7 \pi^4} \frac{L_\alpha}{T^4_\alpha}\frac{E_\nu^2}{e^{E_{\nu}/T-\eta_{\alpha}}+1}
\end{equation}
where the pinching parameter $\eta_{\alpha}$ takes the typical values $\eta_{\nue} \sim 5 - 3$, $\eta_{\nueb} \sim 2.5 - 2$, $\eta_{\nu_x} \sim 0 - 2$.
For $\eta_\alpha = 0$ the normalization factor is $N=1$ and $T=\langle E_\nu\rangle/3.15$.
Based on the recent results of~\cite{keil}, we adopt the following average energies and 
total luminosities for the various neutrino components
at the time of the snapshot of fig.\fig{profiles} (see below)
\beq\label{eq:Tnu}\begin{array}{lll}
\langle E_{\nu_e}\rangle\simeq12 \MeV,\qquad&
\langle E_{\bar\nu_e}\rangle\simeq14 \MeV,\qquad &
\langle E_{\nu_x}\rangle\simeq14 \MeV\cr
L_{\nu_e}\simeq30 \cdot 10^{51} \erg \sec^{-1},&
L_{\bar\nu_e}\simeq30 \cdot 10^{51} \erg \sec^{-1}, &
L_{\nu_x}\simeq20 \cdot 10^{51} \erg \sec^{-1}.
\end{array}
\eeq
 In accordance with numerical calculations, we shall assume 
that the ratios of luminosities do not vary much during the 
whole emission.
The initial flux of sterile neutrinos is assumed to be vanishing, as a consequence of the fact that matter oscillations only take place out of the neutrinosphere.\footnote{To be precise, shortly after the collapse the electron neutrino matter potential in the very center of the core is positive, due to the contribution of the trapped neutrinos themselves. This configuration lasts for a short transient period, until the neutrino diffusion depletes their abundance and carries the potential to negative values, where it stays. A small fraction of the electron (anti)neutrinos produced in the deep core could then oscillate into sterile states and constitute a non vanishing flux injected in the mantle.}

\begin{figure}[t]
\begin{center}
\includegraphics[width=17cm]{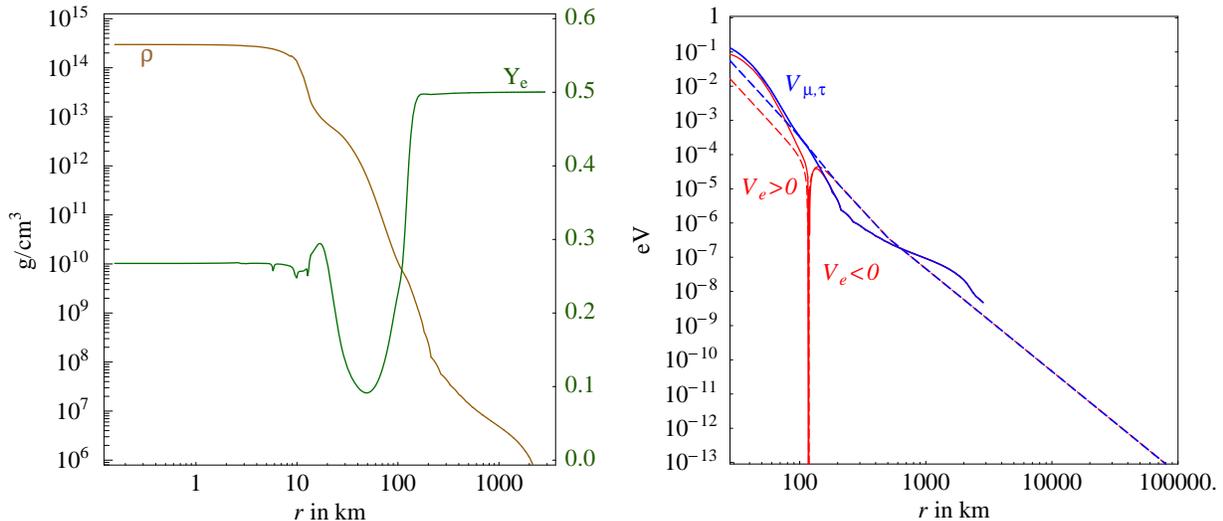}
\caption{\label{fig:profiles}\em {\bf Supernova profile}.
Fig.\fig{profiles}a:
{\color{rosso} Density $\rho(r)$} and {\color{verdes} electron fraction $Y_e(r)$} from~\cite{burrows}.
Fig.\fig{profiles}b: matter potentials in the SN, for {\color{rossos} $\bar\nu_e$} (red solid line) and for {\color{blu} $\bar{\nu}_{\mu,\tau}$} (blue solid line).
The dashed lines are the analytic modelization that we adopt.
}
\end{center}
\end{figure}

\bigskip

Neutrinos traveling in the SN matter experience the MSW potentials~\cite{MSW,KPreview}
\beq\label{sys:potentials}\begin{array}{ll}
V_e =\sqrt{2} \GF \nB \left( 3 Y_e - 1\right)/2,\qquad  & V_\tau =V_\mu + V_{\mu\tau}, \\
\displaystyle
V_\mu = \sqrt{2} \GF \nB \left({Y_e} -1 \right)/2,  & V_{\rm s} = 0, \end{array}
\eeq
where $\nB$ is the baryon number density ($\nB =
{\rho}/{m_N}$ where $m_N \approx 939 \MeV$ is the nucleon mass\footnote{For the very high densities closer to the core, $m_N$ should be replaced by a (quite different) effective nucleon mass.
We can neglect this refinement in the regions of our interest.}) and $Y_e=(N_{e^-} - N_{e^+})/\nB$ 
is the  electron fraction per baryon.
Antineutrinos experience the same potentials with opposite sign.

The difference $V_{\mu\tau}$ in the $\num$ and $\nut$ potentials, which appears at one
loop level due to the different masses of the muon and tau leptons \cite{mutau}, is,
according to the SM
\begin{equation}
V_{\mu \tau}=\frac{3\GF^2 m^2_{\tau} }{2 \pi^2} \left[ 2 (n_p + n_n) \ln \left(
\frac{M_W}{m_\tau} \right) -n_p -\frac{2}{3}n_n \right].
\end{equation}
The effect is not irrelevant in the inner dense regions: for densities above $\rho \sim 10^{8} \gcm$, the $\mu\tau$ vacuum mixing is suppressed.

A crucial point concerns the characteristic of the matter density and of the electron fraction in the mantle of the star. We adopt the profiles represented in fig.\fig{profiles}~\cite{burrows} and we model them with analytic functions that preserve their main features.\footnote{We thank Adam Burrows for having provided us with the data, and for useful discussions.}
Namely, the density profile decreases according to a power law $r^{-4}$ out of the $\sim 10 \km$ inner core (where instead it has a roughly constant, nuclear density value). 
At much larger distances the density profile gets modified in a time-dependent way by the
passage of the shock wave.
Present simulations have difficulties in reproducing this phenomenon
and therefore cannot reliably predict the density profile in the outer region.
Therefore for $r\circa{>}500\km$ 
we assume a power law $\rho = 1.5~10^4(R_\odot/r)^{3}{\rm g/cm}^3$, which roughly
describes the static progenitor star.

The peculiar $Y_e$ profile in fig.\fig{profiles}
is inevitably dictated by the deleptonization process~\cite{burrowlattimer,bethe}: 
behind the shock wave which has passed in the mantle matter, the electron capture on the newly liberated protons is rapid, driving $Y_e$ to low values ($\sim 1/4$). 
In the outer region, where the density is sensibly lower, the efficiency of the capture is much lower, so that $Y_e$ essentially maintains the value $\sim 1/2$ typical of normal matter.
This is important because the matter potential $V_e$ of electron (anti)neutrinos 
flips sign, see eq.~(\ref{sys:potentials}),
 when, in the deep region of the mantle $Y_e$ steeply decreases below $1/3$. 
 On the contrary, at this point the matter potentials of muon and tau (anti)neutrinos are 
 marginally affected.
 Both are plotted in fig.\fig{profiles}.

The data refer to $\sim$0.3 sec after bounce for a typical star of $\sim 11$ solar masses. 
The subsequent evolution is supposed to move the wave of the $Y_e$ profile slightly outwards, maintaining, however, its characteristic shape. 
The slight dependence on the progenitor mass, in turn, is not really relevant~\cite{burrows 2}.

 This SN density profile has been computed~\cite{burrows} in absence of sterile neutrinos effects.
Adding a sterile neutrino, $\nu_e\to \nu_{\rm s}$ conversions can reduce $Y_e$ due to
a non-trivial feedback mechanism on the MSW potential 
experienced by neutrinos~\cite{SNsterile1,fuller r-process}. 
This could even  create a new intermediate region with $Y_e<1/3$,
thus introducing two more level-crossing in the $\bar\nu_e$ channel. 
We neglect these possible extra MSW resonances because they 
do not affect the $\bar\nu_e$ rate when they are both 
adiabatic, or both fully non adiabatic.

In summary, although the profiles that we adopt come from a specific computation and refer to a specific instant in time, they incorporate the peculiar features that are important for our purposes. A more refined treatment of this point (later times behavior of the profiles, fine structures connected with the passage of the shock wave...) would of course require to obtain first a complete simulation of the SN evolution, including the explosion. This could be needed
to describe the signal from a future supernova, but seems unnecessarily complicated for SN1987A.

\subsection{Results}

\begin{figure}
$$\includegraphics[width=17cm]{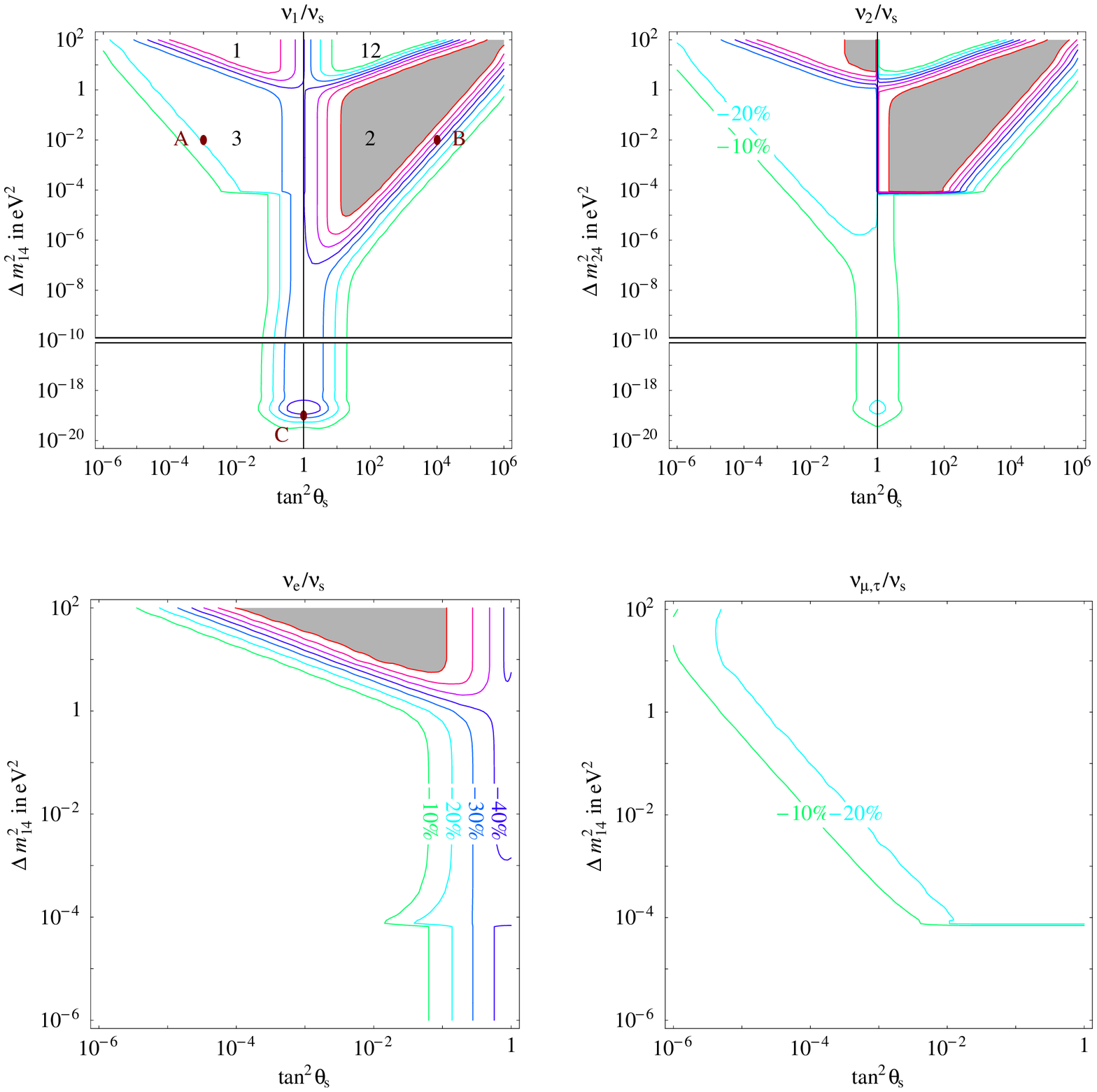}$$
\caption{\label{fig:SN}{\bf Sterile effects in supernov\ae.}
\em 
The iso-contours correspond to a ${\color{verdes}10}$, ${\color{blucc}20},
{\color{blu}30}$, ${\color{blus}40},{\color{viola}50}$, ${\color{viola2}60},{\color{rossos}70}\,\%$ deficit
of the SN $\bar\nu_e$ total rate due to oscillations into sterile neutrinos.
The deficit is measured with respect to the rate in absence of
active/sterile oscillations and in presence of active/active oscillations
(which reduce the no-oscillation rate by $\sim 10\%$).
We shaded as disfavoured by SN1987A data regions with
a deficit larger than $70\%$.
While the qualitative pattern is robust,
regions with MSW resonances can shift by one order of
magnitude in $\Delta m^2$ using different SN density profiles.
$\bar\nu_3/\bar\nu_{\rm s}$ mixing (not plotted) does not give significant effects. 
Fig.\fig{SNsample} studies in detail the sample points here marked as {\rm A, B, C},
and the regions 1, 2, 12, 3 are discussed in the text.}
\end{figure}

Active/sterile mixing significantly affects SN neutrinos for $\theta_{\rm s}\sim1$.
Due to MSW resonances, significant effects can also be present for
small mixing, i.e.\ $\theta_{\rm s}\to 0$ and $\theta_{\rm s}\to  \pi/2$.
The recently established active/active mixings are not taken into
account in older studies~\cite{SNsterile2, ShiSigl} (see however~\cite{SN3+1}).
To understand the main features it is useful to look at the pattern of possible
level crossings, qualitatively depicted in fig.\fig{levels}b at page~\pageref{fig:levels}.
The three mostly active anti-neutrino eigenstates depend on the radius $r$ as dictated
by their measured masses and mixings (we assume $\theta_{13}=0$ and normal hierarchy)
and by the predicted SN density profiles,
and are represented in fig.\fig{levels}b by the three colored curves.
The mostly sterile neutrino is generically
represented by an almost-horizontal line, which is
plotted in  fig.\fig{levels}b in the specific case of small $\bar\nu_e/\bar\nu_{\rm s}$ mixing
and $\Delta m^2_{14}\gg \Delta m^2_{\rm atm}$.
There are three possible kinds of active/sterile MSW resonances~\cite{SNsterile1,SN3+1,SNsterile2, ShiSigl}:
\begin{itemize}
\item[1.] The mostly $\bar\nu_{\rm s}$ eigenstate  crosses the mostly $\bar\nu_e$ eigenstate 
at $r\sim 100\km$, where $V_e$ flips sign.
At this point matter effects dominate over active neutrino masses,
so that active mass eigenstates coincide with flavour eigenstates.
Since $V_e$ flips sign in a steep way this resonance is effective
only if $\Delta m^2_{14}\circa{>} 10^{-1\div0}\eV^2$
(different SN simulations gives values in this range).

\item[2.] If the mostly sterile eigenstate is the lightest one
(in our parameterization this needs $\theta_{\rm s}\circa{>} \pi/4$)
the two eigenstates in 1.\ cross again at larger $r$.
Pictorially,  this second resonance is present when 
the sterile black line is lower than what assumed in fig.\fig{levels}b.
This MSW resonance occurs at large $r$  where $V_e$ is smooth,
so that it is effective down to $\Delta m^2_{14}\circa{>} 10^{-6\div 8}\eV^2$.
Again, the significant uncertainty is due to uncertainties on the SN density gradient.

\item[3.] If instead the mostly sterile eigenstate is
the heaviest or the next-to-heaviest state, it crosses
one or both of two mostly $\bar\nu_{\mu,\tau}$ eigenstates.
This is the case illustrated in fig.\fig{levels}b.
The values of $\Delta m^2_{24}$ and $\Delta m^2_{34}$
 determine at which $r$ these crossings takes place,
 and consequently the  flavour composition of the mostly active states at the resonance.
 Entering in the SN, the small $\bar\nu_e$ component of $\bar\nu^m_{2,3}$
 disappears as soon as 
$V_e-V_\mu$ dominates over $\Delta m^2_{\rm sun}$.\footnote{When
$V_{\mu}-V_{\tau}$ dominates over $\Delta m^2_{\rm atm}$ 
the flavour composition varies from $\bar\nu_\mu\pm\bar\nu_\tau$ to
$\bar\nu_\mu$ and $\bar\nu_\tau$.
This active/active resonance
happens at so high densities that is not relevant for our purposes.}
The color of mass eigenstates in fig.\fig{levels}b illustrates these phenomena.
In any case, active/sterile MSW resonances with the mostly $\bar\nu_{\mu,\tau}$ states
do not significantly affect the $\bar\nu_e$ rate, see eq.\eq{SN3}.
\end{itemize}
These considerations allow to understand fig.s\fig{SN},
where we plot our results for the reduction of the $\bar\nu_e$ rate
due to sterile mixing.

\begin{figure}
$$\hspace{-7mm}\includegraphics[width=18cm]{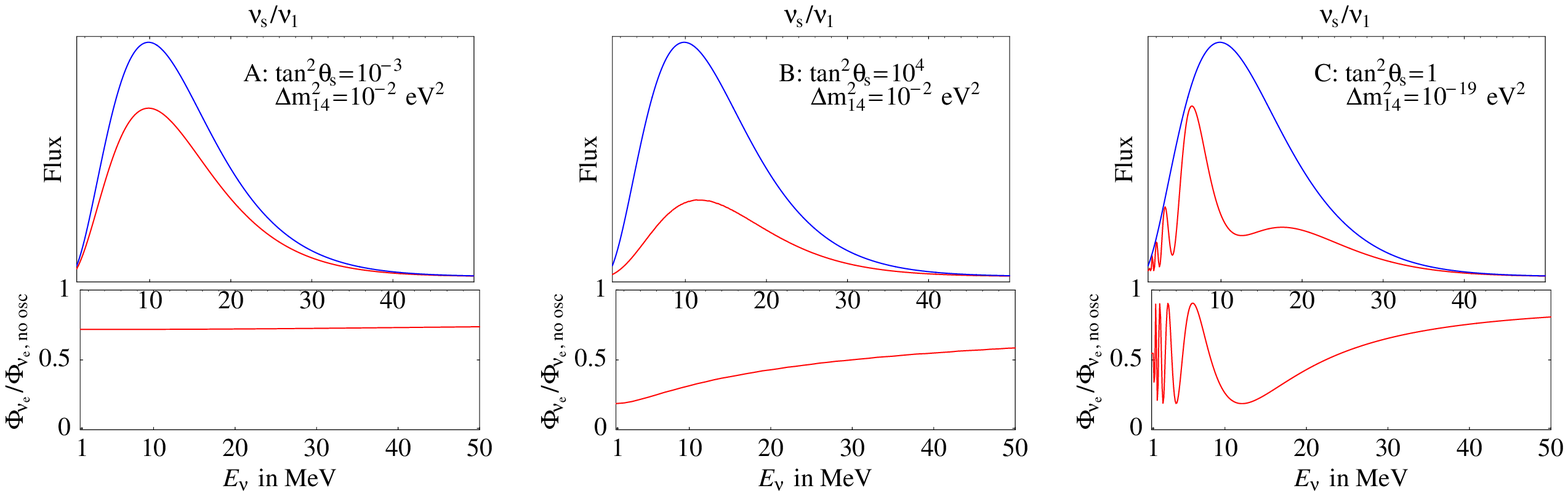}$$
\vspace{-1.5cm}
\caption{\label{fig:SNsample}{\bf Sterile effects in supernov\ae.}
\em Distortion of the $\bar\nu_e$ flux at sample points {\rm A, B, C}.}
\end{figure}

\smallskip

Let us start from fig.\fig{SN}a that studies $\nu_{1}/\nu_{\rm s}$ mixing.
Resonance 1 gives a sizable reduction in region 1 of fig.\fig{SN}a
and resonance 2 gives a sizable reduction in region 2.
Had we ignored solar mixing the maximal deficit would have been $100\%$,
while in presence of solar oscillations the
maximal effect is a $\sim 80\%$ deficit (see also~\cite{SN3+1}).
More precisely, in the interior of region 1 one obtains
$\Phi_{\bar\nu_e} = \sin^2\theta_{\rm sun} \Phi^0_{\bar\nu_e}$
because resonances 1 and 3 are fully adiabatic.
In the interior of region 2 one obtains
$\Phi_{\bar\nu_e} = \sin^2\theta_{\rm sun} \Phi^0_{\bar\nu_{\mu,\tau}}$
because resonance 2 is fully adiabatic and resonance 1 irrelevant.
Therefore,  given the assumed initial fluxes, 
the $\bar\nu_e$ rate gets reduced  slightly more strongly in region 2
than in region 1.

In region 12 both resonances 1 and 2 are effective, 
and tend to compensate among each other:
resonance 1 converts $\bar\nu_e$ into $\bar\nu_{\rm s}$
and resonance 2 reconverts $\bar\nu_{\rm s}$ into $\bar\nu_e$.
In region 3, resonances 3 gives a $20\%$ suppression of the $\bar\nu_e$ rate,
that sharply terminates when $\Delta m^2_{14}< \Delta m^2_{\rm sun}$.
This is due to a strong suppression of the mostly $\bar\nu_{\mu,\tau}$ eigenstates,
which due to solar oscillations would give a $20\%$ contribution to the $\bar\nu_e$ rate
(ignoring solar mixing, there would be no suppression of the $\bar\nu_e$ rate in region 3).
This reduction of the $\bar\nu_{\mu,\tau}$ fluxes induced by resonances 3 
could be better probed by measuring the NC rate (which gets a $\circa{<}40\%$ reduction)
and, if neutrinos cross the earth, by distortions of the $\bar\nu_e$ energy spectrum.
At smaller $\Delta m^2$ and around maximal mixing
vacuum oscillations can reduce the $\bar\nu_e$ rate by $\circa{<}50\%$:
their effect persists down to  $\Delta m^2\sim E_\nu/D\sim 10^{-18}\eV^2$.
The precise value depends on the distance $D$
(we assumed $D=10$ kpc).

The other mixing cases are understood in similar ways.
We remind that our parametrization is discontinuous at $\theta_{\rm s}=\pi/4$:
this is reflected in fig.\fig{SN}b, where we consider $\nu_{2}/\nu_{\rm s}$ mixing.
Resonance 2 sharply terminates when $\Delta m^2_{24}<\Delta m^2_{\rm sun}$.
Up to these differences, this case is quite similar to the previous one, 
because $\bar\nu_1$ and $\bar\nu_2$ get strongly mixed by matter effects.

On the contrary $\nu_{3}/\nu_{\rm s}$ mixing (not shown) does not give a significant reduction
of the $\bar\nu_e$ rate.
Mixing with the flavour eigenstates behaves in a similar way. 
Namely, 
the $\nu_{\mu,\tau}/\nu_{\rm s}$ figure shows the reduction in region 3 due to resonances 3,
and the $\nu_e/\nu_{\rm s}$ figure  shows the reduction in region 1 due to resonance 1.
The additional feature at $\Delta m^2_{14} \sim \Delta m^2_{\rm{sun}}$ is
due to adiabatic conversion (for large sterile angles) between the mostly-sterile state and $\nu_2$, 
that are almost degenerate in this condition. 
For larger $\Delta m^2_{14}$, there is no $\bar\nu_e$ component in $\bar\nu^m_2$ so that the crossing is totally non adiabatic, while for smaller $\Delta m^2_{14}$ the two states are separated. 
As in the solar case, sterile effects persists at all values of $\Delta m^2_{14}$, even if it is small.
Unlike in the solar case, such effects are not sensitive to vacuum oscillations
(i.e.\ $\Delta m^2\sim E_\nu/D\sim 10^{-18}\eV^2$) because in a SN 
neutrinos of all energies experience the adiabatic LMA resonance.
Therefore in the lower row of fig.s\fig{SN}
we only show results at $\Delta m^2_{14}>10^{-6}\eV^2$;
nothing changes at smaller values.

\medskip

SN1987A data can be precisely compared with expectations 
doing an event by event fit~\cite{SNfit}; however the result strongly depends
on the assumed average energies and total luminosities of the initial fluxes.
The data do not permit detailed studies of the energy distribution.
Even more, they cannot discriminate in a significant way between a larger
total flux with smaller average energy and a smaller flux with higher
average energy.
%
In absence of a quantitative estimation of theoretical uncertainties,
and in view of the doubtful aspects of SN1987A data discussed previously,
today we cannot derive precise constraints.
Therefore we simply shaded as `disfavoured' regions where sterile effects reduce the 
$\bar\nu_e$ rate by more than $70\%$.
This is nothing more than a reasonable arbitrary choice.

\medskip

Future data will permit to know precisely the total rate of
$\bar\nu_e$ events, and also its distribution in energy and time.
The most important single observable  could be the average $E_{\bar\nu_e}$ energy,
or more generically the energy spectrum.
Rather than showing iso-contour plots of $\langle E_{\bar\nu_e}\rangle$
we describe their main features.
In absence of sterile oscillations we expect $\langle E_{\bar\nu_e}\rangle \approx 15\MeV$.
Along the `diagonal sides of the MSW triangles' 1 and 2
(e.g. around our sample point B)
the average energy increases up to $\langle E_{\bar\nu_e}\rangle \approx 18\MeV$.
Along the `diagonal sides of the MSW triangle' 12
it can decrease down to $\langle E_{\bar\nu_e}\rangle \approx 11\MeV$.
Vacuum oscillations can give the well known distortions of the spectrum,
as exemplified in fig.\fig{SNsample}C.
These effects seem larger than experimental and theoretical uncertainties.
In all other cases sterile effects give a quasi-energy-independent suppression of
the $\bar\nu_e$ rate, and therefore negligibly affect $\langle E_{\bar\nu_e}\rangle$.
Fig.\fig{SNsample}A gives an example.
Of course, the average energy of positrons generated by $\bar\nu_e p \to n \bar{e}$ scatterings
is higher than  $\langle E_{\bar\nu_e}\rangle$ because the cross section increases with energy.

\bigskip

We  focused on supernova 
antineutrinos.
The analogous plots for neutrinos cannot obtained from
our $\bar\nu$ plots by flipping $\tan\theta_{\rm s}\to 1/\tan\theta_{\rm s}$,
because we are taking into account the effects of active/active oscillations.
Effects in neutrinos are more similar to what happens to solar $\nu_e$.





\medskip

\subsection{Hints and anomalies: supernov\ae}
To conclude, we list hints 
of anomalous effects possibly related to sterile effects in SN neutrinos:
\begin{itemize}
\item In the past, the initial SN neutrino fluxes in different flavours
were believed to follow an 
almost exact equipartition of the energy and have
gaps in neutrino temperatures larger than in eq.\eq{Tnu} 
(typical values were $13, 16, 23 \MeV$~\cite{raffelt book}).
Under these assumptions SN1987A data, which point to a lower average energy,
disfavor solar oscillations with large mixing angle
($\sin^2 2\theta_{\rm sun}<0.9$ at $99\%$ C.L.~\cite{LMAdead})
while solar data established the relatively large mixing $\sin^2 2\theta_{\rm sun}\approx 0.8$.
The tension strongly depends on the initial temperatures,
and disappears assuming
the more recent values in eq.\eq{Tnu}~\cite{LMAdead}.

\item Pulsar motion. It has been proposed that the resonant (or non resonant~\cite{PulsarMotion}) conversion of sterile neutrinos could explain the observed large proper velocity of newly born neutrons stars~\cite{kusenko segre}. Indeed, in presence of very strong and axially oriented magnetic fields, which could be plausible in the NS environment, the MSW potential includes a contribution (at one loop, from neutrino scattering on the polarized medium) which depends on the relative angle between the neutrino momentum and the magnetic field; this leads to an asymmetric neutrino emission, which could be enough to account for the observed velocities.
This mechanism needs sterile neutrinos with keV-scale masses and $\theta_{\rm s}\circa{<}10^{-7}$.

\item r-process nucleosynthesis~\cite{fuller r-process}: $\nu_e \to \nu_{\rm s}$ conversions in the mantle of the star could drive $Y_e$ to low values and have thus been proposed to provide a favorable environment for r-process nucleosynthesis. The relevant range of $\nu_e/\nu_{\rm s}$ 
mixing parameters spans $\Delta m^2 \sim (1 \div 10^2) \eV^2$ and 
$\sin^2 2 \theta_{\rm s} \sim 10^{-3} \div \textrm{few} 10^{-1}$, 
which should be checked against the bounds from cosmological probes that we discussed in the present paper.
In the same region, the ``conversion plus re-conversion'' of antineutrinos instead guarantees that the SN $\bar\nu_e$ is not dramatically affected.

\end{itemize}

\subsection{Neutrinos from other astrophysical sources}
Different kinds of experiments will try to
detect neutrinos emitted by extragalactic sources,
such as active galactic nuclei.
There are no firm expectations.
In particular we do not know if the fluxes of these neutrinos
will be detectably large, and eventually in which energy range.
However, since these neutrinos are presumably
mostly generated by $\pi$ decays,
like atmospheric neutrinos, one expects
similar fluxes of $\nu$ and $\bar\nu$ with
flavour ratio $e:\mu:\tau\sim1:2:0$
(this expectation might be wrong).
Atmospheric oscillations then 
convert the flavour ratio into $1:1:1$, which is blind
to other active/active oscillations.

In presence of an extra sterile neutrino,
the $e:\mu:\tau:{\rm s} \sim 1:1:1:0$ flavour ratio
is not blind to extra active/sterile oscillations.
Sterile oscillations can reduce the fluxes of active neutrinos,
and vary the relative flavour proportion.
Since the total initial fluxes and energy spectra are  unknown, and
since experiments will probably be able of
tagging $\mu$ and maybe $\tau$ neutrinos well~\cite{UHE},
we focus only on the observable $\Phi_\mu/\Phi_\tau$.

We assume that the baseline, about 100 Mpc,
is much longer than all oscillation lengths.
In such a limit the oscillation probability reduces to
multiplication of probabilities (rather than of quantum amplitudes):
\beq \label{eq:degosc}P(\nu_\ell\to \nu_{\ell'})=
P(\bar\nu_\ell\to \bar \nu_{\ell'}) = \sum_{i=1}^4 |V_{\ell i}|^2 |V_{\ell'i}|^2
\eeq
One can verify that sterile mixing with a flavour eigenstate
$\nu_\ell$ ($\ell = e$ or $\mu$ or $\tau$)
mainly gives a depletion of $\Phi_\ell$.
However, as discussed in the rest of this paper,
in such a case large active/sterile 
mixing angles are already disfavoured by other data,
so that it is not possible to get a sizable effect.

Present data allow large active/sterile mixing with mass eigenstates 
$\nu_i$ ($i=1$ or $2$ or $3$)
if $\nu_{\rm s}$ and  $\nu_i$ are quasi-degenerate,
with $\Delta m^2 \ll 10^{-9}\eV^2$.
In view of the long base-line, neutrinos from cosmic sources
are affected by oscillations with $\Delta m^2 \circa{>}10^{-17}\eV^2$.
However each of the mostly active mass eigenstates contains roughly equal component
of $\nu_\mu$ and $\nu_\tau$ (unless the atmospheric mixing angle is non-maximal,
or unless $\theta_{13}$ and the CP-phase significantly differ from zero):
therefore $\nu_{\rm s}/\nu_i$ oscillations do not significantly affect
$\Phi_\mu/\Phi_\tau$.

In conclusion, near-future experiments that
will try to discover neutrinos from cosmic sources
do not seem to allow promising searches of sterile neutrino oscillations.
See also~\cite{SterileUHE}.

\medskip

Eq.\eq{degosc} also allows to study vacuum oscillation effects in other
kinds of cosmological neutrinos:
\begin{itemize}

\item In a near future it seems possible to
detect the $\bar\nu_e$ emitted by {\bf past core-collapse supernov\ae}.
At the moment we only have order-of-magnitude predictions,
but future SN experiments and studies might allow to
better predict their total flux or spectrum.
SK almost reached the apparently necessary sensitivity,
and it seems possible to improve the efficiency 
of tagging neutrons emitted in $\bar\nu_e p \to \bar{e} n$ scatterings~\cite{SKbarnu}.
Relic SN $\bar\nu_e$ are affected by oscillations in the SN (as discussed
in the previous section) and by oscillations in vacuum (down to $\Delta m^2 \circa{>} 10^{-25}\eV^2$).

\item In a far future, it might be possible to directly detect
{\bf CMB neutrinos} by coherent scatterings.
This would allow to probe oscillations down to $\Delta m^2\sim 10^{-30}\eV^2$.

\end{itemize}

\section{Sterile effects in atmospheric, reactor and beam neutrinos}\label{atm}
In this section we discuss how SK, K2K, MACRO, {\sc Chooz}, {\sc Bugey},
CDHS, CCFR, {\sc Karmen}, {\sc Nomad}, {\sc Chorus}
and future experiments of these kinds probe sterile oscillations.
KamLAND data have been studied in section~\ref{solar},
together with solar data.

\subsection{Technical details}\label{AtmDetails}
It is convenient to compute oscillation probabilities using the neutrino
and anti-neutrino density matrices.
We convert the initial value 
(e.g.\ $\rho(E_\nu) = \diag(\Phi_e(E_\nu), \Phi_\mu(E_\nu),0,0)$
for atmospheric neutrinos)
to the instantaneous mass eigenstate basis, $\rho_m = V^\dagger \rho V$.
In this basis evolution in each medium (air, mantle, core) 
is simply given by a diagonal matrix of phases,
$\mathscr{U} =\diag \exp (-2 i \delta)$
where $\delta_i = m_{\nu_{mi}}^2L/4E_\nu$.
Using
the matrix density in the mass eigenstate basis, we can 
analytically average `fast' oscillations to their mean value 
by appropriately inserting a small imaginary
part $\epsilon$ in the oscillation phases $\delta$.
This is achieved by evolving the matrix density as
$\rho^m_{ij}(L)=\rho^m_{ij}(0) e^{2i \delta_{ij} - \epsilon|\delta_{ij}|}$
where $\delta_{ij} = \delta_i - \delta_j$ and $\epsilon$ is an arbitrary positive small number
(we choose $\epsilon =0.01|\delta_{ij}|$).
This makes  computations much faster than usual techniques
which require lengthy  numerical averages of the oscillation factors.
In the simplest case of vacuum oscillations of 2 neutrinos
this amounts to modify the oscillation factor as
$\sin^2 \delta_{12}\to (1-e^{-|\delta_{12} \epsilon|}\cos 2\delta_{12})/2$.

At the air/mantle and mantle/core boundaries
eigenstates change in a non-adiabatic way:
this effect is accounted by the `level-crossing' flavour matrices $P$
already described at page~\pageref{Psun}.

\smallskip

Our fit of atmospheric data takes into account all most recent results:
SK atmospheric data~\cite{SKatm}, the K2K spectrum and total rate~\cite{K2K},
and MACRO~\cite{MACRO} data about trough-going muons.
These events arise from neutrino scatterings in the rock below the detector:
MACRO is competitive with SK because the important parameter
is the surface of the detector, rather than its mass.
We do not include data from
older atmospheric experiments, which studied essentially 
the same kind of observables better measured by SK.
Data are fitted by forming a global $\chi^2$.
Uncertainties are taken into account following~\cite{SKatm},
and systematically working in Gaussian approximation.
More na\"{\i}ve definitions of the $\chi^2$ (e.g.\ fitting only the
zenith-angle spectra of the single classes of events)
do not give significantly different final results.

\smallskip

Atmospheric data contain the evidence for the atmospheric
anomaly. We assume it is due to $\nu_\mu\to\nu_\tau$ oscillations and
take into account the uncertainty on
$\Delta m^2_{\rm atm}$ and on $\theta_{\rm atm}$
using the same technique developed in section~\ref{sunfit}
for marginalizing 
over $\Delta m^2_{\rm sun}$ and on $\theta_{\rm sun}$.
Even in the atmospheric case, this technique now gives an accurate 
analytical approximation to the usual
active-only atmospheric fit (not shown).
As previously discussed, we assume $\theta_{13}=0$.

Finally, we include $\bar\nu_e$ disappearance data from the {\sc Chooz}~\cite{CHOOZ} (14 bins) and  {\sc Bugey}~\cite{Bugey} (60 bins) reactor experiments.
We include $\nu_\mu$ disappearance data from
CDHS~\cite{CDHS}  (15 bins) and CCFR~\cite{CCFR}  (15 bins).
{\sc NuTeV} data could give additional information,
if certain anomalous features will be understood~\cite{NuTeV}.
Although disappearance experiments give the dominant constraint on sterile effects,
we also include $\bar\nu_\mu\to\bar\nu_e$  data from {\sc Karmen}~\cite{Karmen},
using the likelihood computed by the {\sc Karmen} collaboration
on an event-by-event basis.\footnote{We thank K. Eitel and M. Steidl
for giving us the table of the {\sc Karmen} likelihood,
and B. Louis and
G. Mills for the LSND likelihood.}
LSND results are not included and separately discussed in section~\ref{LSND}.
Finally, we include {\sc Nomad} and {\sc Chorus} data on $\nu_\mu\to\nu_\tau$~\cite{NOMADCHORUS},
which are relevant in the case of $\nu_3/\nu_{\rm s}$ mixing
at $\Delta m^2\sim 10\eV^2$.

\subsection{Results}
\begin{figure}[t]
\vspace{-9mm}
$$\hspace{-5mm}\includegraphics[width=18cm]{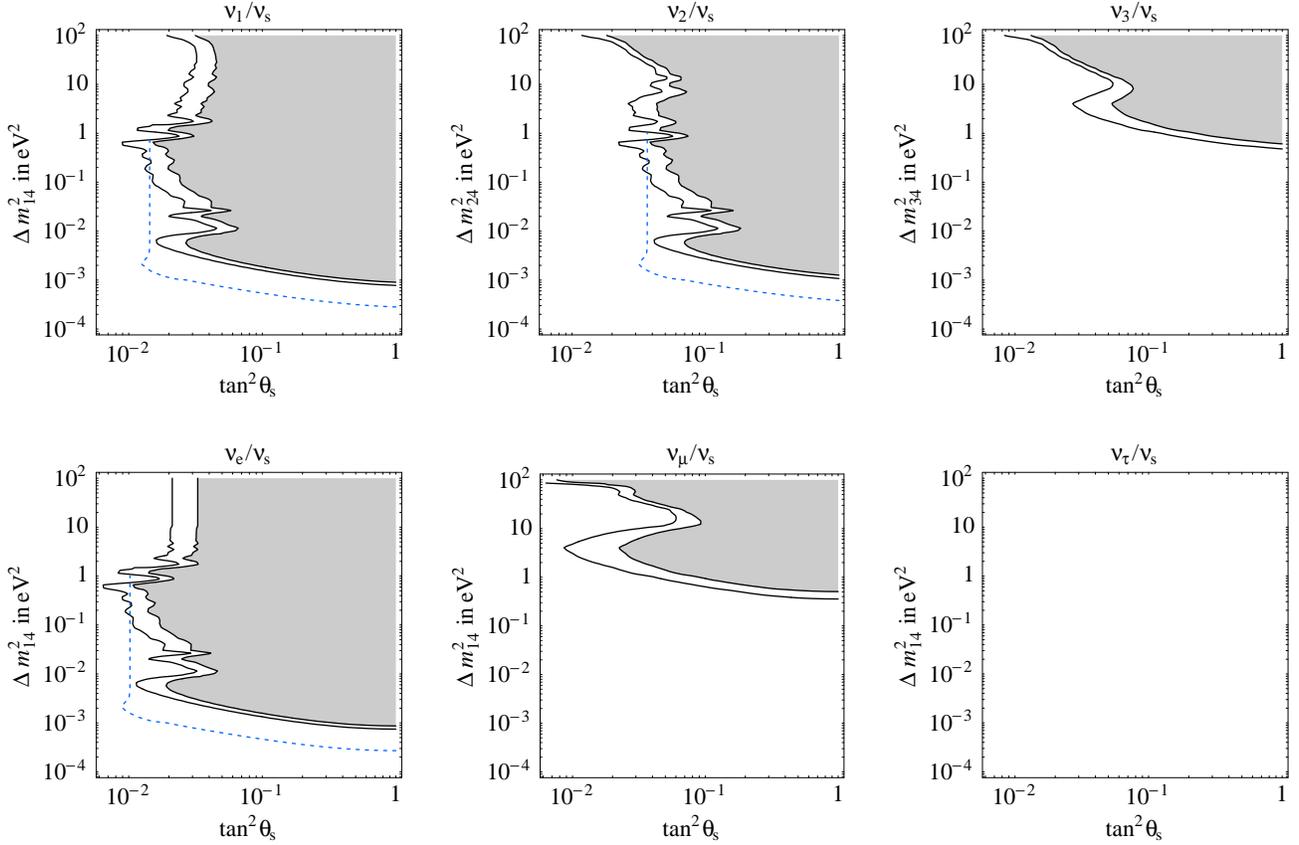}$$
\caption{\label{fig:SBL}\em {\bf Sterile mixing:
effects in short base-line experiments} {\sc Chooz}, {\sc Bugey}, {\sc CDHS}, {\sc CCFR}, {\sc Karmen},
{\sc Nomad}, {\sc Chorus}.
Shaded regions: excluded at $90,99\%$ C.L.
The {\color{blue} blue dashed lines} estimate the region that seems
explorable by a future short-baseline reactor experiment.
The plot is symmetric under
$\tan\theta_{\rm s}\leftrightarrow 1/\tan\theta_{\rm s}$
so that we only show $\tan\theta_{\rm s}\le 1$.}
\end{figure}

\begin{figure}[t]
\vspace{-9mm}
$$\hspace{-9mm}\includegraphics[width=18cm]{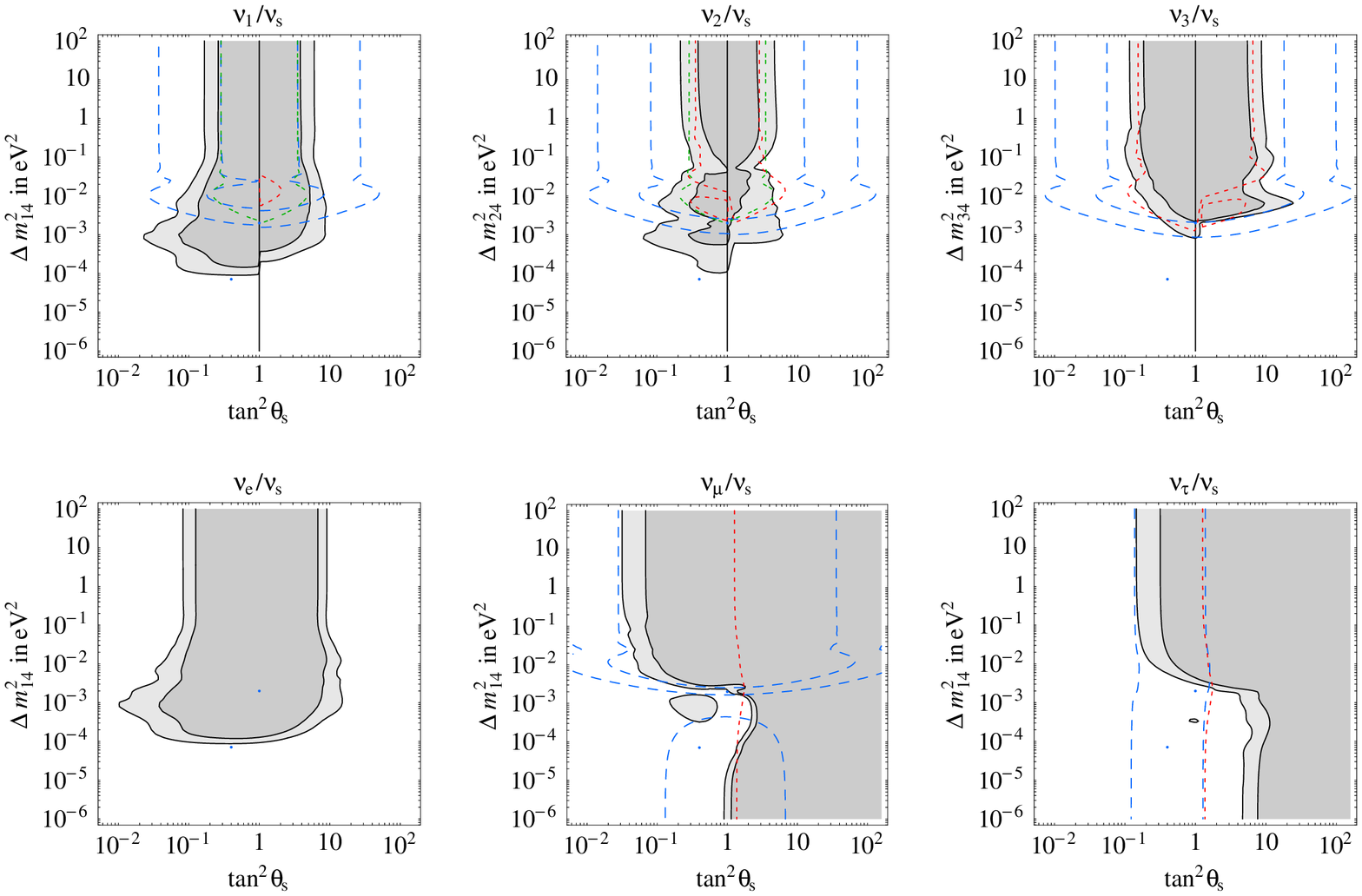}$$
\caption{\label{fig:atm}\em {\bf Sterile mixing:
effects in atmospheric neutrinos (SK, MACRO) and K2K.} 
No statistically significant evidence is found.
Shaded regions: excluded at $90,99\%$ C.L.
Coloured lines are iso-curves of a few promising signals.
{\color{blue} Blue dashed lines: $5\%$ and $1\%$ reduction of the NC rate at MINOS.}
{\color{verdes} Green dot-dashed lines: $P_{e\mu} = 0.01$.}
{\color{rossos} Red dotted lines: $|\Delta P_{\mu\tau}| = 0.01$ at CNGS.}}
\end{figure}

Since there are many relevant experiments,
it is useful to divide them into
two classes, and separately study their impact:
\begin{itemize}
\item[1)] Fig.s\fig{SBL} show the constraints from experiments that are
{\rm not} sensitive to the atmospheric anomaly ({\sc Chooz}, {\sc Bugey}, CDHS, CCFR,
{\sc Karmen}, {\sc Nomad}, {\sc Chorus}).
Sterile oscillation effects in such experiments are simply described by eq.\eq{Pij},
so that we only had to compile their results.
Disappearance experiments provide the dominant constraints.

\item[2)]
Atmospheric neutrinos are a powerful probe of neutrino oscillations
with large mixing angle and $\Delta m^2 \circa{>}10^{-4}\eV^2$.
Fig.s\fig{atm} show how the experiments that see the atmospheric anomaly
(SK, MACRO and K2K) constrain extra oscillations into sterile neutrinos.
\end{itemize}
Analyzing the second class of experiments requires
a non trivial work: these experiments probe
sterile mixing in a significant but indirect way.

The upper limit on the `sterile fraction involved in atmospheric oscillations'
conventionally quoted in the literature corresponds to the following `minimal' configuration:
$\nu_{\rm s}/\nu_{\tau}$ mixing with $\Delta m^2_{34}\gg\Delta m^2_{\rm atm}$.
In this limit, the `sterile fraction' is related to our $\theta_{\rm s}$ angle by
$\eta_{\rm s} = \sin^2\theta_{\rm s}$ and
our analysis reasonably agrees with previous computations.
Atmospheric neutrinos oscillate at a single detectable frequency.
Matter effects become relevant at higher neutrino energies
and suppress active/sterile mixing
without affecting $\nu_\mu\to \nu_\tau$ oscillations.
This effect allows to indirectly discriminate the two channels.
Sterile oscillations are mostly disfavoured by the zenith-angle
spectra of $\mu$-like events with TeV-scale energies.


More general active/sterile oscillation schemes manifest in different ways.
A sterile neutrino mixed with different flavours can give appearance signals.
Active/sterile oscillations with  $\Delta m^2$ comparable to $\Delta m^2_{\rm atm}$
give oscillations at multiple frequencies, distorted by matter effects.
When $\Delta m^2_{34}<0$ matter effects
increase the oscillation length and give a non-standard
energy and path-length dependence of $P(\nu_\mu\to\nu_\mu)$.
SK can best probe such effects with its multi-GeV $\mu$-like sample.
However, SK cannot safely test if an oscillation dip is present.
This kind of studies needs a dedicated future detector~\cite{Monolith}.

The SK collaboration probes sterile effects
in another more direct way: selecting a sample of NC-enriched events~\cite{SKatm}.
In absence of a precise description of the cuts performed
to obtain the NC-enrichment, we have not included this sample
(as also done in previous reanalyses of SK data).
The `minimal' active/sterile configuration
is more strongly constrained indirectly by the zenith-angle spectra.
We estimate that this remains true in most of the parameter space,
and consequently  we do not include in our final results our
 approximate reanalysis of NC-enriched data.

\medskip

As in the solar case, we looked if present data contain some evidence for
sterile effects which correct in a minor way many observables, 
by searching for local minima of the global $\chi^2$.
No statistically significant hint is found: since subleading sterile
effects do not improve the global fit in a significant way
(at most by $\Delta \chi^2\approx 4$) 
our plots only show excluded regions.
The excluded region in fig.s\fig{atm}d, e 
(which correspond to $\nu_\mu/\nu_{\rm s}$ and
to $\nu_\tau/\nu_{\rm s}$ mixing) extends down to
$\Delta m^2_{41}=0$ because even in this limit there are
sterile oscillations at the atmospheric and solar frequencies.

\medskip

It is useful to compare the sensitivity of the two classes of experiments, 1) and 2).
Since there are no MSW resonances,
all these experiments are sensitive only to relatively large sterile mixing,
$\theta_{\rm s} \circa{>} 0.1$.
Sterile mixing with $\nu_e$ (and with the $\nu_1$ and $\nu_2$ mass eigenstates 
that contain a sizable $\nu_e$ fraction)
is better probed by reactor experiments,
although $e$-like events at SK
extend the sensitivity down to smaller values of $\Delta m^2$.
On the contrary atmospheric experiments give more stringent tests
of $\nu_{\rm s}/\nu_\tau$ mixing and of $\nu_{\rm s}/\nu_\mu$ mixing.
Within standard cosmology,
the sterile effects detectable by the experiments discussed in this section
are already disfavoured by measurements of the primordial $^4$He abundancy,
and can be fully tested by future CMB or BBN data.


\bigskip

A detailed analysis of capabilities of future 
beam or reactor neutrinos as probes of sterile neutrinos
seems not necessary.
In fact, there are many proposals motivated by other considerations,
and in each case it is easy to compute sterile effects.
We only make a few general comments.

The blue dashed line in fig.s\fig{SBL} shows what can achieved by
a future high-precision short-baseline reactor experiment
able of detecting a $2\%$ deficit in the $\bar\nu_e$ flux~\cite{SterileReactor}.
Sterile oscillations give a $\bar\nu_e$ deficit,
which might be energy-dependent if the sterile oscillation length at $E_\nu \sim \hbox{few}\MeV$
is comparable to the base-line $L$.
Both are unknown; we assumed $L\sim 2\km$.
Most of the explorable region at small $\Delta m^2$ is already excluded
by solar and atmospheric experiments.
The region with large $\Delta m^2\sim \eV^2$ is more difficult
because even a near detector only sees averaged oscillations;
one has to rely on theoretical predictions for the total flux of reactor $\bar\nu_e$~\cite{ReTh}.

Sterile oscillations with large $\Delta m^2\sim \eV^2$
have a wave-length comparable to the earth radius at energies
$E_\nu \sim \TeV$: experiments such as
AUGER and {\sc IceCube} can study atmospheric neutrinos
in this energy range, and could see an anomalous
zenith-angle dependence if $\theta_{\rm s}$ is large enough~\cite{nuno}.

Future atmospheric experiment such as {\sc Monolith}~\cite{Monolith} could test if
the first oscillation dip is present as predicted by $\nu_\mu\to \nu_\tau$ oscillations.
Sterile effects with $\Delta m^2 \sim \Delta m^2_{\rm atm}$
can significantly distort the expected oscillation pattern.

Distortions not related to earth matter effects
can also be searched by the planned 
$\nu_\mu$-beam experiments.
Furthermore, the MINOS detector can statistically distinguish NC/CC events 
that are tagged as short/ long tracks, and it should be possible to measure the
NC rate with $5\%$ precision~\cite{MinosNC}.
The blue dashed lines in fig.s\fig{atm} correspond to a
$5\%$ and $1\%$ anomaly in the NC total rate
in a $\nu_\mu$ beam experiment with $L=730\km$
and {\sc Minos}-like parent energy spectrum.

A sterile neutrino can also manifest as $\nu_\tau$ appearance:
considering a CNGS-like beam~\cite{CNGS}
in fig.s\fig{atm} we show iso-curves corresponding to 
an average $\nu_\mu\to\nu_\tau$ conversion probability  $\pm 0.01$
different from atmospheric oscillations only
(which give a conversion probability of about $0.02$, depending 
on the precise value of $\Delta m^2_{\rm atm}$ and on the
average energy of the beam).

Constraints from reactor experiments make more difficult,
but not impossible, to have detectable $\nu_e$ appearance
in a $\nu_\mu$ beam (or $\nu_\mu$ appearance in a $\nu_e$ beam)
as signals of active/sterile oscillations.
In experiments sensitive to atmospheric oscillations
a non zero $\theta_{13}$ gives such effects.
Assuming $\theta_{13}=0$,  in fig.\fig{atm} we show iso-curves corresponding to 
a $\nu_\mu\to\nu_e$ conversion probability of $0.01$.

\bigskip

\subsection{Hints and anomalies}
To conclude, we list some possibly anomalous features of present data:
\begin{itemize}
\item The total rate of $e$-like events at SK
is higher than the expected central value.
The $\mu/e$ ratio seems to be too low
(see e.g.~\cite{LoSecco}).
Constraints from solar, atmospheric and short-baseline data show that
sterile neutrinos can reduce the $\mu$ rate
(e.g.\ in the case of  $\nu_\mu/\nu_{\rm s}$ mixing with $\Delta m^2_{14}\sim 10^{-4}\eV^2$)
but cannot increase the $e$ rate.

\item The total rate of trough-going muons at SK
is higher than the expected central value.
Predictions and direct measurements of cosmic ray primaries at
the relevant energies are difficult~\cite{Battistoni}.

\item The LSND result~\cite{LSND}, discussed below.
\end{itemize}

\begin{figure}
$$\includegraphics[width=7cm]{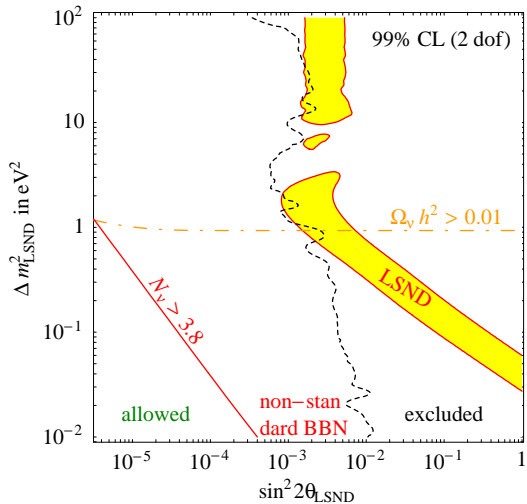}$$
\caption{\label{fig:LSND}\em 
{\bf The LSND anomaly interpreted as oscillations of 3+1 neutrinos}.
Shaded region: suggested at 99\% C.L.\ by LSND.
Black dotted line: 99\% C.L.\ global constraint from other neutrino experiments
(mainly Karmen, Bugey, SK, CDHS).
{\color{rossos} Continuos red line: $N_\nu = 3.8$ thermalized neutrinos.}
{\color{rossoc} Dot-dashed orange line: $\Omega_\nu h^2 =0.01$.}}
\end{figure}

\subsection{LSND}\label{LSND}
The LSND experiment~\cite{LSND} finds an evidence for $\bar\nu_\mu\to\bar\nu_e$,
that ranges between 3 to $7\sigma$
depending on how 
data are analyzed.
It is difficult to conceive new physics that can explain the LSND
result compatibly with all other constraints.
Among reasonably conservative interpretations,
a sterile neutrino with eV-scale mass 
seems to be the less disfavoured possibility~\cite{LSND?,SN3+1,2+2}.
According to this interpretation
the LSND anomaly arises as $\nu_\mu\to\nu_{\rm s}\to \nu_e$
so that the effective $\theta_{\rm LSND}$ $\nu_\mu/\nu_e$ mixing angle
is predicted to be $\theta_{\rm LSND}\approx \theta_{es}\cdot\theta_{\mu s}$.
This formula is valid only for small mixing angles, and eq.\eq{Pij} gives
the general expression.
This prediction gives rise to 3 problems:
\begin{itemize}
\item[1)] $\nu_e$ and $\nu_\mu$ disappearance experiments
imply that $\theta_{es}$ are $ \theta_{\mu s}$ are
somewhat smaller than what suggested by LSND~\cite{LSND?,SN3+1,2+2};
\item[2)] standard BBN predicts that the sterile neutrino thermalizes,
so that primordial abundances should have values corresponding to $N_\nu = 4$~\cite{LSNDBBN};
\item[3)] according to standard cosmology, the sterile neutrino 
gives a contribution to the neutrino density $\Omega_\nu$ 
somewhat larger than what suggested by global fit of CMB and LSS data
(see e.g.~\cite{instant}).
\end{itemize}
Concerning points 2) and 3),
there is not yet general consensus that the sterile neutrino thermalizes,
maybe because this LSND issue has never been analyzed by authors
that performed, at the same time,
a precise study of neutrino data and of cosmology
with mixed neutrinos.
Estimates indicate that the region favoured by LSND lies
well inside the region where the sterile neutrino is thermalized~\cite{LSNDBBN}.
This is confirmed by our analysis, shown in fig.\fig{LSND}.
In the relevant region the constraint on $\Omega_\nu$
is well approximated by the horizontal line
corresponding to $\Omega_\nu h^2 = m_4/93.5\eV$,
as assumed in previous analyses~\cite{instant,2+2}.
The accurately computed constraint starts to be weaker only
at much smaller values of the effective $\theta_{\rm LSND}$ mixing angle.

In fig.\fig{LSND} the BBN constraint has been minimized (when allowed by neutrino data)
setting $\theta_{e\rm s}\approx \theta_{\mu\rm s}\approx \theta_{\rm LSND}^{1/2}$.
We see that short-baseline experiments  sensible to a $P(\nu_\mu\to \nu_e)$ 
about $2$ orders of magnitude smaller than the value suggested by LSND
are needed to probe regions compatible with standard BBN.

\section{Summary}\label{all}
A few years ago active/sterile oscillations were studied as an alternative to active/active oscillations.
For example, 
it was shown that neither the solar 
nor the atmospheric anomaly can be produced by oscillations into sterile neutrinos
compatibly with standard BBN.
These active vs sterile issues  have now been firmly solved by experiments, 
and the new relevant questions become:
\begin{quote}
How large can be the subdominant sterile component possibly present in
solar or atmospheric oscillations?
How can we discover new anomalies due to sterile effects?
\end{quote}
In order to address these issues we systematically  compared oscillation effects
generated by one sterile neutrino 
(including the effects of the now established solar and atmospheric oscillations)
with present experiments
and studied capabilities of future probes,
extending previous analyses in several ways.
Almost all previous analyses that studied these new questions considered
only the peculiar sterile oscillation pattern that gives the simplest physics:
the sterile neutrino mixes with a specific flavour 
($\nu_{\mu,\tau}$ in solar analyses, $\nu_\tau$ in atmospheric ones)
and has a large mass, $\Delta m^2\gg \Delta m^2_{\rm atm,sun}$.
Dropping each one of these assumptions gives quite different physics.
We explored the whole parameter space
(three sterile mixing parameters and one sterile mass)
producing precise results for six representative slices, 
that span the spectrum of the various possibilities.
We considered one extra sterile neutrino with arbitrary mass $m_4$, and allowed it to mix
with
$$\nu_e\quad\hbox{or}\quad \nu_\mu\quad\hbox{or}\quad\nu_\tau\quad\hbox{or}\quad
\nu_1\quad\hbox{or}\quad\nu_2\quad\hbox{or}\quad\nu_3$$
where $\nu_{1,2,3}$ are the mass eigenstates in absence of sterile mixing.
The spectrum of active neutrinos is not yet fully known:
rather than studying all possible cases we focussed
on what we believe is most plausible case:
we assumed normal hierarchy of active neutrinos (i.e. $m_1\ll m_2\approx m_3/6$)
and that $\theta_{13}$ is small enough that we can neglect its effects.
If experiments will contradict these assumptions, it will be easy to update
our results; most of them do not depend on these assumptions.

\medskip

\begin{figure}[p]
$$\hspace{-7mm}\includegraphics[width=18cm]{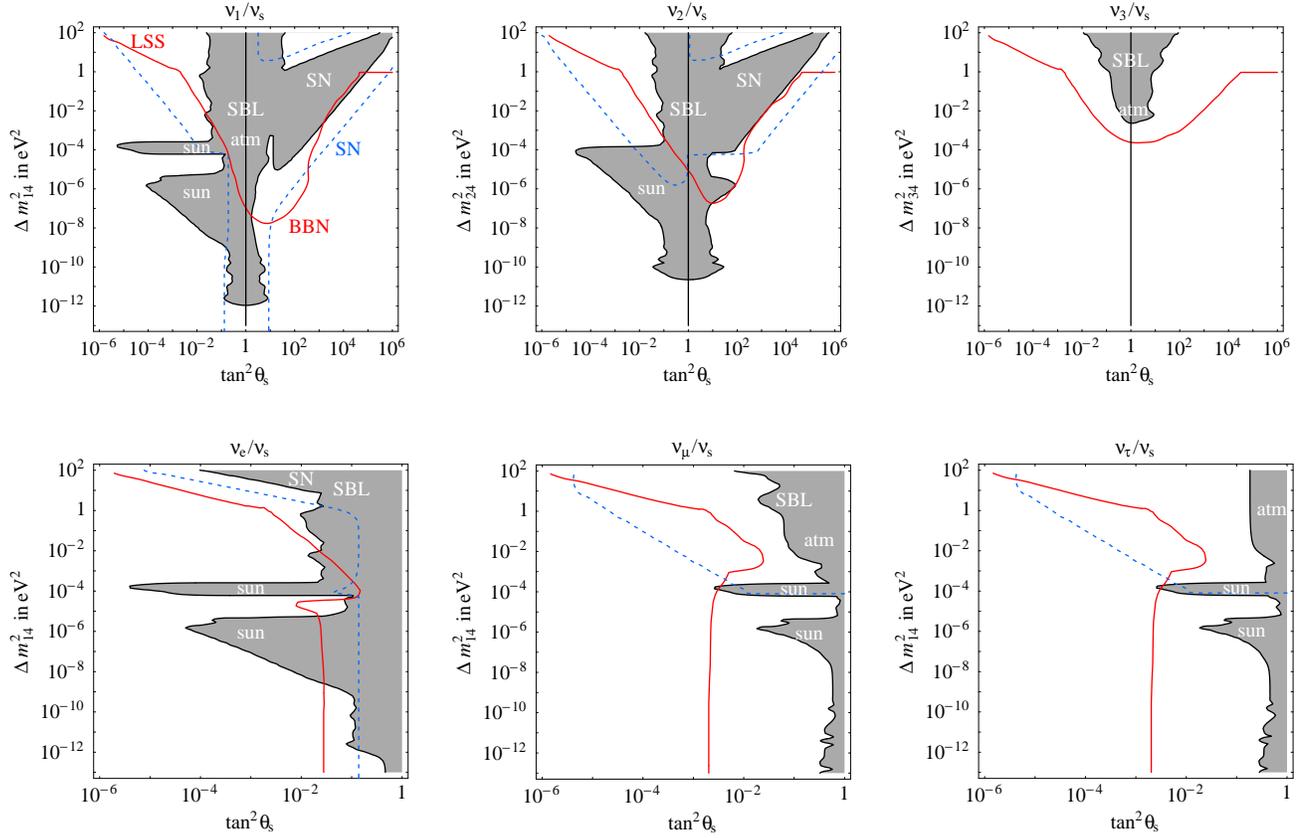}$$
\caption{\label{fig:all}\em {\bf Summary of sterile neutrino effects}.
The shaded region is excluded at $99\%$ C.L.\ (2 dof)
by solar or atmospheric or reactor or short base-line experiments.
We shaded as excluded also regions where sterile neutrinos
suppress  the SN1987A $\bar\nu_e$ rate by more than $70\%$.
This rate is suppressed by more than $20\%$ inside the {\color{blue} dashed blue line},
that can be explored at the next SN explosion if it will be possible to
understand the  collapse well enough.
Within standard cosmology,
the region above the {\color{rossos} red continuous line} is disfavoured (maybe already excluded) by BBN and LSS.
Plots in the various sections show how much each probe can be improved
by future experiments.}
\end{figure}

We considered the most promising ways to probe the existence of 
eV-scale sterile neutrinos.
Most probes are based on a careful study of
natural sources of neutrinos (the universe, the sun, supernov\ae, cosmic rays,...)
which have their own peculiar capabilities and limitations.
We studied how one extra sterile neutrino affects BBN (helium-4, deuterium), CMB, LSS and solar, atmospheric, reactor, beam experiments.
The sensitivity of some of these probes is enhanced by MSW resonances~\cite{MSW}.
In cosmology, active $\nu$ and $\bar\nu$ encounter a 
MSW resonance with sterile neutrinos lighter than active ones.
Roughly the same happens to supernova $\bar\nu_e$
(that also experience less important MSW resonances in the opposite situation).
On the contrary, solar $\nu_e$ encounter a MSW resonance with sterile neutrinos heavier than
active ones.

Fig.s\fig{all} combines present constraints.
Each probe is described in greater detail in its specific series of figures:
fig.s\fig{BBN} for cosmology,
fig.s\fig{sun} for solar experiments, 
fig.s\fig{SN} for supernov\ae, 
fig.s\fig{SBL} for short base-line experiments,
fig.s\fig{atm} for atmospheric experiments.
These figures also show the capabilities of some future experiments,
that we now try to summarize in words.
\begin{itemize}
\item Compatibility with standard BBN constrains 
sterile oscillations occurred at temperatures $T\circa{>} 0.1\MeV$.
It is very important to
improve measurements of the {\bf helium-4} primordial abundancy
(that we parameterize in terms of an effective number of neutrinos $N_\nu^{^4{\rm He}}$,
see eq.~(\ref{sys:HeD}))
until $N_\nu^{^4{\rm He}}=4$ will be safely tested.
This requires overcoming `systematic' uncertainties.
The helium-4 abundancy is sensitive to two different sterile effects: 
increase of the total neutrino density, and depletion of electron-neutrinos.
The second effect makes 
$N_\nu^{^4{\rm He}}$ 
sensitive to sterile  oscillations down to $\Delta m^2\sim 10^{-8}\eV^2$,
while the first effect becomes negligible at $\Delta m^2\circa{<} 10^{-5}\eV^2$.

However, if BBN were non-standard, a modified density of electron neutrinos
could compensate the sterile corrections to $N_\nu^{^4{\rm He}}$:
for example  helium-4 constraints on sterile oscillations can be evaded by
allowing a  large neutrino asymmetry.

\item For all these reasons it is important to measure a second BBN effect.
The {\bf deuterium} primordial abundancy is affected by milder systematic problems:
in the future it might be possible to improve its measurement obtaining an
uncertainty on the effective parameter $N_\nu^{\rm D}$ (precisely defined in eq.~(\ref{sys:HeD})))
significantly below 1, 
possibly making deuterium the most significant BBN probe.
We have computed the ranges of active/sterile oscillation parameters that significantly affect $N_\nu^{\rm D}$:
it is less sensitive than helium-4 to $\nu_e$ depletion and therefore
to  values of $\Delta m^2$ below $10^{-5}\eV^2$  (fig.\fig{BBN}).

\item Future studies of {\bf Cosmic Microwave Background} acoustic oscillations
should allow to precisely measure the  total neutrino density $N_\nu^{\rm CMB}$
at recombination ($T\sim \eV$) with $\pm0.2$ ({\sc Planck}) or
 maybe $\pm0.05$ ({\sc CMBpol}) error~\cite{NuCDM}.
Neutrinos affect CMB in various ways;
neutrino free-streaming offers a clean signature that allows to count neutrinos.
Sterile neutrinos affect  $N_\nu^{\rm CMB}$ only if  $\Delta m^2\circa{>}10^{-5}\eV^2$;
in such a case
$N_\nu^{\rm CMB}\approx N_\nu^{^4{\rm He}}\approx N_\nu^{\rm D}$.
Therefore CMB will cover only a part of the region that 
BBN could probe with fully reliable measurements of
the helium-4 or deuterium abundances.


\item CMB also allows to probe $\eV$-scale $\nu$ masses.
Smaller $\nu$ masses can be probed by measuring how much galaxies are clustered, 
because neutrinos become non relativistic when the observable universe had a size comparable
to present cluster of galaxies.
Relativistic neutrinos freely move and tend to
reduce the amount of clustering.
Recently,  {\bf Large Scale Structure} data (together with precise CMB measurements)
gave a bound on the present energy density in neutrinos
$\Omega_\nu < 0.0076$ at 95 \% C.L.~\cite{WMAP,boundMnu}, dominated by neutrino masses
(rather than neutrino energy).
With only active neutrinos this implies a significant bound on their mass, $m_\nu < 0.23\eV$~\cite{WMAP,boundMnu}.

Sterile neutrinos contribute to $\Omega_\nu$
(a small non-thermal population of relatively heavy sterile neutrinos also
modifies the way $\Omega_\nu$ manifests).
For small active/sterile mixing
LSS constrains sterile oscillations more strongly than BBN (fig.\fig{BBN}).

\item {\bf Solar} $\nu_e$ experiments have explored sterile oscillations not testable by BBN,
thanks to two different effects.
(1) MSW resonances make solar $\nu_e$ sensitive to small active/sterile mixing
and $\Delta m^2\circa{>}10^{-8}\eV^2$.
(2) With large mixing, solar $\nu_e$ are sensitive down to $\Delta m^2\sim 10^{-12}\eV^2$.
Future experiments will explore new aspects of the solar neutrino anomaly,
allowing to measure in a redundant way the active oscillation parameters
or to discover a new anomaly.
We emphasize one qualitative point.
Due to  LMA oscillations, neutrinos exit from the sun as almost pure $\nu_2$
at energies $E_\nu  \circa{>}  \hbox{few MeV}$.
Neutrinos with these energies have been precisely studied by SK and SNO,
but are almost unaffected by sterile oscillations if they involve mostly $\nu_1$.
This could happen either when $\nu_{\rm s}$ mixes with $\nu_1$,
or when $\nu_{\rm s}$ experiences a level-crossing with $\nu_1$.
Therefore {\em there is a whole class of sterile effects which 
manifest only at $E_\nu \circa{<}  \hbox{\rm few MeV}$} ---
an energy range explored so far only by  Gallium experiments.
Future precise measurement of solar $\nu_e$ at sub-MeV energies
will allow to significantly
extend searches for active/sterile effects.
Part of these extended region can be soon 
tested by Borexino, where a sterile neutrino can manifest 
as day/night variations, or as seasonal variations,
or even by reducing the total rate.

\item {\bf Supernova} neutrinos 
will be good probes of sterile oscillations because
have a different pattern of MSW resonances
and a longer base-line than solar $\nu_e$.
Consequently supernova $\bar\nu_e$ are more sensitive than solar $\nu_e$
in two main cases: 
(a) small $\Delta m^2\circa{>}10^{-18}\eV^2$ with large $\theta_{\rm s}$;
(b) $\nu_{\rm s}$ lighter than $\nu_1$ with small mixing.
Oscillations into one sterile neutrino can reduce the $\bar\nu_e$ rate by up to $80\%$
(see fig.s\fig{SN})
and, in a more restricted range of oscillation parameters,
vary the average $\bar\nu_e$ energy by $30\%$.
SN1987A data agreed with expectations.
Future SN experiments can perform quantitative test, but
it is not clear how to deal with theoretical uncertainties.
We also discussed other less promising astrophysical probes.

\item {\bf Atmospheric} experiments (SK, MACRO, K2K) indirectly exclude
active/sterile oscillations with $\Delta m^2\circa{>}10^{-3\div 4}\eV^2$
and $\tan^2\theta_{\rm s}\circa{>}0.1\div0.2$.
Up to minor differences, this applies to all flavours (fig.\fig{atm}).
{\bf Terrestrial experiments} that mainly probed disappearance of
$\bar\nu_e$ and $\nu_\mu$ ({\sc Chooz}, CDHS,\ldots)
exclude active/sterile  mixings with these flavours with $\tan^2\theta_{\rm s}\circa{>}0.03$
and $\Delta m^2\circa{>}10^{-3}\eV^2$  (fig.\fig{SBL}).
Therefore future short-baseline experiments can search for sterile effects with smaller $\theta_{\rm s}$.
Possible signals are $\bar\nu_e$ disappearance in reactor experiments,
a deficit of NC events or $\nu_\tau$ appearance in beam experiments.
Within standard cosmology these effects can be probed by CMB and BBN,
which already disfavour them.

\end{itemize}
We listed present anomalies that can be interpreted as due to sterile neutrinos.
None looks particularly significant, with the possible exception of
the LSND anomaly, that can be due to oscillations of `$3+1$' neutrinos.
Our precise study confirms that the extra sterile neutrino suggested by LSND
thermalizes almost completely before BBN
(in agreement with estimations of most previous analyses).
In fig.\fig{LSND} we plot, as function of the effective LSND oscillation parameters
$\theta_{\rm LSND}$ and $\Delta m^2_{\rm LSND}$, the constraints on
$N_\nu^{^4{\rm He}}\simeq N_\nu^{\rm D}\simeq N_\nu^{\rm CMB}$ 
(mainly from helium-4 data) and $\Omega_\nu$ (mainly from LSS data):
in the relevant region our precise plot negligibly differs from previous estimates~\cite{instant,2+2}.
The 3+1 interpretation of LSND is not compatible with standard BBN, 
and gives a $\Omega_\nu$ which is only marginally compatible with standard cosmology.

\bigskip

\paragraph{Acknowledgments}
We thank S.~Woosley, A.~Burrows, P.~Di Bari, S.~Sarkar, S.~Pascoli, A.~Romanino, K.~Olive.
The work of M.C. is supported in part by the USA department of energy 
under contract DE-FG02-92ER-40704. 
Part of the work of M.C.\ was done while at scuola normale superiore (Pisa).
The authors thank the CERN  theory division, where part of the work was done.

\footnotesize
\begin{multicols}{2}
  
\end{multicols}


\begin{thebibliography}{99}
\bibitem{sterileshock}
\art{J.T. Peltoniemi}{Astronomy and Astrophysics}{254}{121}{1992} first identified the region of large $Delta m^2_s$ (now excluded by the upper bound on the masses by cosmology) 
where the effect on the shock can be positive.
\bibitem{CNGS}
The CNGS project is described at its web site,
proj-cngs.web.cern.ch/proj-cngs.
\end{thebibliography}
\end{document}